\documentclass[11pt, preprint]{aastex}
\usepackage{amsmath}

\usepackage{color} 
\graphicspath{{Classify/}{Plot_sec_4/}{Sec2_Fig/}{Sec3_Fig/}{Fig_fit/}{Sec4/}{KOI868/}{KOI148/}{Upper_KOI97/}{Upper_limit_fig/}{new_fig/}}
\usepackage{pdflscape} 
\usepackage{longtable}
\usepackage{lscape} 

\begin{document}


\shorttitle{Systematic Search for Saturn-like Rings around the {\it Kepler} Planet Candidates}
\shortauthors{Aizawa et al.}



\title{Systematic Search for Rings around {\it Kepler} Planet Candidates: \\
Constraints on Ring Size and Occurrence Rate} 



\author{
Masataka \textsc{Aizawa},\altaffilmark{1}
Kento \textsc{Masuda},\altaffilmark{2,3}
Hajime \textsc{Kawahara},\altaffilmark{4,5} \\
and 
Yasushi \textsc{Suto}\altaffilmark{1,5}
} 


\altaffiltext{1}{Department of Physics, The University of Tokyo, 
Tokyo, 113-0033, Japan}
\altaffiltext{2}{Department of Astrophysical Sciences, Princeton University, Princeton, NJ 08544, USA}
\altaffiltext{3}{NASA Sagan Fellow}
\altaffiltext{4}{Department of Earth and Planetary Science, 
The University of Tokyo, Tokyo 113-0033, Japan}
\altaffiltext{5}{Research Center for the Early Universe, School of Science,
The University of Tokyo, Tokyo 113-0033, Japan}

\bibliographystyle{apj}

\begin{abstract}
We perform a systematic search for rings around 168 {\it Kepler} planet candidates
with sufficient signal-to-noise ratios that are selected from all the
short-cadence data.  We fit ringed and ringless models
to their lightcurves, and compare the fitting results to search for the
signatures of planetary rings.  First, we identify 29 tentative systems,
for which the ringed models exhibit statistically significant improvement over the
ringless models.  The lightcurves of those systems are individually
examined, but we are not able to identify any candidate that indicates
evidence for rings. In turn, we find out several mechanisms of
false-positives that would produce ring-like signals, and the null
detection enables us to place upper limits on the size of rings.
Furthermore, assuming the tidal alignment between axes of the planetary
rings and orbits, we conclude that the occurrence rate of 
rings larger than twice the planetary radius is less than 15 percent. 
Even though the majority of our targets are short-period planets, 
our null detection provides statistical
and quantitative constraints on largely uncertain theoretical models of
origin, formation, and evolution of planetary rings.
\end{abstract}



\section{Introduction}

One of the numerous breakthroughs that Galileo Galilei achieved with his
own telescope in 1610 is the discovery of Saturn's ``ears''. First he
thought that Saturn is a three-body system, but later he was very much
confused of the interpretation of his discovery. Even after Christiaan
Huygens correctly pointed out in 1655 that Saturn has a ring, the nature
and origin of the ring has remained largely unknown. Nevertheless the
rings of Saturn have attracted people over many generations. Also, many small 
rings have been identified for Jupiter, Uranus, Neptune, Chariklo 
\citep{2014Natur.508...72B}, Chiron \citep{2015A&A...576A..18O}, and 
Haumea \citep{2017Natur.550..219O}. Therefore, the presence of rings is now
supposed to be fairly universal in the Solar system. 

This naturally raises the question ``Are planetary rings also common in
planetary systems outside our Solar system?''  For more than 20 years
since the discovery of an exoplanet around a Sun-like star, 
photometric and spectroscopic accuracies of observations have
significantly improved, and we are now potentially in a position to answer
the question in a quantitative and statistical manner.  Indeed the
precise photometry with the {\it Kepler} mission has already reached the
sensitivity to detect transiting ringed planets if any
\citep[e.g.][]{2004ApJ...616.1193B,2009ApJ...690....1O,2017AJ....153..193A}.

Several observational techniques have been proposed for the detection
and characterization of exoplanetary rings. Lightcurves of transiting
ringed planets should leave characteristic signatures that cannot be
produced by ringless planets \citep{1999CRASB.327..621S}.  Reflection
light from rings just before and after the transits is also identifiable
in principle \citep[]{2004A&A...420.1153A,2005ApJ...618..973D}.  The
spectroscopic Rossiter-McLaughin effect can be used to increase the
reliablity of the ring-like photometric signal candidates
\citep{2009ApJ...690....1O}.  Anomalous stellar density and planetary
radii may select the possible candidates for ringed planets
\citep{2015ApJ...803L..14Z}.

In addition to those theoretical proposals, there are several previous
attempts to search for rings and/or put constraints on their parameters
from real data. For instance, \cite{2001ApJ...552..699B} analyzed 4
transit lightcurves of the first transiting system, HD 209458, with the
Hubble Space Telescope (HST), and concluded that HD 209458b cannot be
accompanied by an opaque ring with its radius exceeding 1.8 times the
planetary radius.

 \cite{2015A&A...583A..50S} pursued the possibility that the anomalously
large reflection light and rotational velocity of Peg 51 b indicates the
presence of a ring, but the required configuration for such ring systems
was found to be unstable due to the strong tidal interaction with the host
star. Therefore the ring interpretation of Peg 51 b is excluded.
\cite{2017A&A...603A.115L} searched for rings around the long-period
exoplanet CoRoT-9b ($P=95.3$ days) using the Spitzer photometry. They
did not find any signatures of a ring, and instead derived constraints
on the inclination of a possible ring. \cite{2018NewA...60...88H} tested 
the ring hypothesis for one of the longest-period Kepler planets KOI-422.01, and 
they excluded the possible rings with obliquity angles $90^{\circ}$, $60^{\circ}$, 
$45^{\circ}$, or $20^{\circ}$. 

 Furthermore, there are several observational claims of possible circumplanetary rings or disks on the basis of the transit method or direct imaging. \cite{2008Sci...322.1345K} interpreted the anomalously large optical flux of Fomalhaut b in terms of a possible circumplanetary disk. \cite{2012AJ....143...72M} found a series of interesting photometric variations during a single transit of a sub-stellar object orbiting around J1407, which can be explained by a gigantic planetary ring ($\sim$ 1 au). \cite{2017MNRAS.471..740O} also found the similar features that repeated during two eclipses of an object around PDS 110, 
which are interpreted as a giant ring. \cite{2017ApJ...850L...6A} claimed 
a marginal signal (4$\sigma$ level) at 1.5 au around Proxima Centauri 
with ALMA. Among several possibilities, one interesting scenario is a planet 
with a sufficiently large ring. 

Instead of the constraints on the specific exoplanets, there are a
couple of systematic attempts to search for rings around transiting
planets from the {\it Kepler} archive data.  \cite{2015ApJ...814...81H}
examined 21 {\it short-period} planets with 1 ${\rm day} <P< 51$ days, but found
no plausible candidate.

\cite{2017AJ....153..193A}, on the other hand, focused on the 89 {\it
long-period} planets ($P>200$ days for most systems) and planet candidates that exhibit up to three transits so as to search for Saturn-like icy rings.  They
discovered one possible candidate whose anomaly of the lightcurve during
a single transit is consistent with the signature of a ring similar to
that of Saturn (but also consistent with a binary-planet model, and a
circumstellar disk around a dwarf star if the host star is a giant star; see
\cite{2017AJ....153..193A} for further discussion).  Unfortunately the
orbital period of the system is fairly uncertain because of the lack of
the multiple transits, and thus the follow-up observation is
very challenging.

The important lesson learned from those early attempts, however, is the
encouraging fact that the detection of rings around exo-planets, if any,
is close to within reach even though not yet easy obviously. Therefore we decide
to extend our previous search to all {\it Kepler} transiting
planets with sufficiently high photometric accuracy in their 
short-cadence data.

More specifically, we select 168 {\it Kepler} planet candidates with high
signal-to-noise ratios using the short-cadence data, so that we are able
to probe tiny and short-duration characteristic signatures of rings.
Because of those selection criteria, majority of our targets turned out
to be short-period planets. Thus our survey is preferentially designed
for rocky, instead of icy, rings in practice, but we can test the
robustness of possible ring signatures at separate transit epochs.  From
this point of view, the present work is very complementary to our
previous work \citep{2017AJ....153..193A}, and regarded as a significant
extension of \cite{2015ApJ...814...81H}.

While we believe that some fraction of exoplanets should accompany
rings, the required condition and the nature of those rings are largely
unknown both theoretically and observationally. Even though we have not
identified any candidate for a ringed planet in the analysis of the
present paper, we found several cases that mimic signature of rings,
which are useful examples of false-positives for future ring
searches. Also we are able to constrain the ring parameters from our
null results for the targets.
Our statistical and observational constraints would add insights into
the origin and evolution of rings in a completely different environment
than those in our Solar system. The approach of our current methodology
will eventually answer the question to what extent our Solar system is a
typical (or atypical) planetary system in the Galaxy, hopefully
affirmatively.

The rest of the paper is organized as follows. Section 2 describes our
selection of target planets. Section 3 explains the data reduction
and analyses of lightcurves with transiting ringless or ringed planets
in detail.  Section 4 presents the results and implications of our
analysis.  Finally, Section 5 concludes and discusses the future
prospects for exoplanetary ring search.

\section{Target selection \label{method}}

Since signatures of planetary rings are tiny, we have to carefully
select target systems with sufficient signal-to-noise ratios for
detailed analysis before performing a time-consuming individual
analysis. We adopt the signal-to-noise ratio $(S/N)$ of transiting
systems as a measure of a rough potential detectability of their rings:
\begin{equation}
\label{mes_planet}
(S/N) = \sqrt{\frac{T_{\rm obs}}{P_{\rm orb}}}
\frac{ \delta_{\rm TD}}{ \sigma_{\rm TD}}.
\end{equation}
In the above equation, $T_{\rm obs}$ is the total duration of the
observed lightcurve in the short-cadence data (1 month $\leq T_{\rm obs} \leq $ 4 years), 
$P_{\rm orb}$ and $\delta_{\rm TD}$ denote 
the the orbital period and transit depth, 
and finally $\sigma_{\rm TD}$ is the effective uncertainty
of the data on the transit depth. To estimate $\sigma_{\rm TD}$, 
we interpolate or extrapolate the photometric uncertainty corresponding to the 
transit duration $\tau_{\rm TD}$ using values of the robust root-mean 
square (RMS) combined differential photometric precision (CDPP) 
in the {\it Kepler} Stellar Table. 

In the present paper, we focus on the {\it Kepler} short-cadence (1 min)
data alone. The long-cadence data (29.4 mins) are not suitable for
searching for signatures of rings, which are identifiable only for short
timescales around the egress and ingress of the transit.  We first
retrieve parameters from the Q1--Q17 Data Release 25 catalog of all {\it
Kepler} Objects of Interests (KOIs) \citep{2017arXiv171006758T}, and
calculate $(S/N)$ of those KOI planets that have short-cadence data.  We
exclude the systems whose dispositions are ``FALSE POSITIVE'' in the
catalog.

The total duration $T_{\rm obs}$ corresponds to the observed duration of
the system in the {\it Kepler} short-cadence data. Roughly speaking,
$(S/N)=1$ corresponds to the 1$\sigma$-detection of the transit of a
planet, not of a planetary ring.  Since a typical amplitude of the photometric
anomaly due to a Saturnian ring is less than 1 percent of the
planetary transit depth, we select all {\it Kepler} planet
candidates with $(S/N)>100$ as our targets.

\begin{figure}[htpb]
\begin{center}
\includegraphics[width = 0.68 \linewidth]{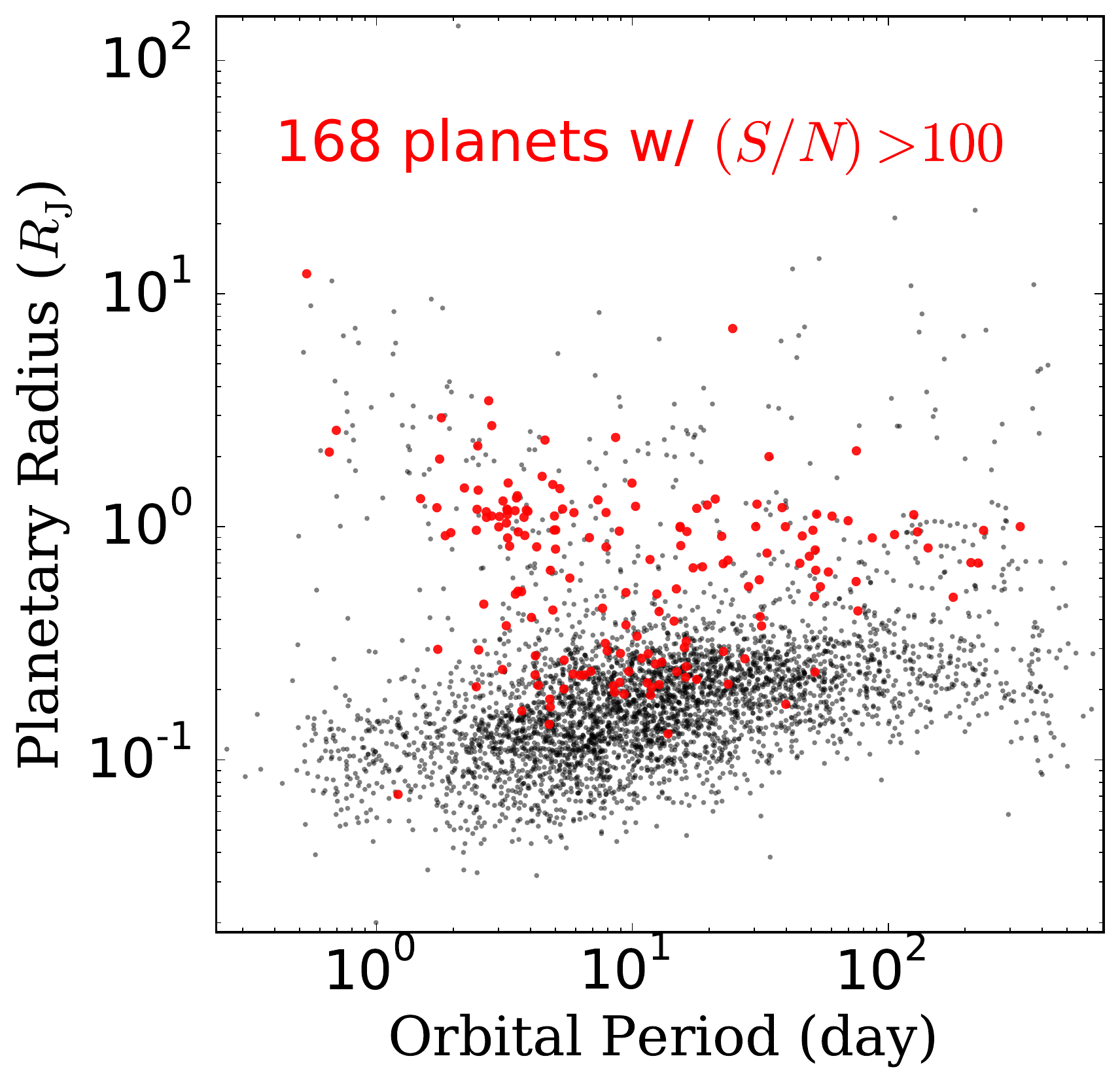}
\caption{Planetary radii of 4029 KOIs against their orbital periods.
Red points indicate the 168 targets with $(S/N)>100$ that are examined
 closely in the present paper. } 
\label{period_mess_rad}
\end{center}
\end{figure}

Orbital periods and planetary radii of all 4029 KOIs with short-cadence data 
are shown in Figure \ref{period_mess_rad}.  The majority of the KOI planets
have insufficient $(S/N)$ to detect possible rings, and 168 KOI
planets satisfy $(S/N)>100$ (plotted in red circles). We note that our targets include all systems  in \cite{2015ApJ...814...81H} except for KOI-398.02 with $(S/N)= 97.1$ (20 out of 21).

\section{Ring survey method: 
data reduction and fits of ringless and ringed planet models }
 \label{sec_data}

This section describes our analysis method of ring survey, including
lightcurve data reduction and fit to the parametrized templates of a planet
with and without a planetary ring. The method is largely based on 
our previous paper \citet{2017AJ....153..193A}. 

We approximate the stellar intensity profile $I(x,y)$ with $(x,y)$ being
the coordinates with respect to the stellar center as
\begin{equation}
I(x,y)\propto
[1-2 q_{2}\sqrt{q_{1}}(1-\mu) - \sqrt{q_{1}}( 1 - 2 q_{2})
(1-\mu)^{2}],
\end{equation}
where ($q_{1}$, $q_{2}$) are limb-darkening parameters, $\mu =
\sqrt{1-(x^{2} + y^{2})/R_{\star}^{2}}$, and $R_{\star}$ is the stellar
radius \citep{2013MNRAS.435.2152K}. 
Throughout the present analysis, we adopt the circular orbit of
all the planets for simplicity.

Our ringless planet model is specified by seven parameters: the planet
to star radius ratio $R_{\rm p}/R_{\star}$, the impact parameter $b$,
the semi-major axis normalized by the stellar radius $a/R_{\star}$, the
time of a transit center $t_{\rm 0}$, limb-darkening parameters $q_{1}$
and $q_{2}$, and the normalizing factor of the light curve $c$.

Our ring model is specified by additional five parameters; inner ring
radius $R_{\rm in}$, outer ring radius $R_{\rm out}$, shading rate $T$,
and orientation angles for ring axes $\theta$ and $\phi$. If $T=1$, a
ring is fully opaque, and if $T=0$, the ring is completely
transparent. To increase the efficiency of numerical fitting, we employ
$r_{\rm out/in} = R_{\rm out}/R_{\rm in}$ and $r_{\rm in/p} = R_{\rm
in}/R_{\rm p}$, instead of $R_{\rm out}$ and $R_{\rm in}$.  Thus our
ringed planet model is specified by 12 parameters in total.  Further
details of the model are found in \cite{2017AJ....153..193A}.

\subsection{Making phase-folded lightcurves \label{phase_fold}}
If a transiting planet has a ring, the ring signature should be
imprinted equally in each lightcurve at different transit epochs. Since, the
ring parameters, in particular the orientation angles of the ring, are
supposed not to vary for the timescale of $T_{\rm obs}$, the
signal-to-noise ratio of the signature should increase by stacking all
the lightcurves properly.  To produce such precise phase-folded
lightcurves requires an accurate determination of both the transit
center and baseline of each lightcurve at different transit epochs.

We use the short-cadence Pre-search Data Conditioned Simple Aperture
Photometry (PDC-SAP) fluxes of the target objects from the Mikulski
Archive for Space Telescopes (MAST).  We adopt the transit model $F(t)$
implemented by the Pytransit package \citep{2015MNRAS.450.3233P} for
transiting ringless planets, which generates the lightcurve based on the
model of \cite{2002ApJ...580L.171M} with the quadratic limb darkening
law.

We first apply the ringless model separately to each transit by varying
the transit centers and baseline functions alone. Here, we take the
fourth-order polynomials as the baseline functions, and we retrieve the
transit duration and the other parameters of transiting planets from the
MAST pipeline with the help of the Python interface {\it kplr}
(http://dan.iel.fm/kplr/). We extract the lightcurve during the epoch of
$\pm 2$ times the transit duration with respect to each transit center
for the subsequent analysis.

After fitting, we exclude outliers exceeding $5\sigma$ amplitude in the
flux so as to determine the baseline of the lightcurve accurately.  We
repeat the fitting procedure and removal of outliers until no outliers
are left. Then we visually check each transit in order to exclude
inappropriate transits that may be strongly affected by instrumental
systematics. 

 Several transits exhibit large transit timing variations, which our
pipeline cannot automatically deal with. In such cases, we appropriately 
choose the initial transit centers before fitting so as to correctly
identify the transits. Finally, we obtain the best baseline using the
out-of-transit (outside $\pm 0.6\times$ transit duration around the
transit center) data alone, and normalize the lightcurve with the fitted
baseline. Our fit to the transit model lightcurve is performed with the
public code {\it mpfit} \citep{2009ASPC..411..251M} that is based on the
Levenberg-Marquardt (LM) algorithm. 

We stack the obtained normalized lightcurve at each transit, and make
the phase-folded lightcurve.  We derive the transit duration by applying
the ringless model to the phase-folded lightcurve.  With the updated
transit duration, we repeat the above procedure to obtain the final 
phase-folded lightcurve.

We extract the phase-fold lightcurve during an epoch within $\pm 1$
transit duration around the transit center. To finish the fitting procedures in 
realistic time, the lightcurve is divided into 500 bins with an equal time interval. 
Here, we require one bin to accommodate at least 10 points to guarantee
the appropriate binning. So, for systems with the number of the 
phase-folded data less than 5000, we choose the bin width for 
one bin to have 10 data points. 

Finally, we have phase-folded lightcurves for 168 planets, which are
analyzed for ring search in the next subsection.

\subsection{Separate fitting to planetary solutions 
with and without a ring \label{optim_sol}}

Our search for ring signatures is based on the comparison between the
separate best solutions for a planet with and without a ring for all our
targets.

In order to find the best solution in the 7 parameter space for a
ringless planet model, we randomly generate 1000 different initial sets of
parameters from the homogeneous distribution in a finite range.  Then, we
use the LM method to find the local minima starting from each of initial
values, and we choose the best solution among the solutions.  In
fitting, we use the binned data that are produced in Section
\ref{phase_fold}.  We confirm that generally 100 initial sets of
parameters are sufficient to find the minimum for our purpose.

Finally, we calculate the chi-squared value:
\begin{equation}
\chi_{\rm ringless}^{2} = \sum_{\rm i}( [d(t_{\rm i}) 
- m(t_{\rm i})]/\Delta d(t_{\rm i}))^{2} \label{chi_ringless}
\end{equation}
from the binned data.  Here, $d(t_{\rm i})$ , $m(t_{\rm i})$, and
$\Delta d(t_{\rm i})$ are the observed flux, the expected flux of the
model, and the uncertainty in observed flux at $t = t_{\rm i}$,
respectively. We assume $\Delta d(t_{\rm i})$ to be 
a standard deviation of the normalized flux of each lightcurve 
estimated from its out-of-transit epoch. 

The same procedure is performed for a ringed planet model. In this case,
we have 12 free parameters $t_{0}$, $b$, $R_{\rm p}$, $r_{\rm out/in}$,
$r_{\rm in/p}$, $\theta$, $\phi$, $T$, $a/R_{\star}$, $c$, $q_{1}$ and
$q_{2}$.  We calculate the chi-squared value $\chi_{\rm ring}^{2}$,
which has the definition similar to $\chi_{\rm ringless}^{2}$.

One fit of the ringed model takes about a few minutes in a lap-top, and the fits to the
entire datasets were carried out with PC clusters in The Center for
Computational Astrophysics (CfCA) in National Astronomical Observatory,
Japan.

\subsection{Searching for ring signatures via comparison between 
ringless and ringed planet models \label{p_search_even_odd}}

Our next procedure is to create a list of tentative ringed-planet
candidates from the comparison between the best-fit values for the two
models,  $\chi_{\rm ringless, \;min}^{2}$ and
$\chi_{\rm ring,  \;min}^{2}$. Specifically for this purpose, we adopt a
$F$-test with $F$ statics \citep[e.g.][]{2011Natur.470...53L}, and
define
\begin{equation}
F_{\rm obs} = \frac{ (\chi^{2}_{\rm ringless, \; min} 
- \chi^{2}_{\rm ring,  \;min})
/(N_{\rm ring}-N_{\rm ringless})}
{\chi^{2}_{\rm ring,  \;min}/(N_{\rm bin} - N_{\rm ring}-1)},
\label{F_value}
\end{equation}
where $N_{\rm bin}$ is the number of in-transit bins of the phase-folded lightcurve (typically 500),
and $N_{\rm ring} = 12$ and $N_{\rm ringless} = 7$ are the number of
free parameters in the planetary models with and without a ring,
respectively.

The numerator of the right-hand side of Eq. (\ref{F_value}) corresponds
to the improvement in $\chi^{2}$ of the ring model 
divided by the number of the additional degrees of freedom characterizing a ring.  The denominator is the $\chi^2$ per degree of
freedom for the ringed model. Thus, $F_{\rm obs}$
represents a measure of relative improvement of the fit by introducing
the ring. The large $F_{\rm obs}$ prefers the ringed planet model.
Note, however, that $F_{\rm obs}$ is defined simply through the ratio of
the minimum values of $\chi^{2}$ for the two models. Therefore it is
nothing to do with the goodness of the fit for either model, which needs
to be checked separately.

According to the $F$-test, the measure of the the null hypothesis that
our ringed model does not improve the fit relative to the ringless model
is given by the $p$-value defined as
\begin{equation}
p = 1- \int_{0}^{F_{\rm obs}} 
F(f| N_{\rm bin} - N_{\rm ring}-1, N_{\rm ring} - N_{\rm ringless}) {\rm d}f, 
\end{equation}
where $F(f| N_{\rm bin} - N_{\rm ring}-1, N_{\rm ring} - N_{\rm
ringless})$ is the $F$-distribution with the degrees of freedom
$(N_{\rm bin} - N_{\rm ring}-1, N_{\rm ring} - N_{\rm ringless})$.

The larger value of $F_{\rm obs}$, therefore the smaller value of $p$ 
disfavors the null assumption, i.e., the ringed model better fits the data 
than the ringless model. In this paper, we adopt the
condition of $p<0.05$ for the rejection of the null hypothesis.
For those tentative candidates of ringed planets, we 
attempt to understand the origins of anomalies by examining 
individual lightcurves and statistics (e.g. $\chi^{2}_{\rm ring,\; min}$) further. 

We also test the robustness of possible ring signatures
by dividing the multiple transits into those at even and odd 
transit numbers, creating the phase-folded lightcurves separately, and 
computing the $p$-values $(p_{\rm even}, p_{\rm odd})$.
Unlike the other analyses, we use the non-binned data here in order 
to evade the additional uncertainties in the lightcurves due to 
the extra binning step, especially for systems with the low number of 
the data. For the calculation of ($p_{\rm even}$, $p_{\rm odd}$), we approximate the
best-fit model of the binned data as that of the non-binned data, and
then we calculate $F_{\rm obs}$ in Eq (\ref{F_value}) for non-binned
data. If rings mainly account for signals in lightcurves, we expect 
$p_{\rm even}$ to be close to $p_{\rm odd}$ because of the consistency 
of the signals.  

Finally, we comment on the validity of applying the $F$-test to our ring
search. The $F$-test needs to satisfy two conditions
\citep[e.g.][]{2002ApJ...571..545P}. One is that the two models are
nested in a sense that the more complicated model reduces to the simpler one
if the additional parameters in the former model are removed.  This is
trivially satisfied in the present case. 
The other condition is that the simpler model should not be located at
the edge of the parameter space of the more complicated model. 
Strictly speaking, this condition may not hold because our ring model
reduces to the ringless model in the limit of $R_{\rm out}\rightarrow
R_{\rm p}$.  Nevertheless, $F$-test gives us a practically useful
criterion, and we decide to use it in selecting tentative candidates for further analysis.

\subsection{Obtaining upper limits on the outer radius of a ring
 \label{result_const}}

Even for planetary systems without any detectable signatures of a ring,
we may constrain the property of a possible ring within the
observational detection limit. To proceed realistically, we need to
reduce the number of free parameters charactering the ring. Thus we fix
the inner radius of the ring as $R_{\rm in}=R_{\rm p}$, and set the
opacity of the ring as $T=1$ just for simplicity. Furthermore, we focus on 
two cases for the orientation angles of the ring as we describe in the next subsections. 
Thus we are left with a single parameter, the outer radius of the ring $R_{\rm out}$. In practice, we place upper limits on the 
ratio $R_{\rm out}/R_{\rm p}$ from the fit to the lightcurves.

\subsubsection{Aligned with the planetary orbit \label{align_cond}}

Under the strong tidal interaction with the star, the ring becomes aligned
to the orbital plane of the planet. Indeed \cite{2001ApJ...552..699B}
gave the upper limit on the ring size of a Hot Jupiter, HD 209458 b, as
$1.7 R_{\rm p}$ assuming the alignment.

In a similar manner, we place upper limits on the ring size assuming
the tidal alignment.  The tidal alignment leads to the orientation of
$\theta = \arcsin(b/(a/R_{\star}))$ and $\phi=0$.  In addition, the small
value of $\theta$ enhances the effective optical depth viewed from the 
observer, relative to that from the top-view.  Thus, we assume $T=1$
even though rings can be very thin like Jupiter's rings.

 In summary, we fix $\phi=0^{\circ}$, $\theta = \arcsin( b R_{\star}/a)
$, $T=1$, and $R_{\rm in} = R_{\rm p}$ for fitting. Assuming these
conditions, we fit the ringed model to the data using at least 100 sets
of randomly chosen initial parameters, and we pick up the best solution
among the local optimum solutions.

After obtaining the best solutions with fixed values of $R_{\rm
out}/R_{\rm p} $, we define the 3$\sigma$ limit $(R_{\rm out}/R_{\rm p}
)_{\rm upp}$ where
\begin{equation}
\Delta \chi^{2}(R_{\rm out}/R_{\rm p} )\equiv (\chi^{2}_{\rm ring, \; min}
(R_{\rm out}/R_{\rm p} ) - \chi^{2}_{\rm ringless, \; min})/ (\chi^{2}_{\rm
ringless, \; min}/{\rm dof})
\end{equation}
becomes 9. In practice, we compute $\Delta \chi^{2}(R_{\rm out}/R_{\rm
p})$ at 11 values of $R_{\rm out}/R_{\rm p}$: 1.1, 1.3, 1.5, 2.0, 2.5,
3.0, 4.0, 5.0, 6.0, 8.0, and 10.0. Then we interpolate them to find
$(R_{\rm out}/R_{\rm p} )_{\rm upp}$.

\begin{figure}[htpb]
\begin{center}
 \includegraphics[width = 0.68 \linewidth]{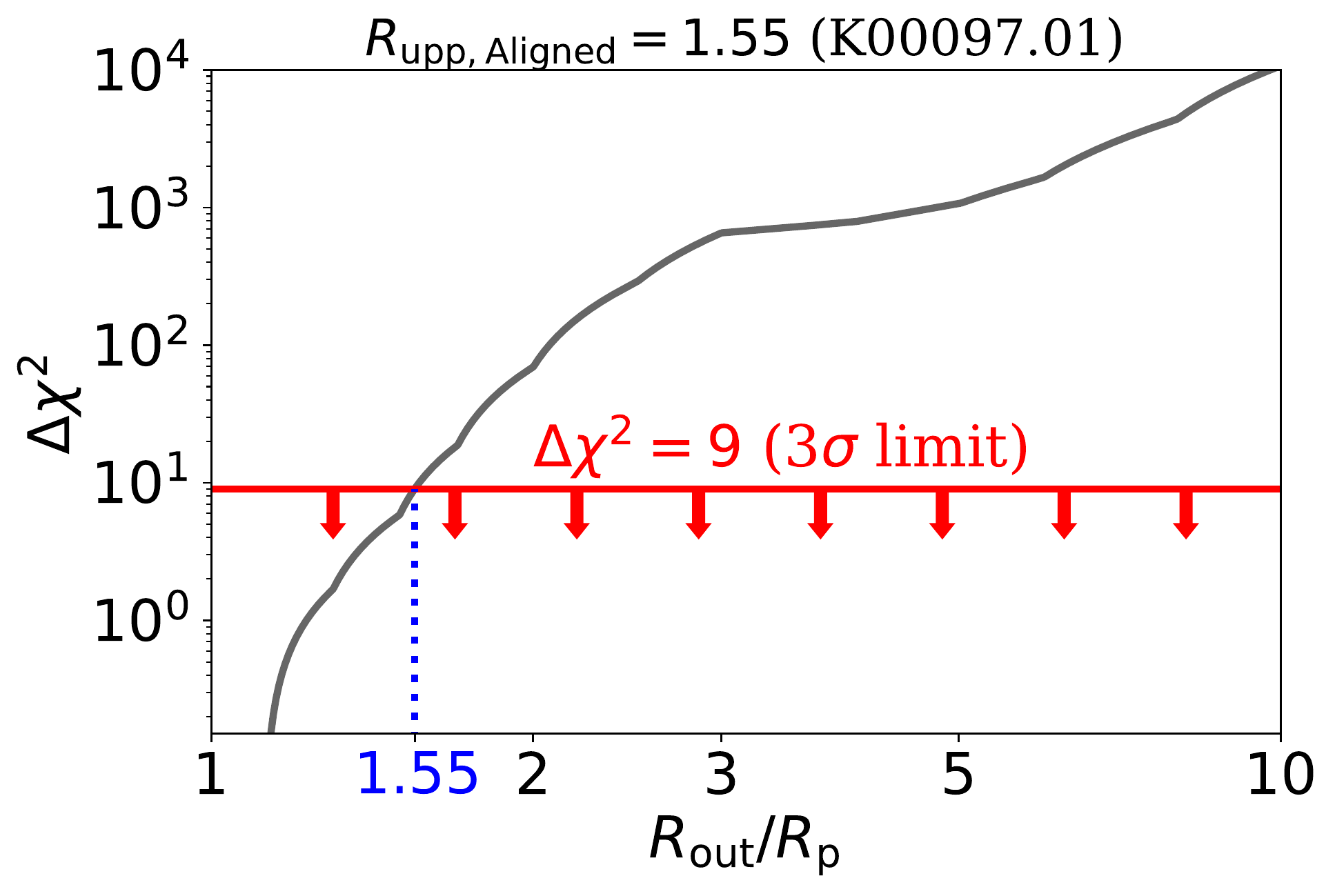}
\caption{An example illustrating how to set an upper limit on $R_{\rm
out}/R_{\rm p}$. Black curve shows $\Delta\chi^2$, eq.(6), of an aligned
ring model for KOI-97.01.  The value of $R_{\rm out}/R_{\rm p}=1.55$
where $\Delta\chi^2=9$ is defined as our $(R_{\rm out}/R_{\rm p} )_{\rm
upp, \;Aligned}$.  } \label{KOI97_upper}
\end{center}
\end{figure}

\begin{figure*}[htpb]
\centering \includegraphics[width =0.80 \linewidth]{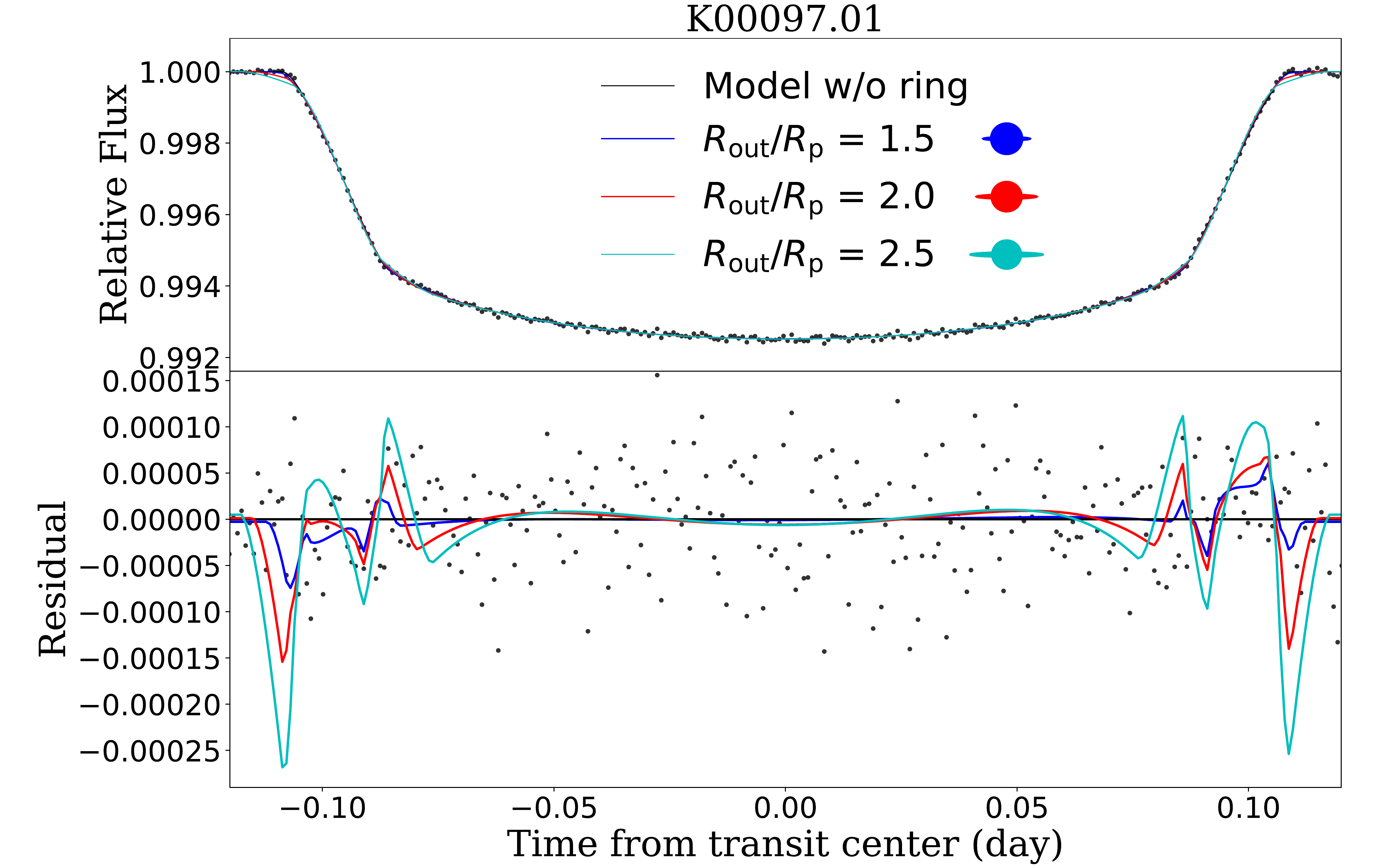}
\caption{Lightcurves for KOI-97.01. Gray points are the binned data of
KOI-97.01. Blue, red, and cyan curves correspond to the best-fits
of the ringed model 
with $R_{\rm out}/R_{\rm p}=1.5$, 2.0, and 2.5, respectively.
The lower panels indicate the
residuals with respect to the best-fit of the ringless model.
} \label{KOI97_curve}
\end{figure*}

Our procedure to setting the upper limit is illustrated in Figure
\ref{KOI97_upper} for KOI--97.01.  In this example, the interpolated
curve crosses the $\Delta \chi^{2}=9$ threshold at $R_{\rm out}/R_{\rm
p} =1.55$. Thus we obtain $(R_{\rm out}/R_{\rm p} )_{\rm upp,
\;Aligned}=1.55$ for KOI--97.01.  Figure \ref{KOI97_curve} plots three
corresponding fitting curves with $R_{\rm out}/R_{\rm p}=1.5$, $2.0$,
and $2.5$ along with the curve of the ringless model. 

If $\Delta \chi^{2}<9$ for $R_{\rm out}/R_{\rm p}
=10.0$, we do not place upper limits $(R_{\rm
out}/R_{\rm p})_{\rm upp}$. These cases are marked as \nodata in Tables
\ref{final_table1} to \ref{final_table3} below. 

The alignment condition is determined by the tidal dissipation function
$Q_{\rm p}$ and the Love number $k_{\rm p}$, the planet/star 
mass ratio, the dimensional moment of the inertia of the planet $C$, the orbital period 
$P_{\rm orb}$, and the normalized semi-major axis $a/R_{\star}$. 
  As discussed in Appendix A, 154 out of the 168 planetary systems are supposed to become aligned
within a timescale of 1Gyr, if we adopt a fiducial values, $Q_{\rm
p}=10^{6.5}$, $k_{\rm p}=1.5$, $C=0.25$, and the mass-radius relation (Eq (8) in 
\cite{2013ApJ...768...14W}).  We compute the upper limit on 
$R_{\rm out}/R_{\rm p}$ for all 168 systems
in any case even if their alignment timescale is long.

\subsubsection{Orientation of the Saturnian ring \label{sat_upp}}

As another model for the ring orientation, we simply adopt the Saturnian
case $\phi=0^{\circ}$ and $\theta = 26.7 ^{\circ}$, in addition to
$T=1$ and $R_{\rm in} = R_{\rm p}$ as before. Although 
the values of $T$ and $R_{\rm in}$ are adopted just for simplicity, 
the derived upper limits are mainly sensitive to 
$R_{\rm out}$, and can be scaled with the different values of $T$. 
An additional small signal due to an inner gap 
may be extracted if $R_{\rm in}>R_{\rm p}$, while 
it is not important in the present analysis 
\citep[e.g][]{2004ApJ...616.1193B, 2018A&A...609A..21A}.

With fixed values of $R_{\rm out}/R_{\rm p}$, we search for the optimal
solutions by varying other parameters in the similar manner as in Sec
\ref{align_cond}.  In the analysis, we vary $R_{\rm out}/R_{\rm p}$ up
to $1/\sin(26.7^{\circ}) \simeq 2.22$, above which a shape of an assumed
ring is not distinguishable from an oblate planet with the same
oblateness. Practically, we use 8 fixed values of $R_{\rm out}/R_{\rm
p}$: 1.1, 1.2, 1.3, 1.4, 1.6, 1.8, 2.0, and 2.22. Then, we obtain the
3$\sigma$ limit $(R_{\rm out}/R_{\rm p} )_{\rm upp, \;Saturn}$ by
interpolating the values of \{$R_{\rm out}/R_{\rm p} $, $\Delta
\chi^{2}(R_{\rm out}/R_{\rm p} )$\}.  If $\Delta \chi^{2}<9$ for $R_{\rm
out}/R_{\rm p} = 2.22$, we do not give the upper limits $(R_{\rm
out}/R_{\rm p} )_{\rm upp, \; Saturn}$.

In addition to the limits on $R_{\rm out}/R_{\rm p}$, we also place
upper limits on the ratio of the outer radius of the ring and the
stellar radius, $(R_{\rm out}/R_{\star})_{\rm upp}$.  Qualitatively this
is simply given by $(R_{\rm out}/R_{\rm p})_{\rm upp} \times (R_{\rm
p}/R_{\star})_{\rm ringless}$, but not exactly because the best-fit
planet radius may be different if the ring model is assumed instead.  To
evaluate $(R_{\rm out}/R_{\star})_{\rm upp}$ correctly, we estimate
$R_{\rm p}/R_{\star}$ corresponding to $(R_{\rm out}/R_{\rm p})_{\rm
upp,\; Saturn}$ by interpolating the values of \{$R_{\rm p}/R_{\star}$,
$(R_{\rm out}/R_{\rm p})_{\rm upp,\; Saturn}$\}. Then, we obtain
$(R_{\rm out}/R_{\star})_{\rm Saturn, upp} = (R_{\rm out}/R_{\rm p})
_{\rm upp, \; Saturn} \times (R_{\rm p}/R_{\star})$ using the
interpolated values. For simplicity, we only give $(R_{\rm 
out}/R_{\star})_{\rm upp}$ for systems with $(R_{\rm out}/R_{\rm
p})_{\rm upp,\; Saturn}$.

Incidentally the damping timescales of the 13 systems with $p<0.001$
turned out to be significantly less than 1 Gyr except for KOI-868. Thus
the possible rings for the 12 systems are likely to be aligned with the
planetary orbital plane. Thus we do not compute $(R_{\rm out}/R_{\rm p}
)_{\rm upp, \;Saturn}$ for all the systems with $p<0.001$.

\section{Result of the ring survey \label{result}}

\subsection{No Convincing Candidate for a Ringed Planet}

We have performed a ring search following the method described
in Section \ref{sec_data}. The result for all the 168 Kepler objects is
summarized in Tables \ref{final_table1} to \ref{final_table3}.  

\begin{figure}[htpb]
\begin{center}
 \includegraphics[width =10cm]{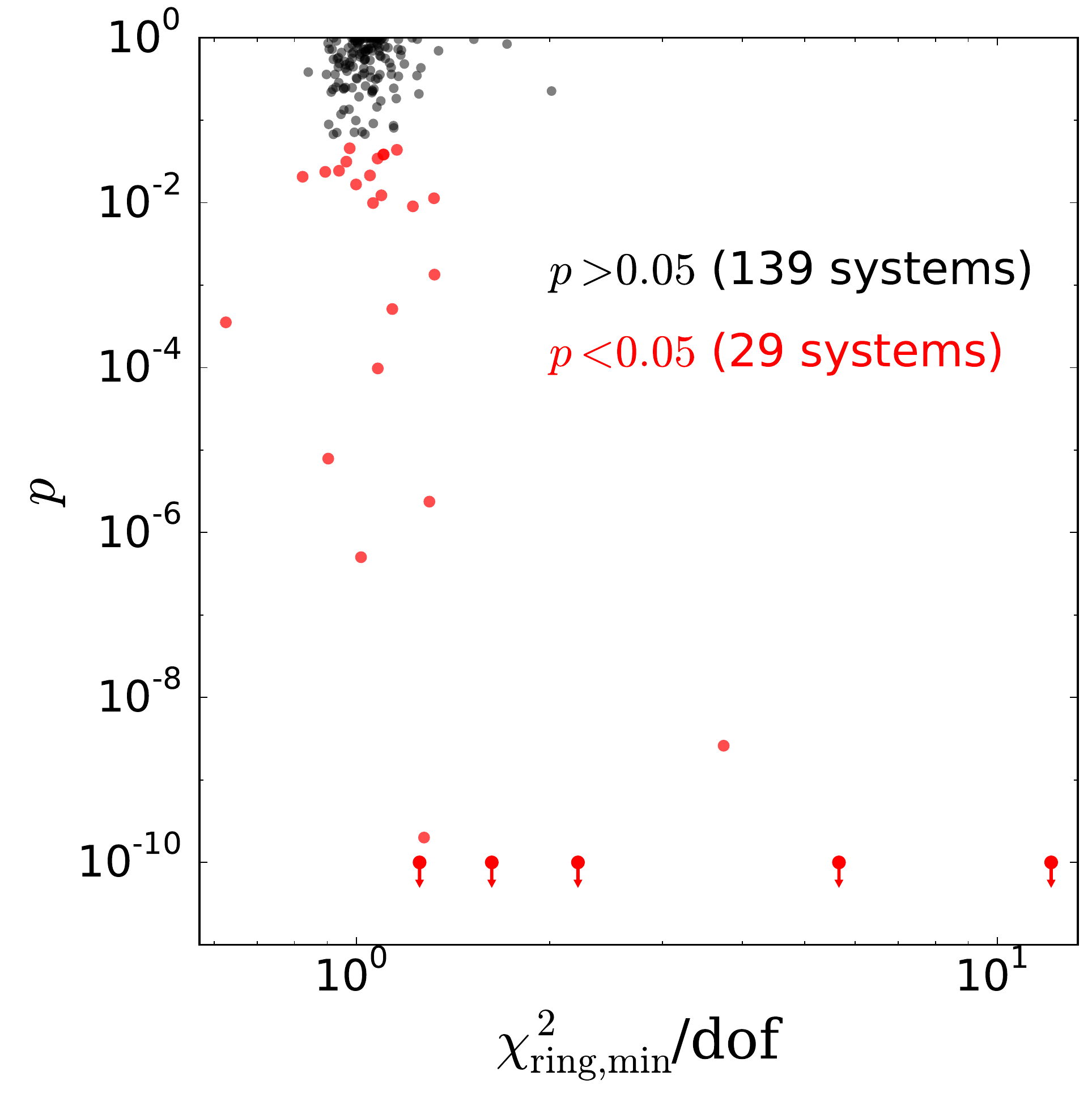} \caption{The
$p$-values against $\chi^{2}/$dof for our 168 targets.}
\label{p_chi_dof}
\end{center}
\end{figure}

We identify 29 candidate objects with $p$-values less than the threshold
value of $0.05$. For most of these systems, 
the ring model yields $\chi^{2}_{\rm ring}/$dof $\sim 1$ (Figure
\ref{p_chi_dof}). However, after inspecting individual lightcurves of
these systems, we conclude that none of them is a viable candidate for a
ringed planet. The 11 of the 29 candidates do not exhibit any convincing
ring-like signatures in the lightcurves, and so are excluded. The other
18 systems do show anomalous features in the lightcurves, but they are
most likely ascribed to other mechanisms: gravity darkening (2 systems),
spot-crossing (9 systems), disintegration of a planet (1 system), artifacts
generated during the folding process (3 systems), and stellar activity
(3 systems). See Appendix B for details of this process, as well as for
individual lightcurves.

\subsection{Upper Limits on the Ring Size}
\subsubsection{Result}

Given the null detection, we derive upper limits on the outer radius of
the possible ring following the method described in Sec
\ref{result_const}.  The resulting upper limits, $(R_{\rm out}/R_{\rm
p})_{\rm upp, \;Aligned}$, $(R_{\rm out}/R_{\rm p})_{\rm upp,
\;Saturn}$, and $(R_{\rm out}/R_{\star})_{\rm upp}$, are listed in Table
\ref{final_table1} to \ref{final_table3}. If we cannot obtain upper
limits due to poor signal-to-noise ratios, we leave those values blank
in the tables. The following discussions exclude 18 systems for $(R_{\rm
out}/R_{\rm p})_{\rm upp, \;Aligned}$ and 7 systems for $(R_{\rm
out}/R_{\rm p})_{\rm upp, \;Saturn}$ that are identified as possible
false positives in the {\it Kepler} Community Follow-up Program
(CFOP) webpage.\footnote{https://exofop.ipac.caltech.edu/cfop.php}

\begin{figure}[htpb]
\begin{center}
 \includegraphics[width =8cm]{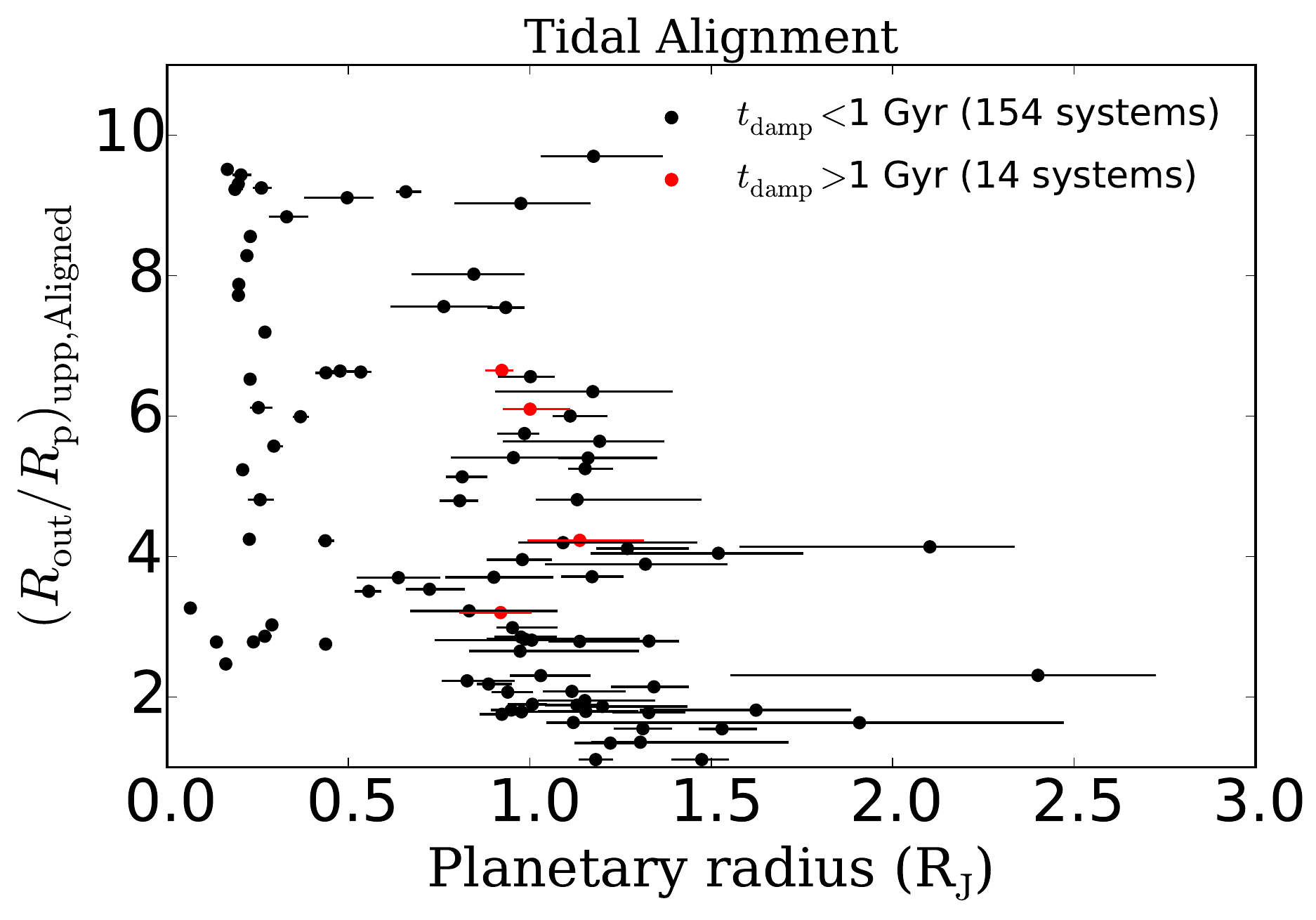}
 \includegraphics[width =8cm]{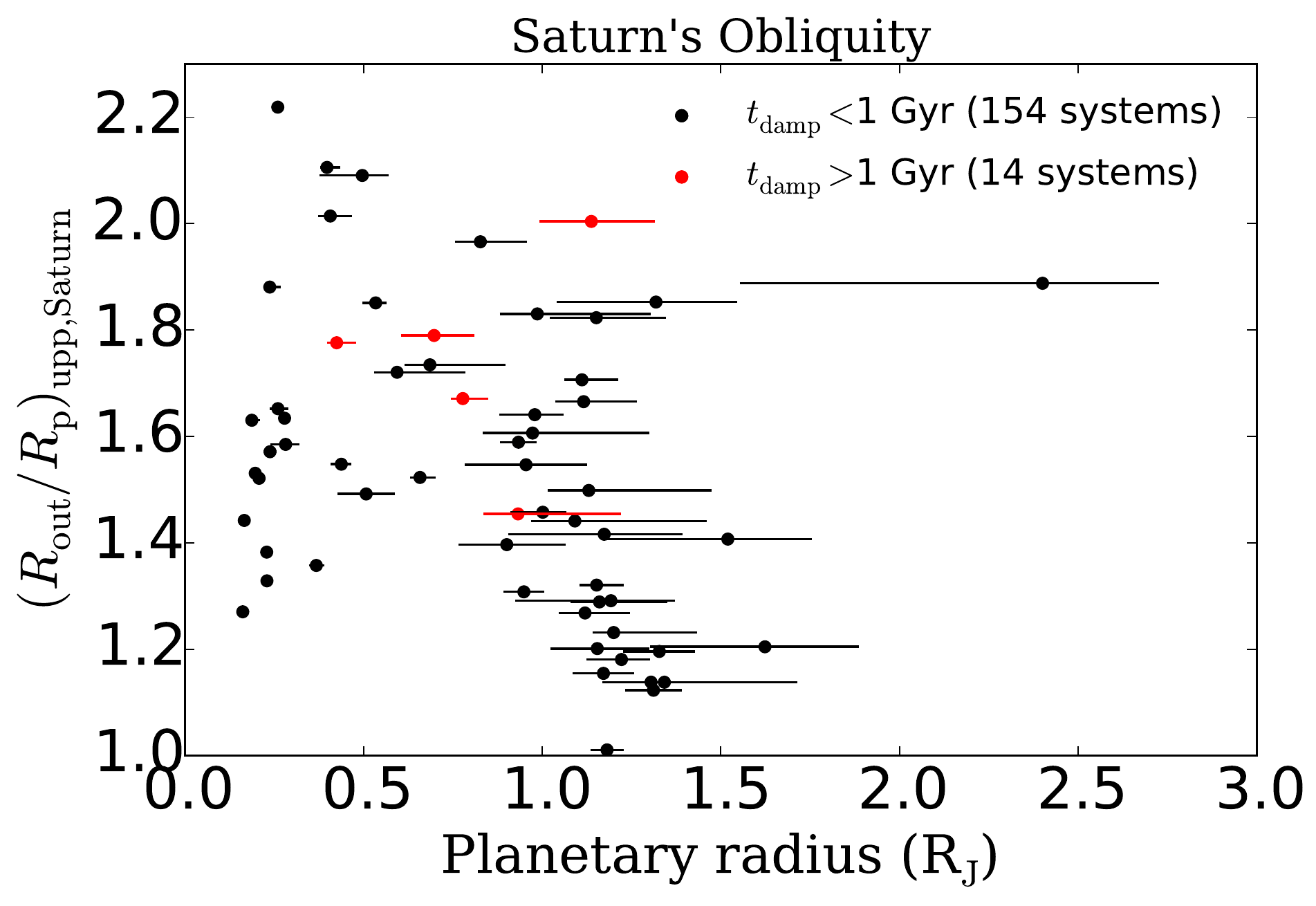}
\caption{\small Upper limits on $R_{\rm out}/R_{\rm p}$ as a function of
$R_{\rm p}$. Left and right panels correspond to the tidally aligned
 ring, and a ring with Saturn's obliquity, respectively. 
Black points ($t_{\rm damp} < 1$ Gyr) are likely candidates for the
 aligned systems. The numbers of systems in panels count all targets with and without limits.}  
\label{prad_upp}
\end{center}
\end{figure}

Figure \ref{prad_upp} compares upper limits $(R_{\rm out}/R_{\rm
p})_{\rm upp}$ for the aligned and Saturn-like configurations against
the physical planetary radii.  The latter values are computed as
$(R_{\rm p}/R_{\star})_{\rm ringless}$ $\times R_{\star}$, where the
values of $(R_{\rm p}/R_{\star})_{\rm ringless}$ are obtained from the
ringless model and the stellar radii are taken from the Kepler
catalog. Even assuming the ring aligned with the orbital plane, we find
fairly tight limits on the ring size (several times $R_{\rm p}$) for a
few tens of systems.

\begin{figure}[htpb]
\begin{center}
 \includegraphics[width =8cm]{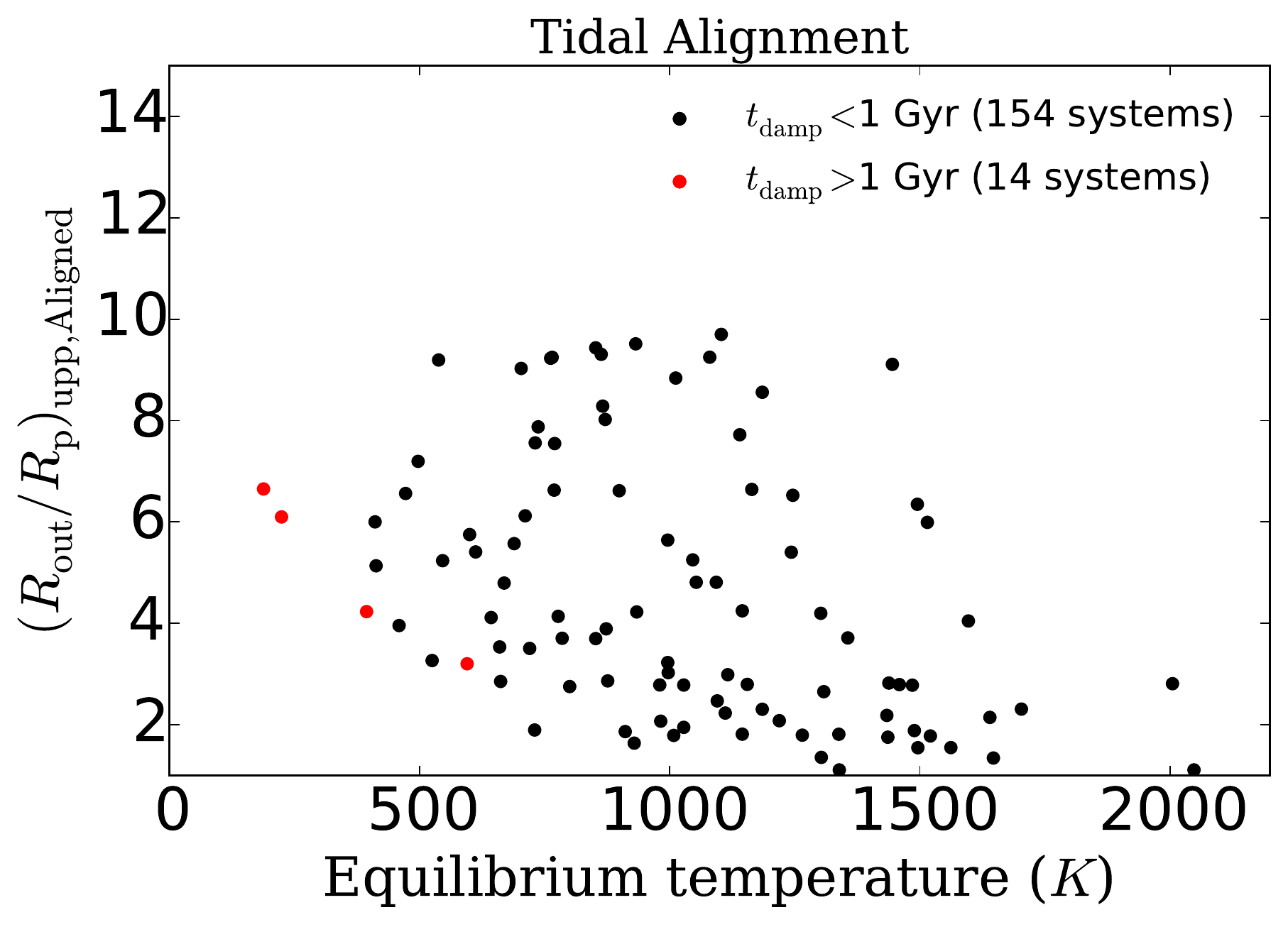}
 \includegraphics[width =8cm]{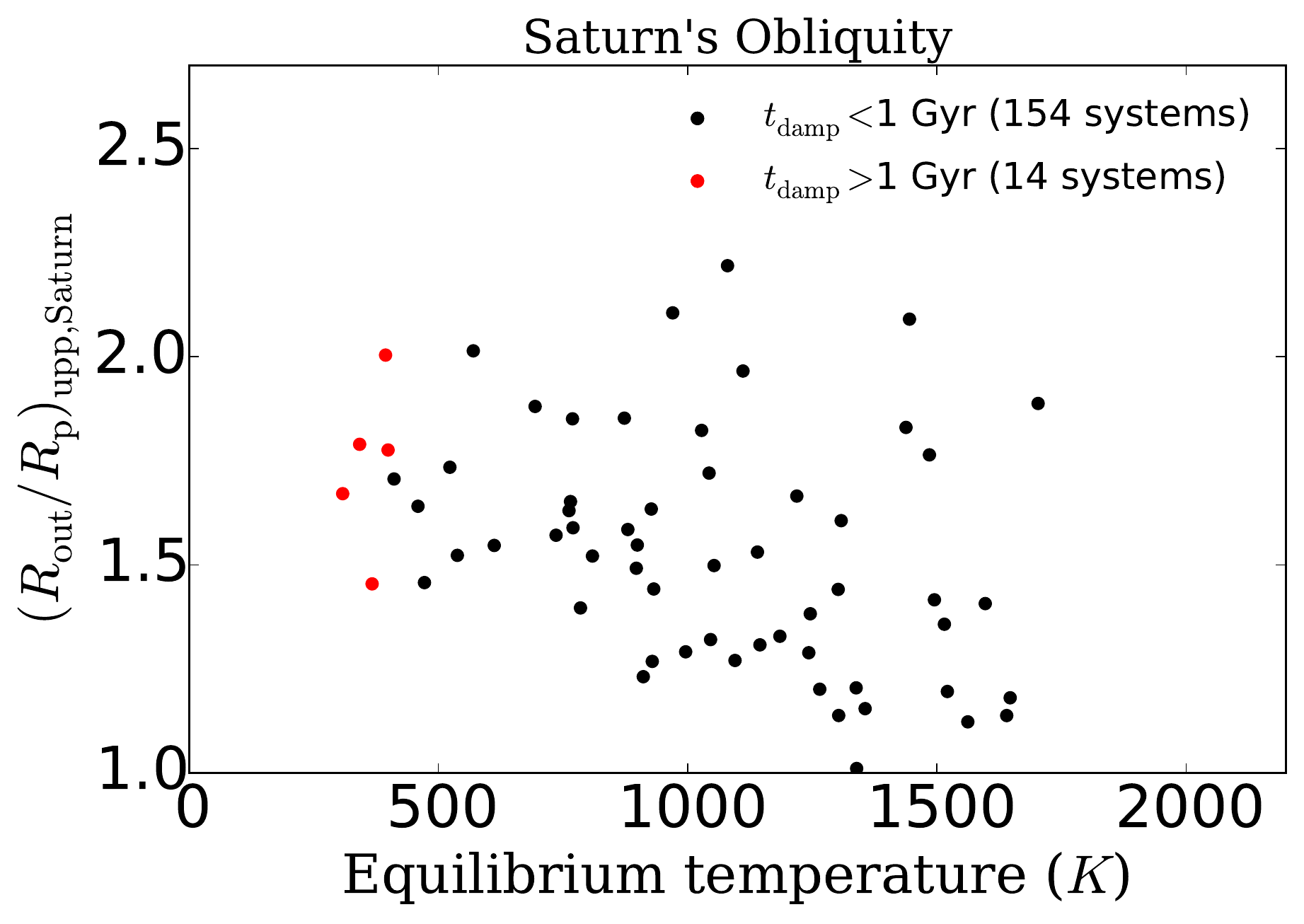}
\caption{\small Same as Fig. \ref{prad_upp}, but plotted against
the equilibrium temperature of the planets.
} \label{limits_teq}
\end{center}
\end{figure}

Figure \ref{limits_teq} is a similar plot to Figure \ref{prad_upp}, but
against the equilibrium temperatures $T_{\rm eq}$ of the planets.  The
exhibited pattern does not reflect the physical dependence of $(R_{\rm
out}/R_{\rm p})_{\rm upp}$ on $T_{\rm eq}$, but simply comes from the
fact that the hotter planets have shorter orbital periods, and hence
larger signal-to-noise ratios of the phase-folded lightcurve. With
sufficient signal-to-noise ratios for future data, however, such plots
would provide interesting constraints on the physical properties of
rings as a function of melting temperature of different compositions.

As mentioned in Section 4.1, the lightcurves of some of the 29 systems
with $p<0.05$ include contributions from the effects other than rings,
such as gravity darkening and spot crossing. Nevertheless, we neglect
them in deriving the upper limits on $R_{\rm out}/R_{\rm p}$.  If we fit
and remove those effects from the lightcurve, the upper limits may
become more stringent. In this sense, the upper limits on $R_{\rm
out}/R_{\rm p}$ listed in Tables \ref{final_table1} and
\ref{final_table2} would be a bit conservative.

\subsubsection{Comparison of Roche radius and upper limits}

To understand implications of the upper limits physically, we compare
the limits $ (R_{\rm out}/R_{\rm p})_{\rm upp}$ with the Roche
radii. If we consider ring formation by tidal destruction of
incoming objects (e.g. satellites), the outer radius of the ring may be set by the
Roche radius:
\begin{equation} 
R_{\rm out} \sim
2.45\,R_{\rm p} \left(\frac{\rho_{\rm p}}{\rho_{\rm s}}\right)^{1/3}
=1.6\,R_{\rm p} \left(\frac{\rho_{\rm p}/1\,{\rm g\,cm^{-3}}}{\rho_{\rm s}/3.5\,{\rm g\,cm^{-3}}}\right)^{1/3},
\end{equation} 
where $\rho_{\rm p}$ is the planetary density and $\rho_{\rm s}$ is that of the incoming object. 
Here we scale the result using $\rho_{\rm p} = 1 \;{\rm
gcm^{-3}}$ and $\rho_{\rm s}= 3.5 \;{\rm gcm^{-3}}$,
which are the typical values for rocky components in the Solar System. 

This implies that, if the inferred upper limit on the ring size $R_{\rm out}/R_{\rm p}$ is much smaller than $1.6$, the ring is unlikely to exist even inside that limit --- unless $\rho_{\rm s}$ is unreasonably large.
In our sample, six systems satisfy $t_{\rm damp}<1 $Gyr and 
$(R_{\rm out}/R_{\rm p})_{\rm upp,\; Aligned} < 1.6$, and 
one satisfies $t_{\rm damp}>1 $Gyr and 
$(R_{\rm out}/R_{\rm p})_{\rm upp,\; Saturn} < 1.6$. 
We may exclude possible rings around these systems.

\subsection{Upper Limits on the Ring Occurrence}

The above limits on $R_{\rm out}/R_{\rm p}$ translate into the
 the upper limit on the occurrence rate of rings $q[>x]$
as a function of $x\equiv R_{\rm out}/R_{\rm p}$. Here 
$q[>x]$ is the probability that a planet has a ring larger than $x$ times the planetary radius. For example, 
$q[>x=1]$ is simply the occurrence rate of rings, and 
$q[>x=2]$ is that of rings larger than twice the planetary radii. 

We attempt to estimate the upper limit on $q[>x]$ as follows. 
For a given value of $x$, consider $n$ samples extracted from systems with $q[>x]$, for which the rings with $R_{\rm out}/R_{\rm p}>x$ should have been readily detectable --- so this may be chosen to be $N[<x]$, the number of systems with $(R_{\rm out}/R_{\rm p})_{\rm upp}<x$.
Then the probability that we
detect $n_{\rm obs}$ rings with $R_{\rm out}/R_{\rm p}\geq x$ out of
the $n$ samples is given simply by the binominal distribution:
\begin{equation}
{\rm Prob}(n_{\rm obs}|q[>x], n) = _{n}C_{n_{\rm obs}} \;q[>x]^
{n_{\rm obs}}\; (1-q[>x])^{n- n_{\rm obs}}. 
\end{equation}
Without any prior knowledge of $q[>x]$ nor $n_{\rm obs}$, we
assume the uniform distribution for $\mathrm{Prob}(q[>x])$ and $\mathrm{Prob}(n_{\rm obs})$ with proper normalizations:
\begin{align}
\int ^{1}_{0} {\rm Prob}\left(q[>x]|n\right) d q[>x] &= 1 \rightarrow\;\;\;{\rm Prob}\left( q[>x]|n\right)= 1 \\
\sum_{n_{\rm obs} = 0}^{n} {\rm Prob}(n_{\rm obs}|n) &= 1 \rightarrow {\rm Prob}(n_{\rm obs}|n) = 1/(n+1)
\end{align}
According to Bayes' theorem, we obtain 
\begin{align}
{\rm Prob}(q[>x]|n_{\rm obs}=0, n)&= 
\frac{{\rm Prob}(q[>x]|n) {\rm Prob}(n_{\rm obs}=0|q[>x], n)}
{{\rm Prob}(n_{\rm obs} = 0|n) } \nonumber\\
&= (n+1)(1-q[>x])^{ n}
\end{align}
The corresponding cumulative distribution function for $q[>x]$ is given by: 
\begin{equation}
{\rm CDF}(q[>x]) = 1- (1-q[>x])^{n+1}.  \label{eq_cdf}
\end{equation}
Here, we would like to obtain the 95\% upper limits of $q[>x]$. Thus, the above equation gives 
\begin{equation}
q[>x]_{\rm upp} = 1 - (0.05)^{\frac{1}{n+1}}. \label{q_upp}
\end{equation}
Now, we substitute the values of $N[<x]$ plotted in 
Fig \ref{limits_teq} into $n$, and obtain the upper limits of 
$q[>x]$ as a function of $x$.

\begin{figure}[htpb]
\begin{center}
 \includegraphics[width =8cm]{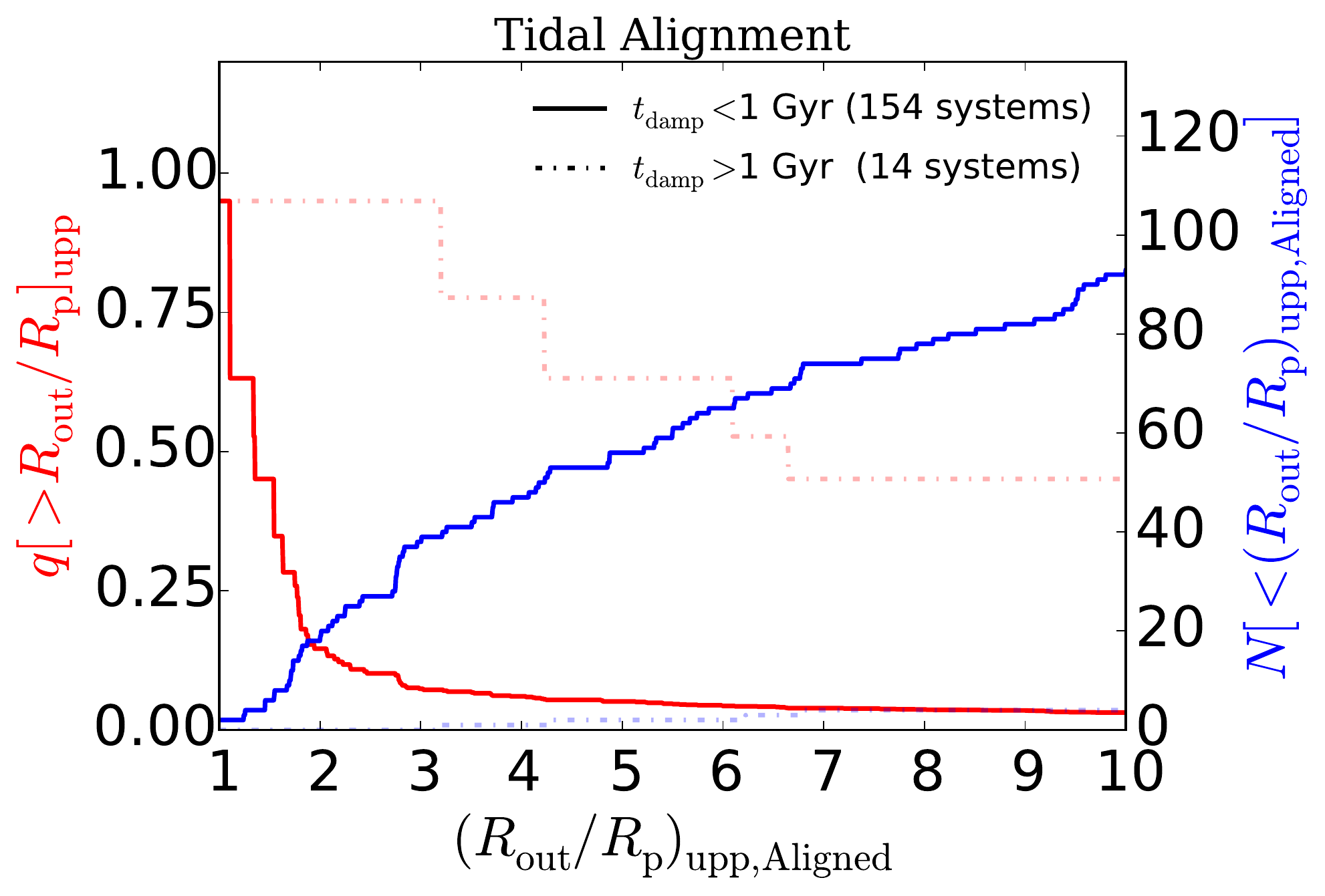}
  \includegraphics[width =8cm]{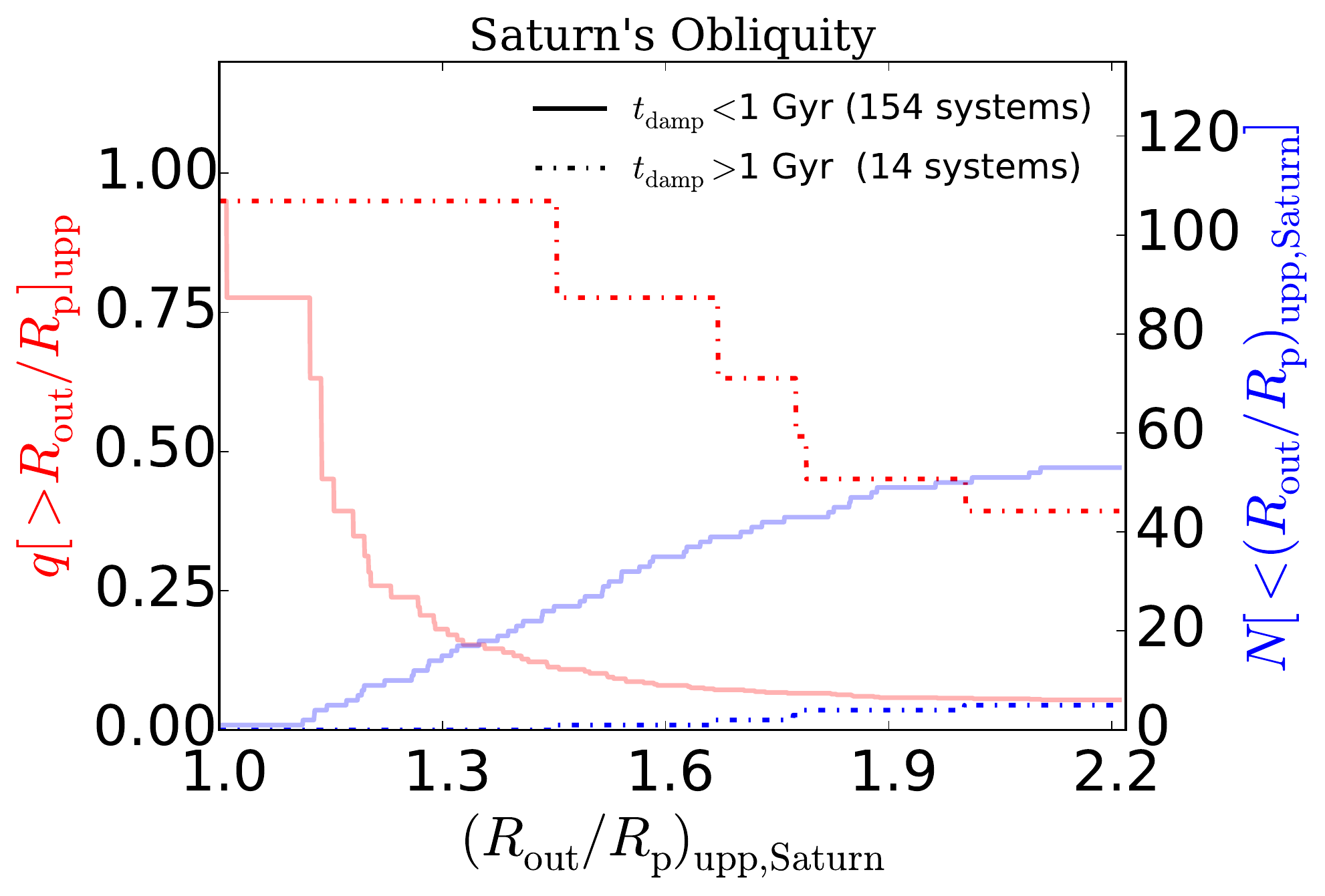}
  \caption{\small Cumulative frequency of upper limits 
  $N[<(R_{\rm out}/R_{\rm p})_{\rm upp}]$  and 
  upper imits on the occurence rate of rings $q[>x]_{\rm upp}$ in Eq (\ref{q_upp}). 
  The left panel assumes the tidal alignment, while the right panel 
  assumes the Saturn's obliquity. Thick lines in the left panel correspond to 154 systems with 
  $t_{\rm damp}<1 $Gyr, while those in the right panel to 14 systems with 
  $t_{\rm damp}>1 $Gyr. } 
  \label{limits_tau}
\end{center}
\end{figure}

Figure \ref{limits_tau} shows $q[>x]_{\rm upp}$ using $N[<(R_{\rm
out}/R_{\rm p})_{\rm upp, \;Aligned}]$, and $N[<(R_{\rm out}/R_{\rm
p})_{\rm upp, \;Saturn}]$. Physically speaking, the limit $(R_{\rm
out}/R_{\rm p})_{\rm upp, \;Aligned}$ is appropriate only for systems with small values of $t_{\rm damp}$, which have likely achieved tidal alignment. On the other hand, $(R_{\rm out}/R_{\rm p})_{\rm upp, \;Saturn}$ may be more relevant for those with large values of $t_{\rm damp}$. 
Therefore, we distinguish the systems with $t_{\rm damp}<1\,\mathrm{Gyr}$ and $t_{\rm damp}>1\,\mathrm{Gyr}$ in the plot.
The more relevant subset is shown with thick lines in each panel. 

\section{Summary and Discussion}

We have performed a systematic and intensive search for exo-rings among
the 168 {\it Kepler} planet candidates.  The targets are homogeneously selected
from all the KOIs that have the signal-to-noise ratio of the
phase-folded lightcurves exceeding 100. As a result, a majority of our
targets are short-period planets. This sample is complementary to that
of long-period planets analyzed by \citet{2017AJ....153..193A}, and
significantly larger than the 21 short-period planet samples by 
\cite{2015ApJ...814...81H}.

For all the targets, we obtained the best-fit ringless and ringed model
parameters from their individual phase-folded lightcurves following
\cite{2017AJ....153..193A}. Then, we compare the two best-fits, and we select
29 systems as tentative candidates for which the ringed-model fit better 
explained the data than the ringless-model fit.

Those 29 systems are further examined individually and visually, and we
conclude that none of them exhibits clear signature of a planetary ring.
Instead, we derive upper limits on the ratio of the outer radius of the
possible ring and the planetary radius assuming two different
configurations; the tidally aligned ring and the Saturn's ring.
The derived upper limits for individual systems are summarized in Tables
\ref{final_table1} to \ref{final_table3}.

The distribution of those upper limits can be used to derive the
statistical upper limits on the occurrence rate of planetary rings as a
function of $R_{\rm out}/R_{\rm p}$. We found that
${\rm Prob}(R_{\rm out}/R_{\rm p}>2)$ should be less than
15 percent for tidally aligned ring systems. 

Given that our targets are mainly in close-in orbits, the null detection
of rings may not be so surprising \citep[e.g.][]{2011ApJ...734..117S}.  This
is also consistent with the fact that dense planetary rings in our Solar
System are discovered exclusively at temperatures close to 70 K
\citep{2015ApJ...801L..33H}.

Nevertheless, our current result clearly indicates that the existing {\it
Kepler} data are already accurate and precise enough to probe the
planetary rings of a comparable size to the planet itself.
This is quite encouraging, and the future effort towards the discovery
of ring would likely be rewarding as we have witnessed numerous
unexpected surprises in the history of astronomy, and especially
exoplanetary science. 

We also believe that the current methodology and examples of
false-positives would be very useful in such future searches for
planetary rings with improved datasets.

Having said so, it is important to emphasize other independent approaches
to the ring survey.  For instance, \citet{2015ApJ...803L..14Z} pointed
out that KOIs flagged as ``FALSE POSITIVES'', which we intentionally
exclude from our current targets, may be promising because they could
include possible ringed planets that are misinterpreted as anomalously
large planets.  Also a precession of planetary rings may induce a
detectable level of transit depth variation
\citep[e.g.][]{2010ApJ...709.1219C,2015ApJ...814...81H}. 
In addition, scattering and diffraction of the star light by the ring
particles may be observable depending on the size of ring particles,
especially through multi-band photometry in space.

Therefore, we expect that the upcoming observations with TESS and PLATO will
substantially improve the observational searches for and understanding
of the exoplanetary rings combined with the current result of the {\it
Kepler} data.

\section*{Acknowledgements}

We are grateful to the {\it Kepler} team for making the revolutionary
data publicly available.  This work is supported by JSPS Grant-in-Aids
for Scientific Research No.14J07182 (M.A.), No.17K14246 (H.K.), and
No.24340035 (Y.S.) as well as by JSPS Core-to-Core Program
``International Network of Planetary Sciences''. M.A. is supported by the
Advanced Leading Graduate Course for Photon Science (ALPS).  This work
was performed in part under contract with the Jet Propulsion Laboratory
(JPL) funded by NASA through the Sagan Fellowship Program executed by
the NASA Exoplanet Science Institute.

\appendix

\section*{Appendix}

\section{Timescale for tidal alignment of planetary ring}

The deviation of the lightcurve due to a ring relative to a ringless
planet model prediction crucially depends on the size, opacity and
orientions of the ring. In turn, a useful constraint on the size of the
ring is placed only if the orientation of the ring is well specified.
The ring axis is most likely aligned with the planetary spin axis.  In
the case of close-in planets as we mainly consider in the present paper,
the planetary spin axis is expected to be tidally aligned with that of
the planetary orbit. Therefore the ring axis in such tidally aligned
systems can be specified physically.

The damping timescale, which is comparable to 
the spin-orbit synchronization timescale, is 
given by 
\begin{equation}
t_{\rm damp} \simeq \frac{ 2 C Q_{\rm p}}{3 k_{\rm p}} 
\left(\frac{M_{\rm p}}{M_{\star}} \right)\left(\frac{a}{R_{\rm p}}\right)^{3} 
\left(\frac{P_{\rm orb}}{2 \pi}\right), \label{damp_2}
\end{equation}
\citep[e.g.][]{2011ApJ...734..117S}.  In the above equation, 
$a$ is the semi-major axis of the planetary orbit, 
$R_{\rm p}$ is the planetary radius, $P_{\rm
orb}$ is the planetary orbital period, 
$M_{\rm p}$ is the planetary mass, $M_{\star}$ is
the stellar mass, $C$ is the dimensionless moment of inertia of the
planet (i.e., divided by $M_{\rm p}R^{2}_{\rm p, eq}$ with $R_{\rm p,
eq}$ being the equatorial radius of the planet), $Q_{\rm p}$ is the
tidal dissipation function of the planet, and $k_{\rm p}$ is the Love number. 

We estimate $t_{\rm damp}$ for our target systems using the parameters
from the Q1--Q17 Data Release 25 catalog of KOIs
\citep{2017arXiv171006758T}, and list the values in Tables \ref{final_table1} to
\ref{final_table3}. In doing so, we adopt typical values of $Q_{\rm p} =
10^{6.5}$, $C=0.25$, and $k_{\rm p}= 1.5$
. The adopted value of $Q_{\rm p}$ is supposed to be typical for gas
giants, but that for rocky planets would be substantially smaller.
Thus the values listed in Tables \ref{final_table1} to
\ref{final_table3} would be significantly over-estimated for
rocky planets.

For the majority of systems, the planetary mass $M_{\rm p}$ is not
directly measured. Thus we adopt Eq. (8) of \cite{2013ApJ...768...14W},
and rewrite it as
\begin{equation}
\frac{M_{\rm p}}{M_{\oplus}} = 
0.337  \left(\frac{R_{\rm p}}
{R_{\oplus }}\right)^{1/0.53}  
\left(\frac{F}{  {\rm erg s}^{-1} {\rm cm}^{-2} } \right)^{0.03/0.53}, 
\label{mp_weiss}
\end{equation}
where $M_{\oplus}$ and $R_{ \oplus}$ are the mass and radius of Earth,
and $F$ is the incident flux of the host star received at the location
of the planet:
\begin{equation}
F = \frac{\sigma_{\rm SB} T_{\rm eff}^{4}R_{\star}^{2}}{ 4 \pi a^{2}},
\end{equation}
with $\sigma_{\rm SB}$ is the Stefan-Boltzmann constant.  For example,
if we consider the Hot Jupiter ($a$=0.05 AU) around the Sun, we obtain $F =
5.46 \times 10^{8} \;{\rm erg s}^{-1} {\rm cm}^{-2}$.

We compute $M_{\rm p}$ from $R_{\rm p}$ in the Kepler catalog 
for 155 systems.  According to Eq. (\ref{mp_weiss}), the remaining 13
systems have $M_{\rm p}>M_{\rm J}$ and we set $M_{\rm p}=M_{\rm J}$ for
such systems, since Eq. (8) of \cite{2013ApJ...768...14W} cannot be
applied for that range. Because we use the values of $M_{\rm p}$ only in
computing their $t_{\rm damp}$, that simple estimate does not change our
result.

\section{Closer consideration of individual systems with $p<0.05$
\label{search_result}}

The analysis described in Section \ref{result} leaves 29 systems with
$p<0.05$.  Their lightcurves are carefully examined and compared with
the expected ring signature. It turned out that they are not caused by
the presence of a ring. We describe the origin of those anomalies
individually here. They are interesting objects themselves, and also
provide useful examples of possible false-positives for future ring
searches.

\subsection{Gravity darkening: KOI-2.01 and 13.01}

Fast rotating stars have higher (lower) effective surface temperature in
the polar (equatorial) regions because of the stronger centrifugal force
along the equatorial plane. Thus the transit lightcurve becomes
asymmetric with respect to the central transit time depending on the path
of the planet. The anomaly due gravity darkening is not confined
preferrentially around the ingress or egress phases unlike the ring
signature (see Fig. \ref{KOI97_curve} for example), and can be
distingushed easily by eye.

\begin{figure}[htpb]
\begin{center}
 \includegraphics[width = 0.68 \linewidth]{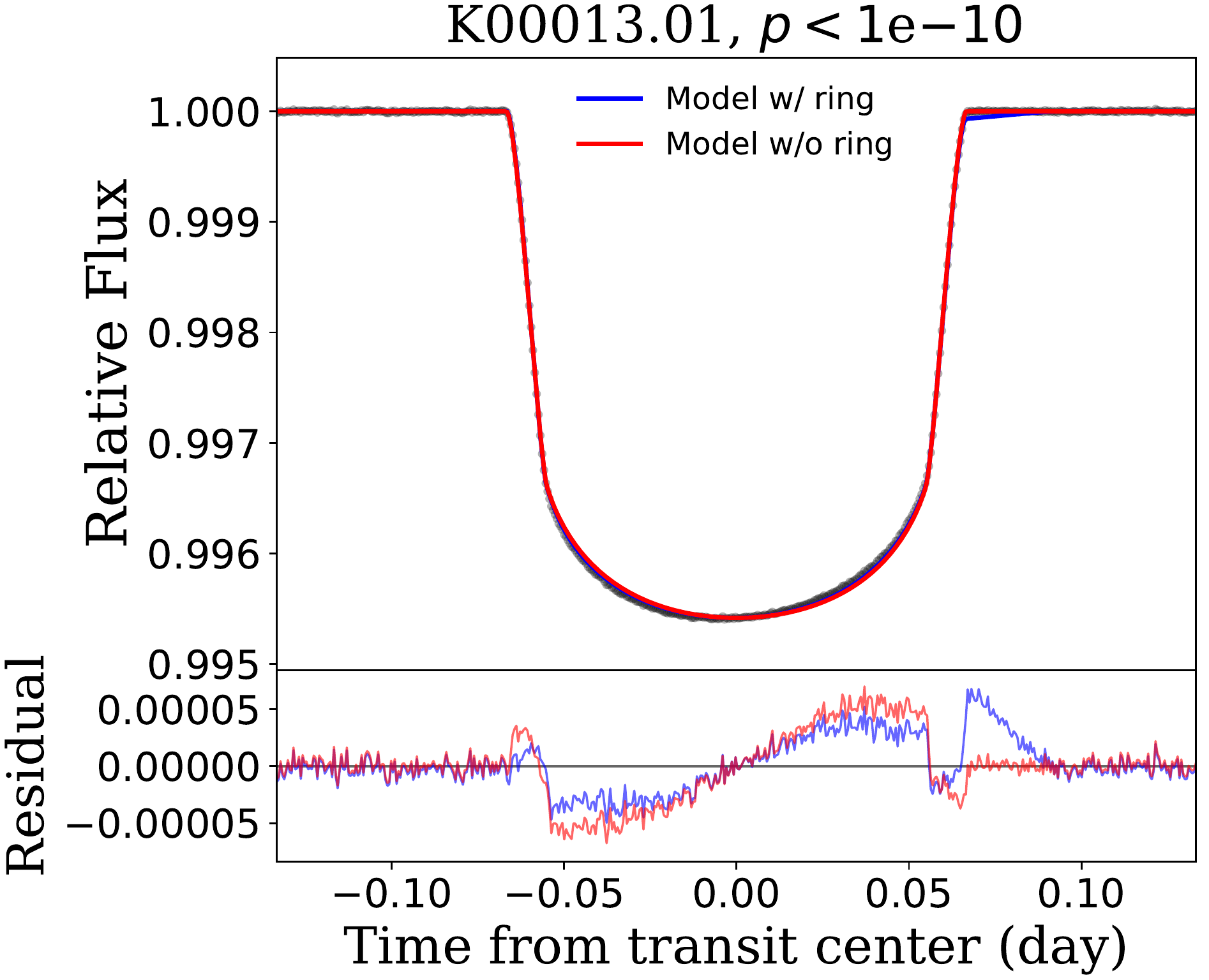}
\caption{Lightcurve of a gravity darkened system, KOI-13.01 (Kepler 13 b).} \label{KOI_13}
\end{center}
\end{figure}

Figure \ref{KOI_13} shows a lightcurve of our tentative candidate
KOI-13.01 (Kepler-13 b), which cannot be well fitted anyway even by
adding a ring. This system was analysed first by
\cite{2011ApJS..197...10B}, who found that the lightcurve is very well
explained by gravity darkening.  \citet{2015ApJ...805...28M} presented a
further elaborated analysis of KOI-13 (Kepler-13), as well as another
gravity darkened system, KOI-2, in our targets.

\subsection{Evaporation of atmosphere: KOI-3794.01}

\begin{figure}[htpb]
\begin{center}
 \includegraphics[width = 12cm]{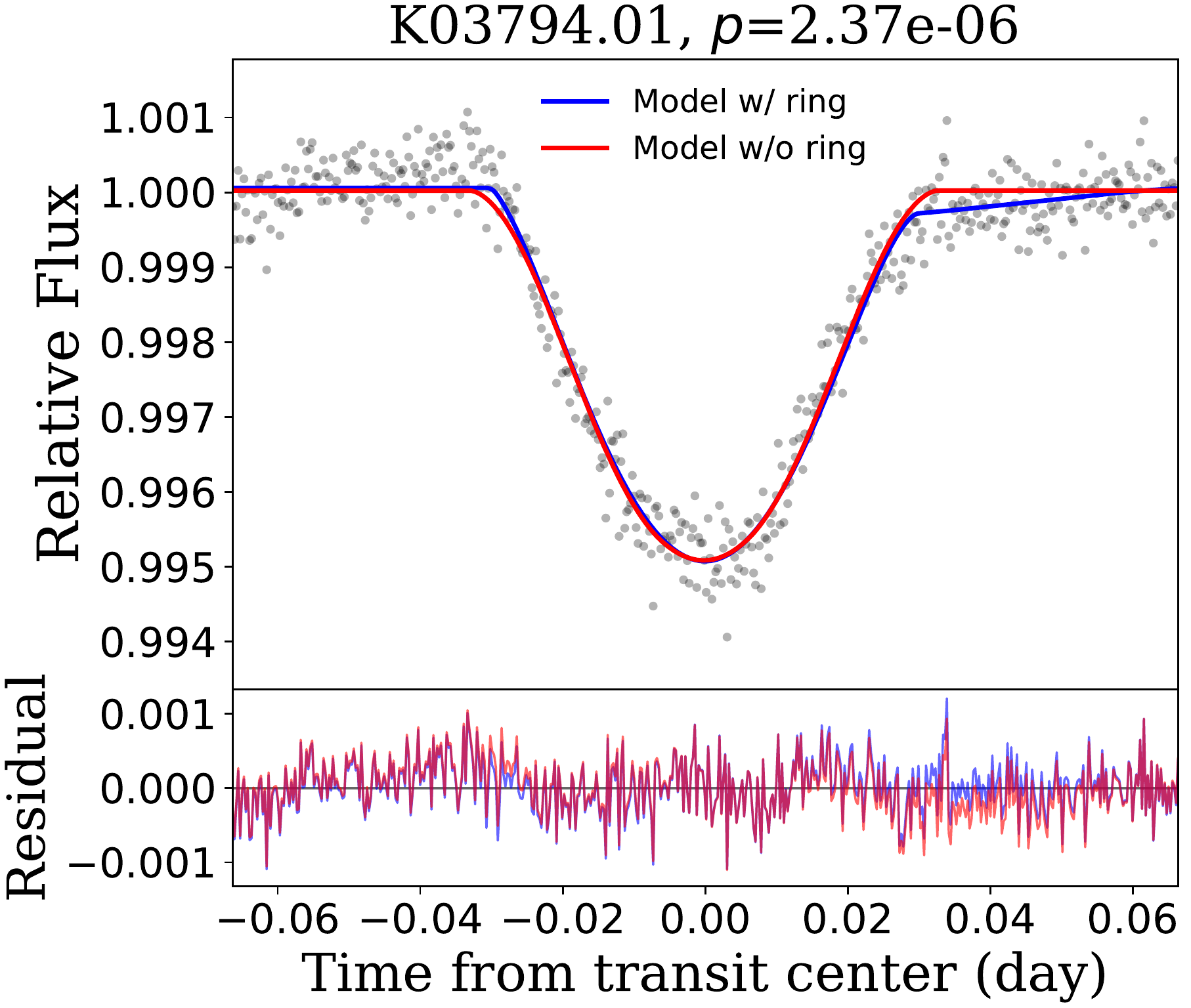}
\caption{Lightcurve of an evaporating planet KOI-3794.01 (Kepler-1520 b)
} \label{KOI_3794}
\end{center}
\end{figure}

Another tentative candidate, KOI-3794.01 (KIC 12557548, {\it
Kepler}-1520 b), is known as an evaporating planet
\cite[e.g.][]{2012ApJ...752....1R}, whose lightcurve is shown in Figure
\ref{KOI_3794}.  Indeed, the transit depth of the lightcurves at
different epochs (before phase-folded) exhibits significant
time-variation, which is inconsistent with the ring hypothesis.

\subsection{Spot crossing during transit: KOI-3.01, 63.01, 676.01, 1353.01, 1416.01, 1539.01, 1714.01, 1729.01, and 6016.01}

\begin{figure}[htpb]
\begin{center}
 \includegraphics[width =12cm]{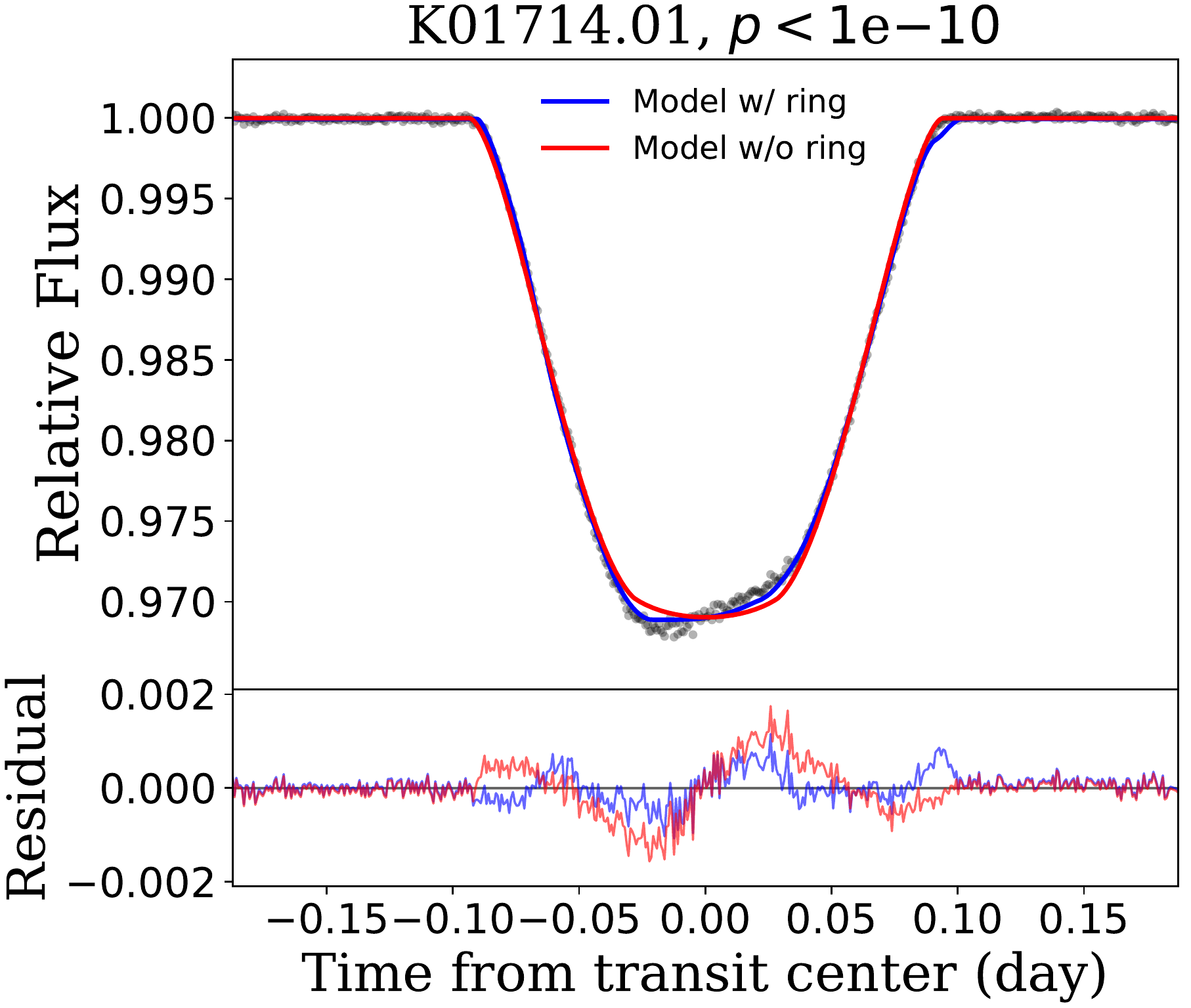}
\caption{Lightcurve of a spot-crossing event, KOI-1714.01. 
} \label{KOI_1714}
\end{center}
\end{figure}

Stellar spots add non-negligible anomalous features in the transit
lightcurves. Among the 29 tentative candidates with $p<0.05$, we find
that 9 systems are likely explained by spot-crossing events, not by a ring. 
As a significant example, we show the phase-folded lightcurve of 
KOI-1714.01 in Figure \ref{KOI_1714}, where the entire flux is strongly 
affected by by spot-crossing events. 

Spot-crossing features have been already reported for four systems out of 
9 systems; KOI-3.01 ({\it Kepler}-3b) show frequent spot-crossing anomalies at
fairly similar phases, and its planetary orbit is estimated to be
misaligned relative to the stellar spin
\citep{2011ApJ...743...61S}. 
Combining the spot anomalies and the Rossiter-McLaughlin effect of
KOI-63.01 ({\it Kepler}-63 b), \cite{2013ApJ...775...54S} concluded that
the system has a large spin-orbit misalignment of $\Psi = 104^{\circ}$.
Also the variability of lightcurves due to spot-crossing events have
been reported for KOI-676.01 ({\it Kepler}-210 c) by
\citet{2013ApJ...775...54S}, and for KOI-1353.01 ({\it Kepler}-289 c) by
\citet{2014ApJ...795..167S}.

The other five systems KOI-1416.01({\it Kepler}-850 b), 
1539.01, 1714.01, 1729.01, and 6016.01 are classified 
as possible false positives in {\it Kepler} CFOP webpages, 
and we confirmed that there are no ring-like signatures. 

\subsection{False anomalies due to an inaccurate choice 
of a transit center: KOI-70.02, 102.01, and 148.01 \label{wrong}}

Phase-folded lightcurves of KOI-70.02, 102.01, and 148.01 show anomalous
features around egress and ingress phases.  The transit depth of those
three systems is very small, and we suspect that the anomalies are
simply caused by inaccurate central transit epochs in phase-folding.

\begin{figure}[htpb]
\begin{center}
 \includegraphics[width =7cm]{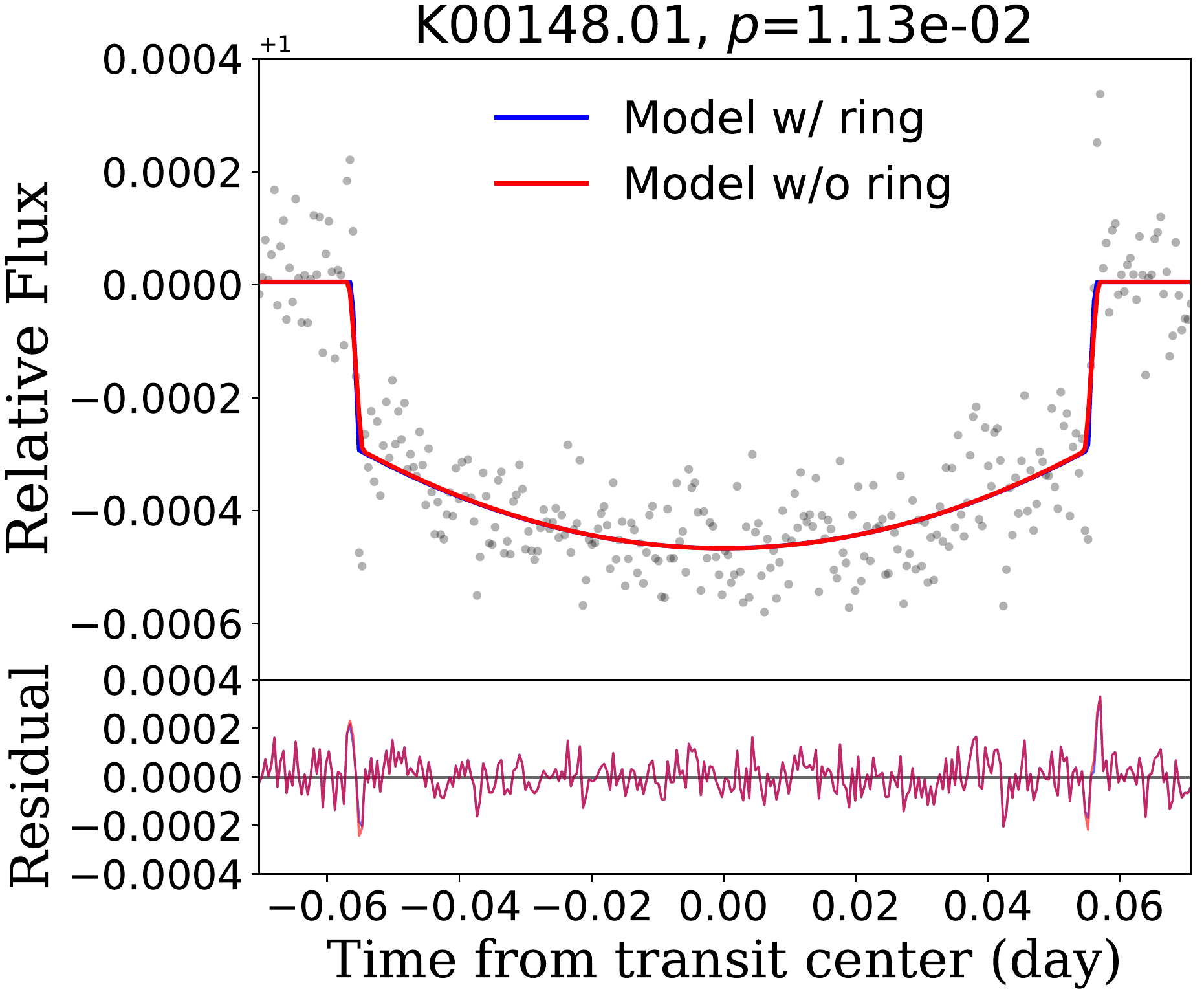}
  \includegraphics[width =7cm]{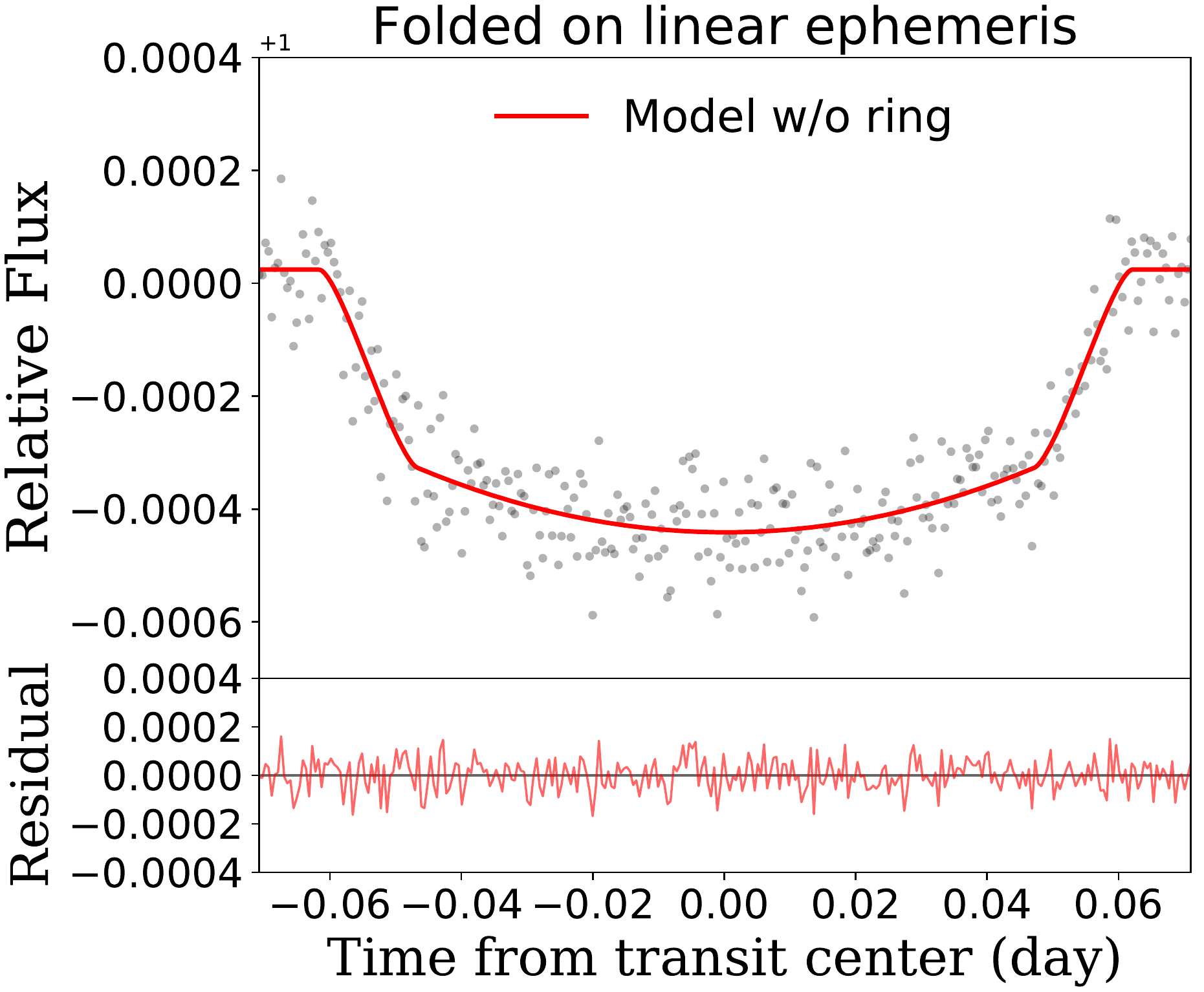} 
\caption{Two different phase-folded lightcurves of KOI-148.01.
}
\label{KOI_148}
\end{center}
\end{figure}

Figure \ref{KOI_148} shows an example for KOI-148.01. In the left panel, 
we show the lightcurve, which is folded as described in \S 3.1. As shown in 
the left panel, the anomalous features appear around the egress and ingress. Then, 
to find out the origin of the anomaly, we create a phase-folded lightcurve 
using a linear ephemeris. Specifically, when we fit the individual transit, 
we fix each transit center to $t_{\rm cen, i} = t_{\rm cen, 0} + {\rm i} P_{\rm orb}$, where 
$t_{\rm cen, i}$ is the transit center at the i-th transit. Here, we retrieve  
$t_{\rm cen, 0}$ and $P_{\rm orb}$ from the {\it Kepler} catalog. 
The refolded lightcurve is plotted in the right panel, which show that the
anomalous features disappear. We made sure that this is also the case for the other two
systems, KOI 70.02 and 102.01.

\subsection{Systems without statistical significance:
 KOI-4.01, 5.01, 212.01, 214.01, 257.01, 423.01, 433.02, 531.01, 686.01,
 872.01, and 1131.01}

\begin{figure*}[htbp]\begin{center}
\caption{\small Lightcurves of 11 systems without statistical significance listed in B.5 \label{weak_transits}}
\includegraphics[width=0.49 \linewidth]{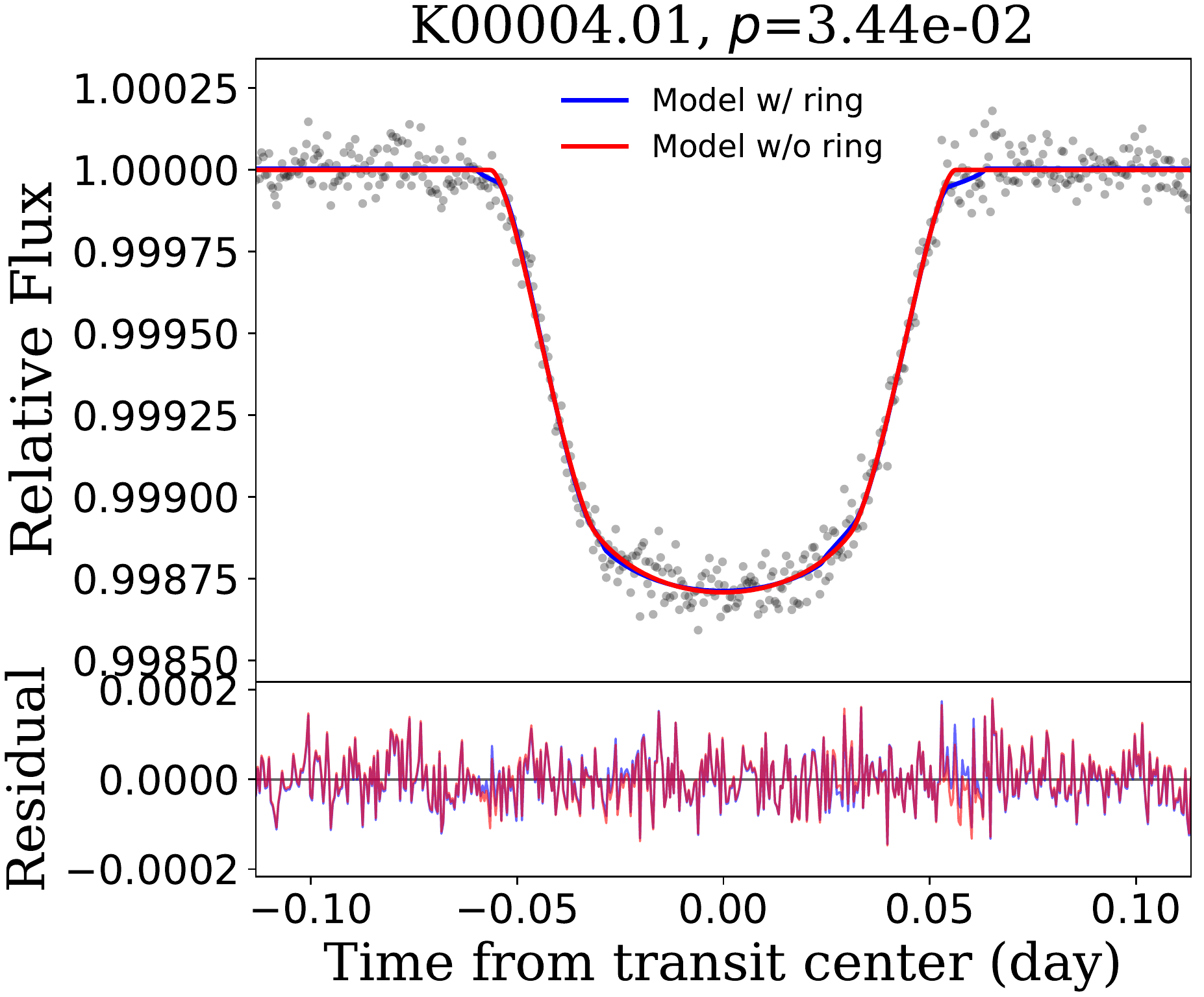}\hfill
\includegraphics[width=0.49 \linewidth]{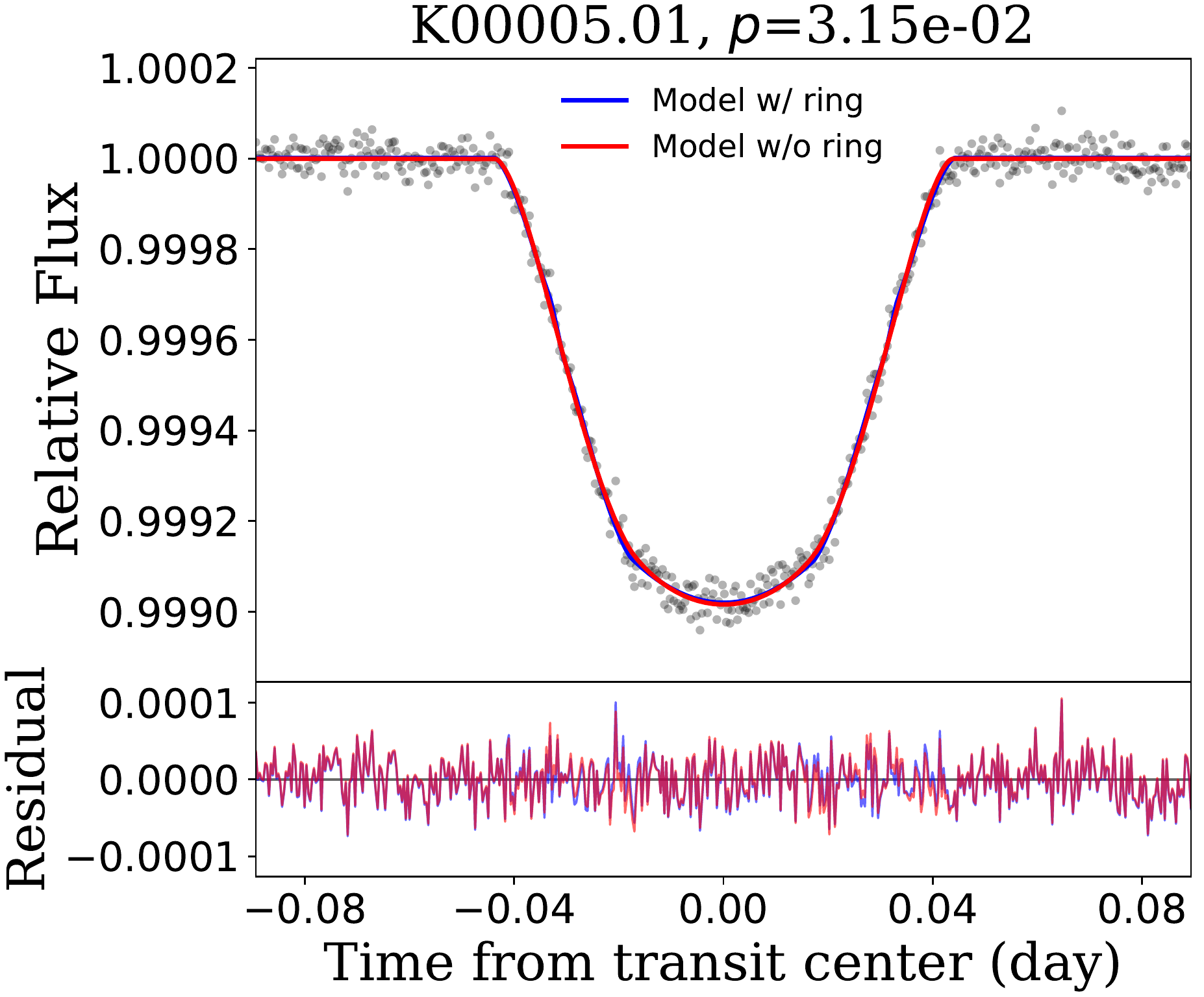}
\includegraphics[width=0.49 \linewidth]{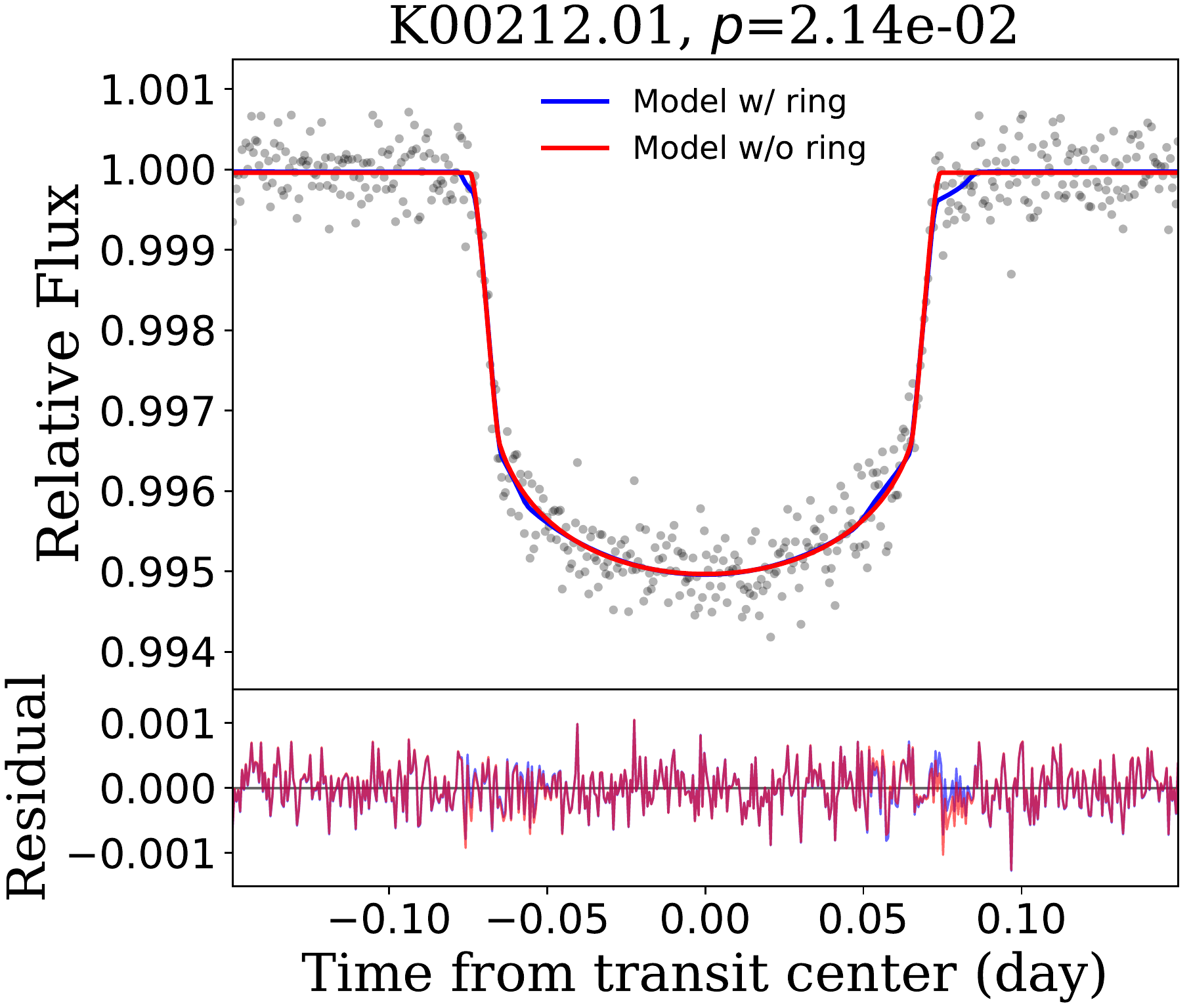}\hfill
\includegraphics[width=0.49 \linewidth]{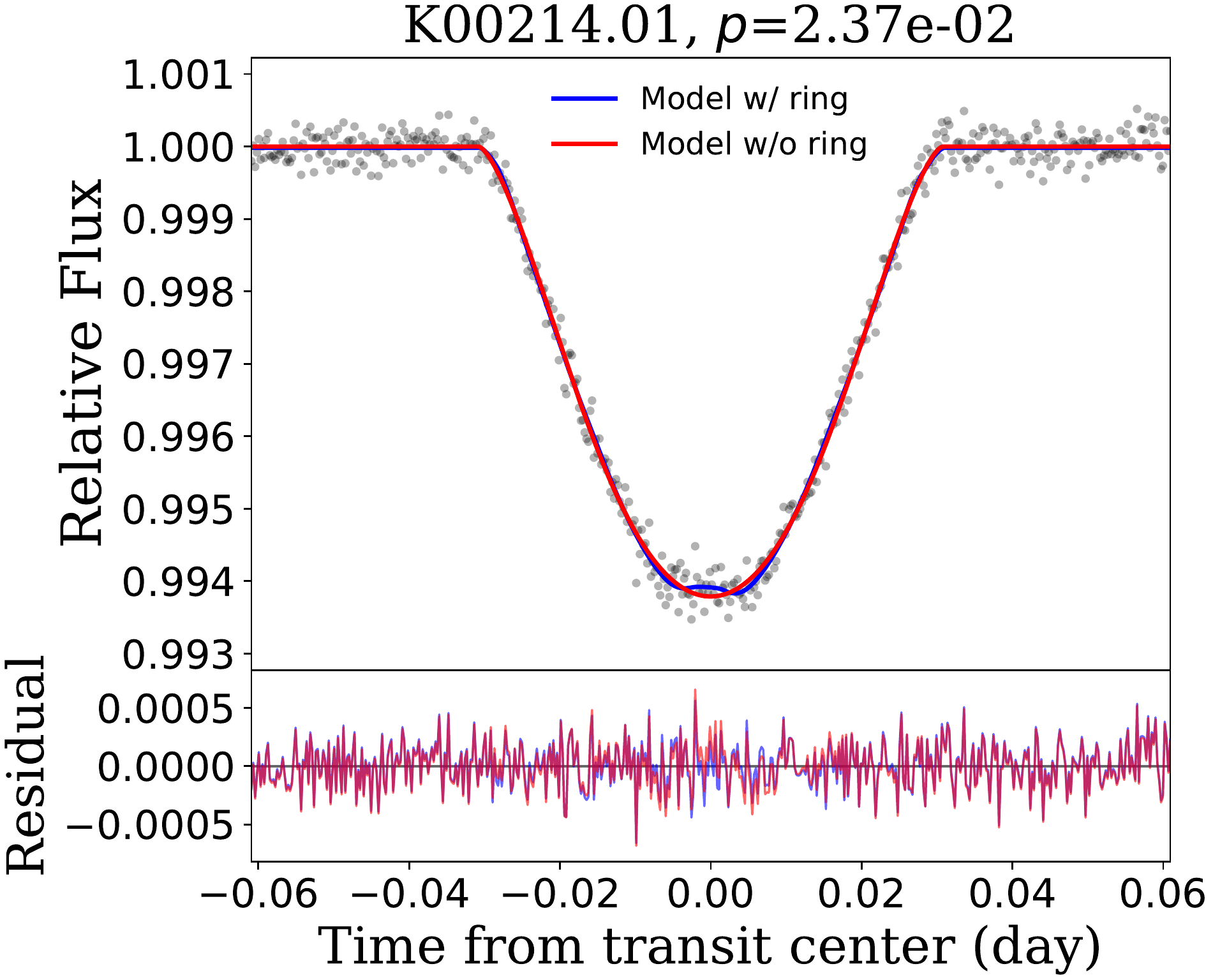}
\includegraphics[width=0.49 \linewidth]{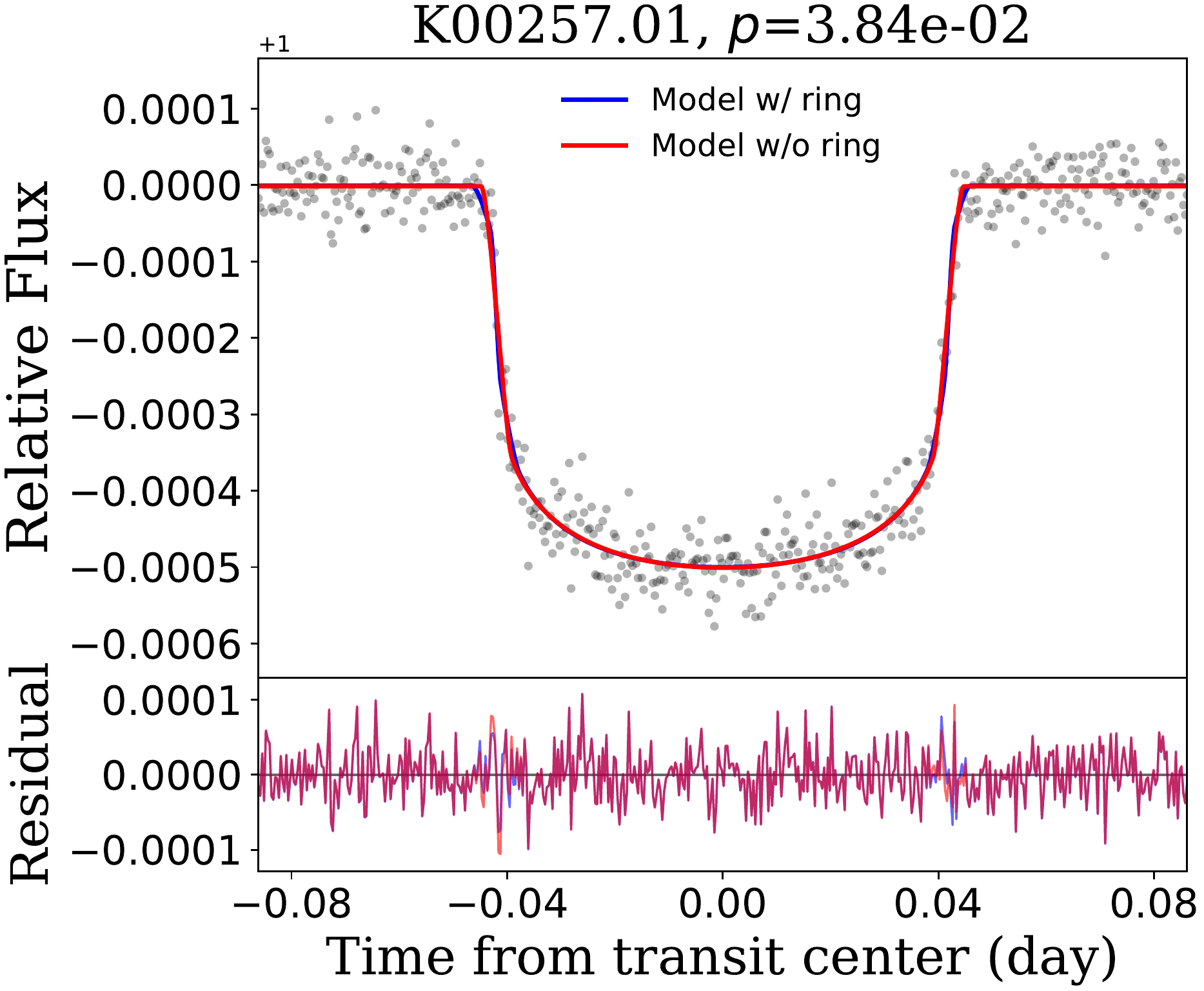}\hfill
\includegraphics[width=0.49 \linewidth]{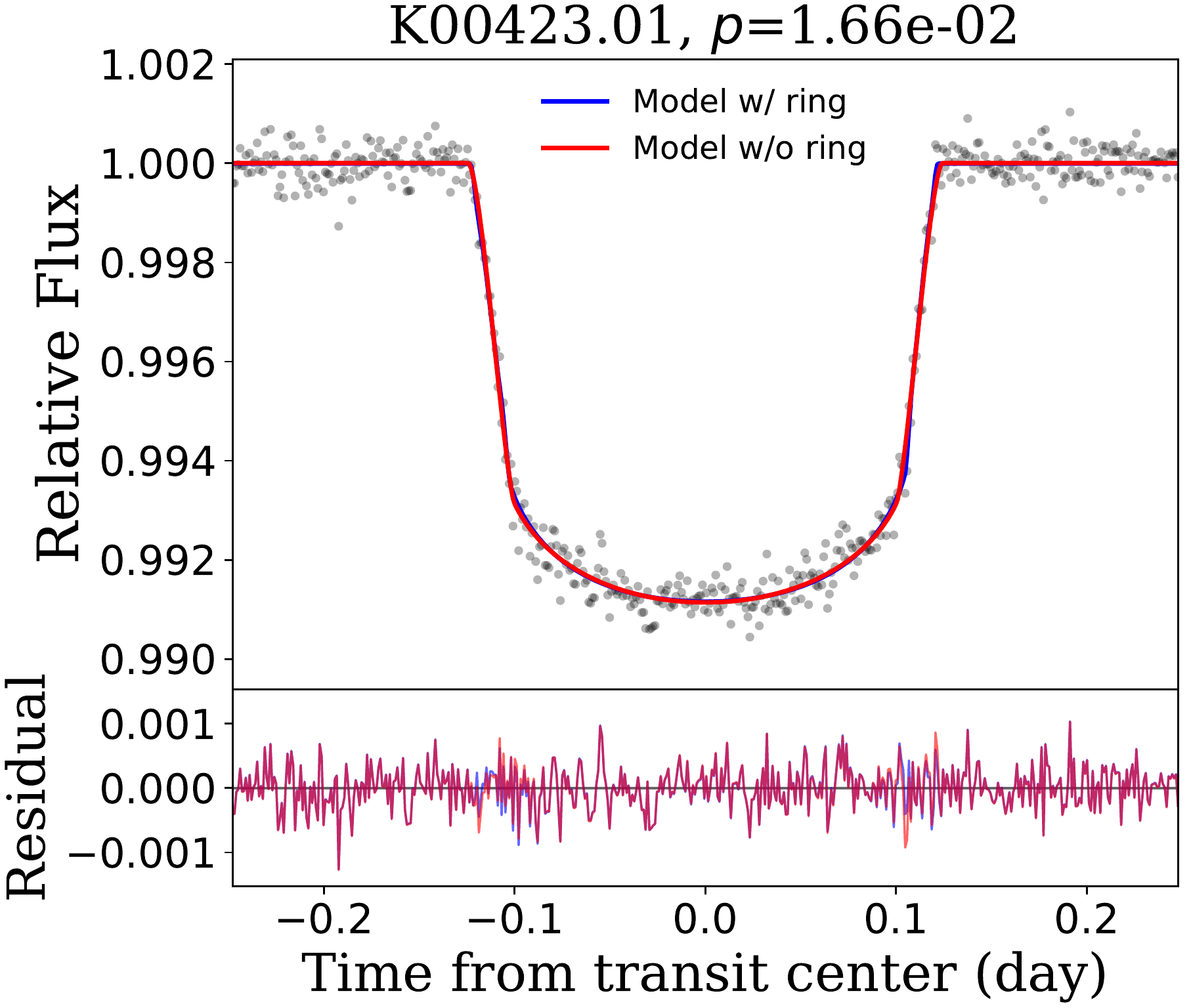}
\end{center}\end{figure*}

Out of the 168 targets, we find 11 systems that marginally favor the
ring model at $2-3\sigma$ levels: KOI-4.01, 5.01, 212.01, 214.01,
257.01, 423.01, 433.02, 531.01, 686.01, 872.01, and 1131.01.  Figure
\ref{weak_transits} shows their lightcurves as well as the best-fit
model with and without a ring (in blue and red lines, respectively)).

To examine their significance, we divide their individual transit
lightcurves into two groups as described in subsection
\ref{p_search_even_odd}. If the anomaly is really caused by a ring,
both $p_{\rm even}$ and $p_{\rm odd}$ should remain small.

We find that 9 systems have both of  $p_{\rm even}$ and $p_{\rm odd}$ 
larger than 0.32 (i.e., 1$\sigma$), and two systems
have both $p_{\rm even}$ and $p_{\rm odd}$ with merely 1$\sigma$
significance: $(0.11, 0.12)$ for KOI-4.01 and (0.029, 0.037) for
KOI-257.01. Even though the two systems are likely to be statistical
flukes, we examined the lightcurve visually in any case.  The lightcurve
of KOI-4.01 seems marginally consistent with the ring signature, but the
amplitude is so tiny and can be easily produced by random noise.  The
features of KOI-257.01 are likely to produced by the folding procedure
as we discussed in Sec \ref{wrong} because the transit depth is so
small.

We note that the rejection of the null hypothesis of a ring with the
level of $p=0.05$ implies that $168\times 0.05 = 8.4$ systems are
expected to show 2$\sigma$ signals even if there is no ring at all.
Thus 11 marginal systems even if there is no ring system are fairly
consistent with our choice of the threshold.

\begin{figure*}[htbp]\begin{center}
\includegraphics[width=0.49 \linewidth]{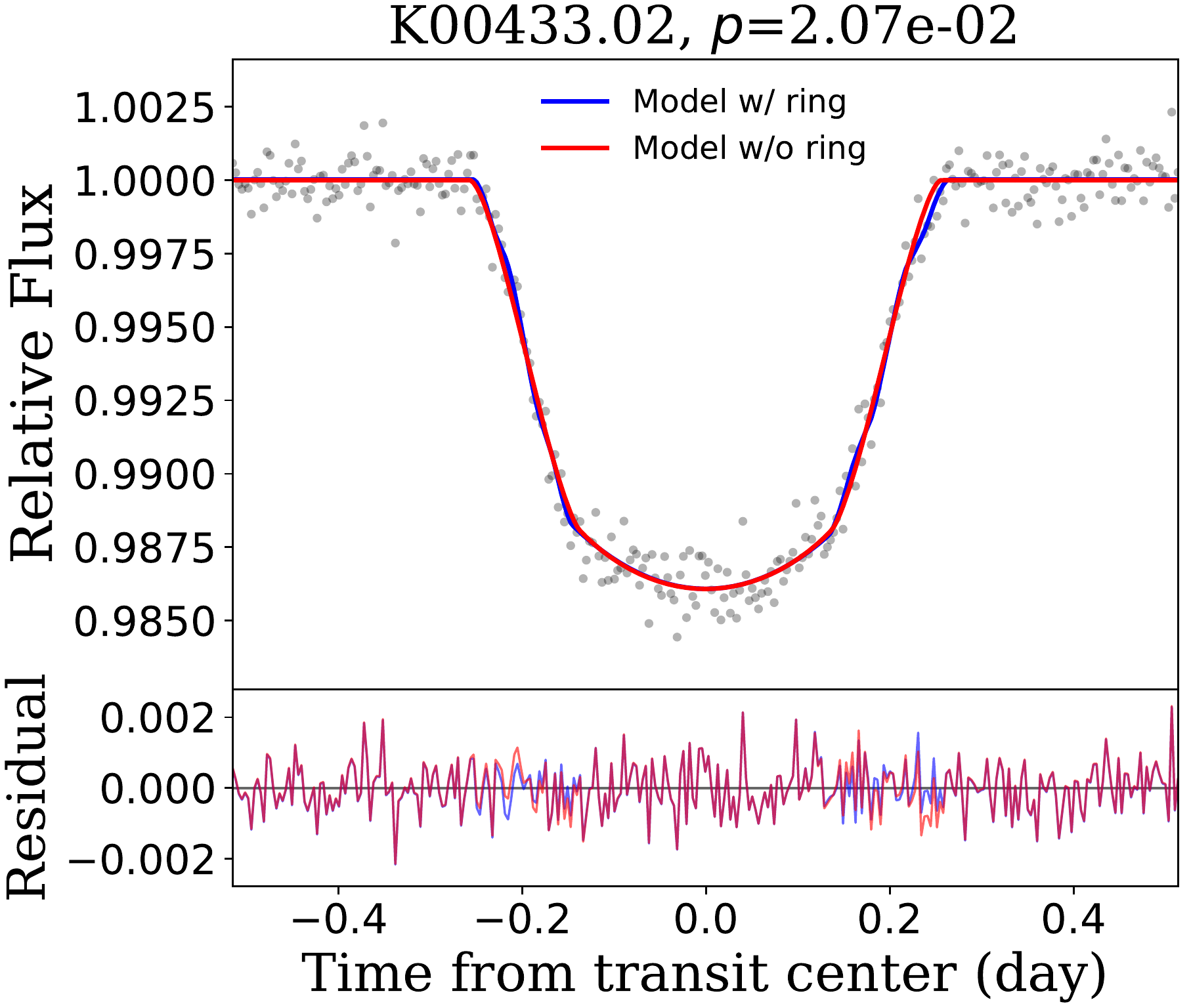}\hfill
\includegraphics[width=0.49 \linewidth]{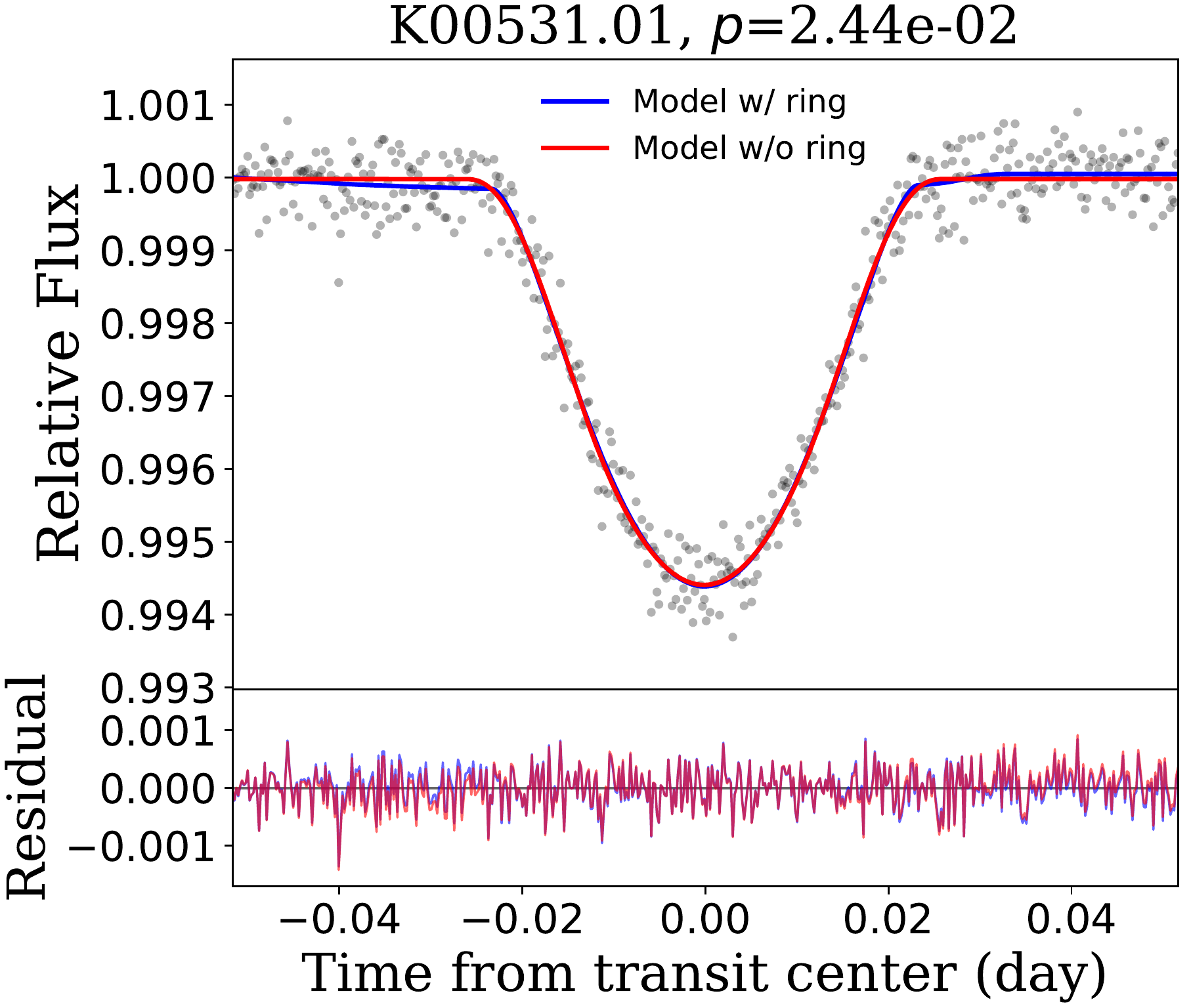}
\includegraphics[width=0.49 \linewidth]{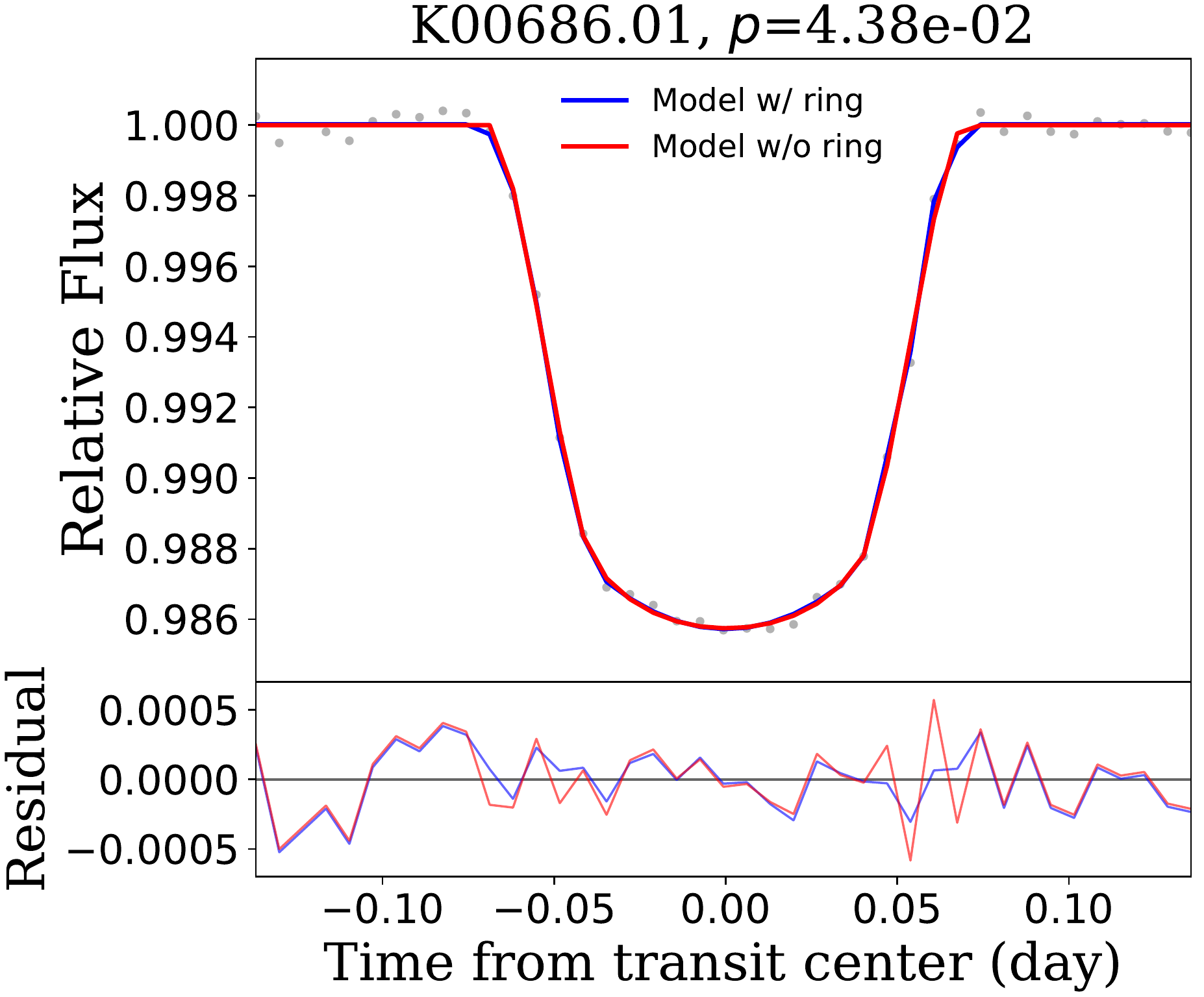}\hfill
\includegraphics[width=0.49 \linewidth]{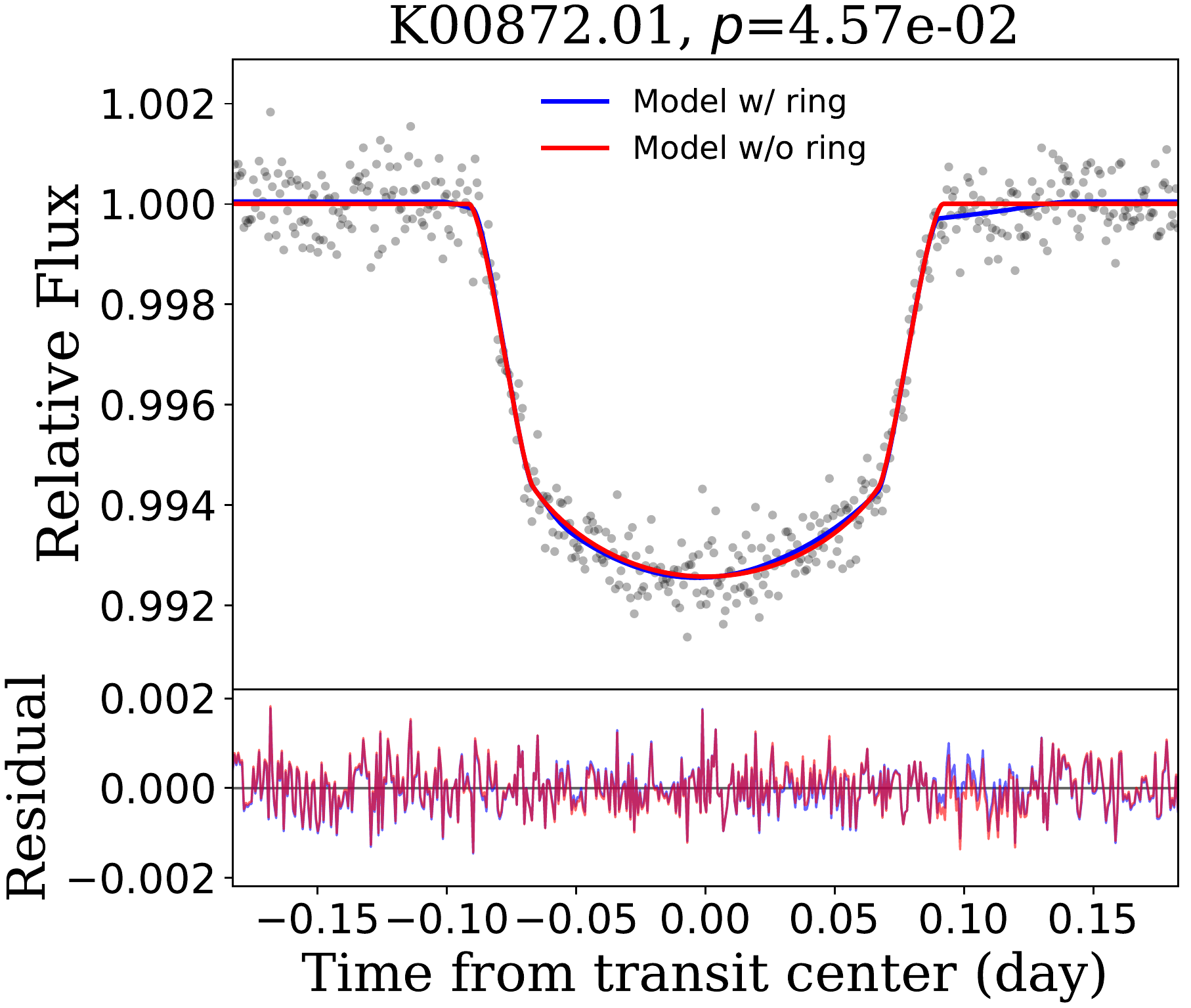}
\includegraphics[width=0.49 \linewidth]{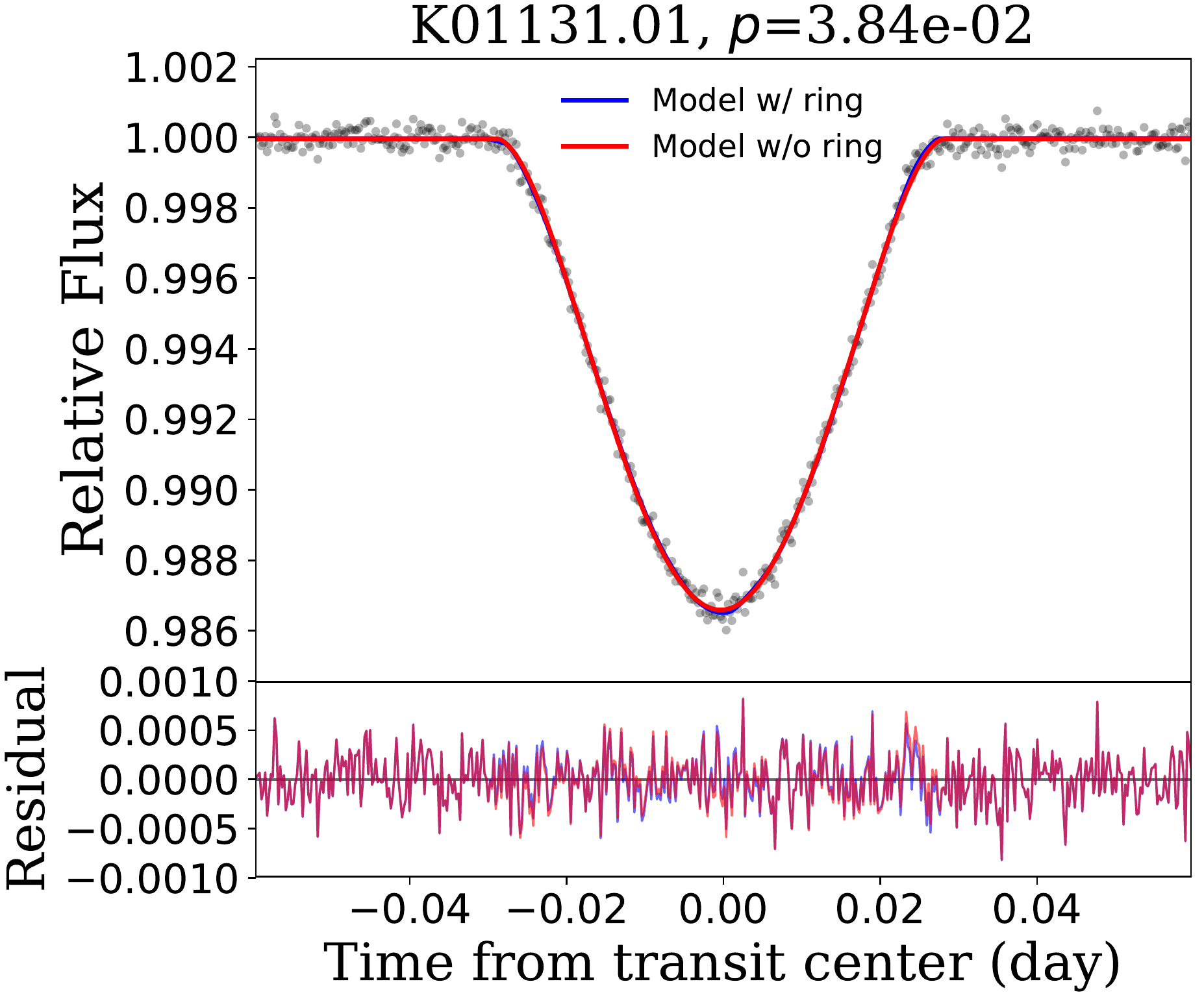}\hfill
\end{center} \end{figure*}

\subsection{The remaining systems: KOI-12.01, 868.01, and 971.01}

Finally, we consider the remaining three systems that have not been
discussed.

The lightcurve of KOI-12.01 ({\it Kepler}-448 b) shows anomalous
features during the transit, which are significant during -0.03 days to
0.1 days with respect to the central transit epoch. 
We find that such large pulse-like signals appear also during
out-of-transit. Thus they are likely due to
stellar activities.

The lightcurves of KOI-971.01 (KIC 11180361) show 
strong stellar activities, which are typical for multiple star systems 
\citep{2015MNRAS.450.2764N}, and CFOP
webpages also identify this system as false positive. Thus, 
the planetary rings are not origins of the signals. 

\begin{figure}[htpb]
\begin{center}
\includegraphics[width = 7cm]{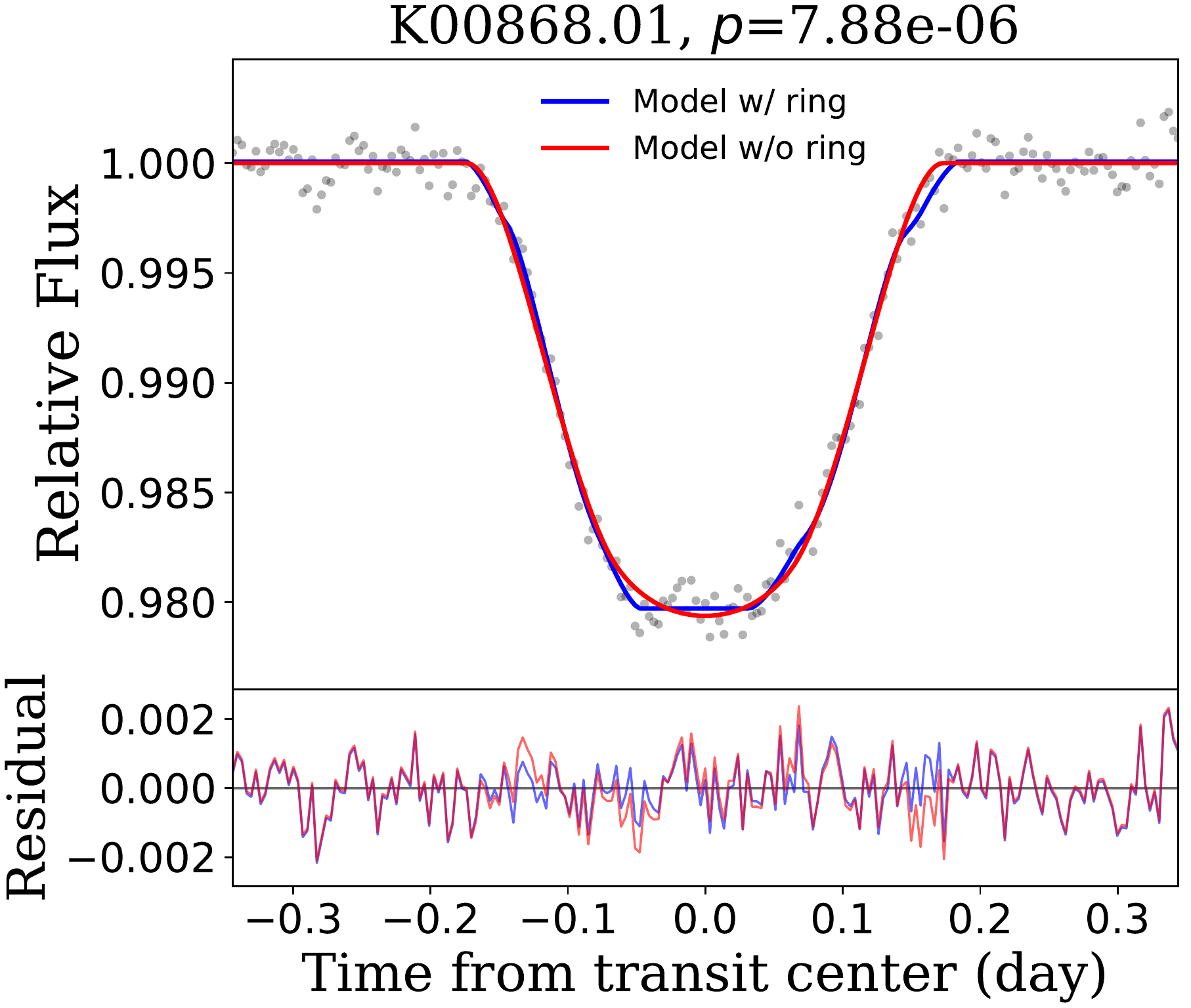}
\includegraphics[width = 7cm]{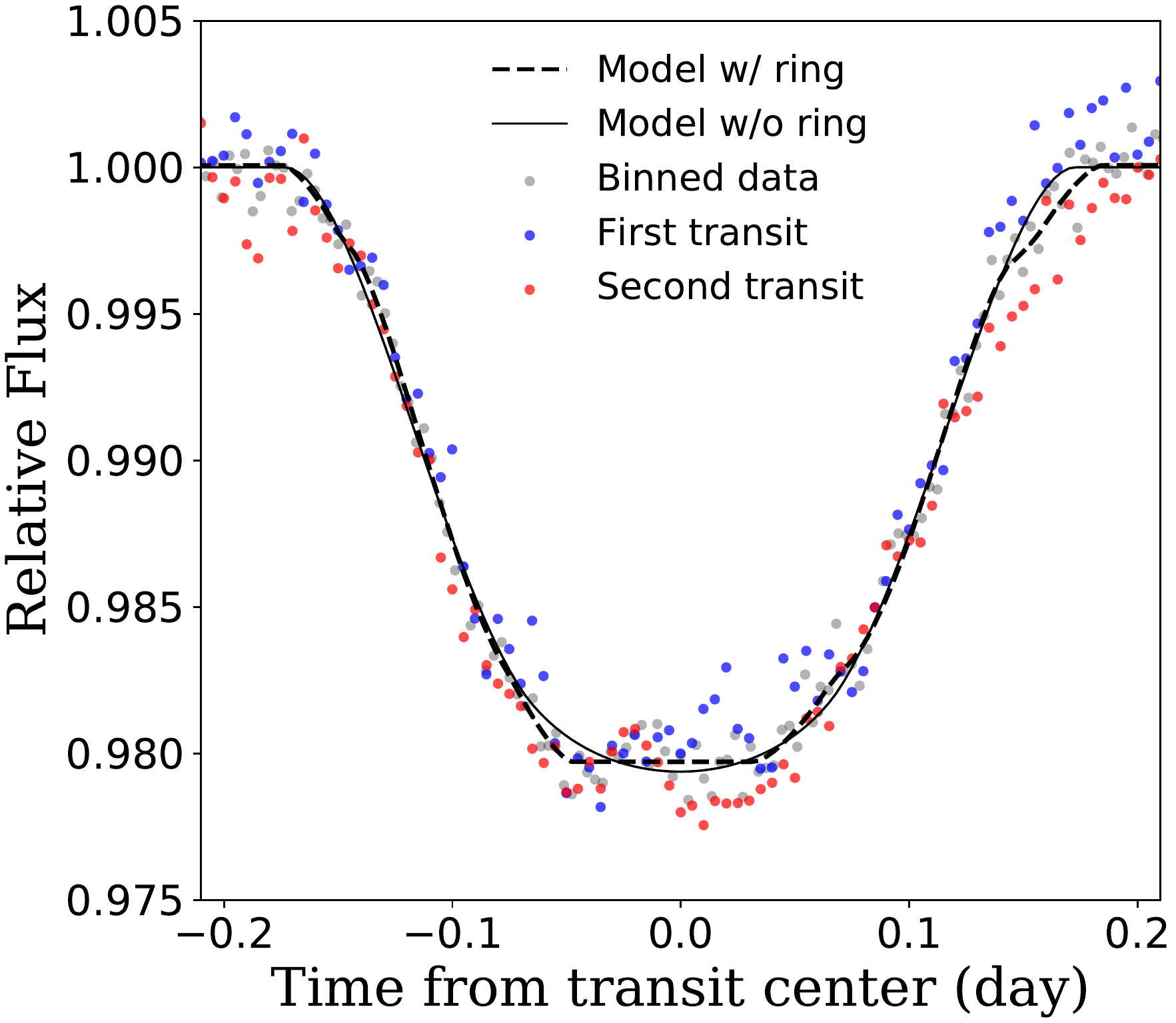}
\caption{Lightcurves of KOI-868.01.
Left panel indicates binned lightcurve (black circles) along wih
the best-fits of the ringed and ringless models.
Right panels shows the comparison for two different transits.
} \label{KOI_868}
\end{center}
\end{figure}

The lightcurve of KOI-868.01 shows an anomaly during the
egress, which is shown in the left panel of Figure \ref{KOI_868} along
with the best-fit models.  The fit yields $\chi^{2}_{\rm ringless,\;min}/{\rm
dof} = 202.0/190$, $\chi^{2}_{\rm ring, \;min}/{\rm dof} = 171.6/195$, and
$p=7.88\times10^{-6}$.  The analysis based on the binned data supports a
Neptune-sized ringed planet of an orbital period of 236 days.  The
best-fit ring model gives $\theta = 25.5\pm 10.0^{\circ} $, $\phi = 12.4
\pm 3.7^{\circ} $, $T = 0.46 \pm 0.18$, $r_{\rm in/p} = 1.88\pm0.36$,
and $r_{\rm out/in} = 1.63 \pm 0.43$.  The radius ratio $R_{\rm
p}/R_{\star} = 0.099\pm 0.012$ gives $R_{\rm p}/R_{\rm J} = 0.63 \pm
0.08$ assuming the stellar radius $R_{\star} = 0.657
^{+0.022}_{-0.032}\, R_{\odot}$.  The non-vanishing obliquity is 
consistent with the long alignment timescale $t_{\rm damp} =
2.95$ Gyr.

In order to check the consistency of signals, we calculate the
$p$-values for the two transits in the short-cadence data separately. As
a result, we find $p=0.76$ and $7.4$e-07 for the first and second
transits, respectively. Indeed as indicated in the right panel
of Figure \ref{KOI_868}, the lightcurves at the first and second
transits are systematically different.  Therefore KOI-868.01 is unlikely
to be a ringed planet.  We do not understand the origin of the anomalies
because there are only two transits, but suspect that temporal stellar
activities or spot-crossing events are responsible.

\newpage

\bibliography{IOPEXPORT_BIB}

\begin{thebibliography}{}
\expandafter\ifx\csname natexlab\endcsname\relax\def\natexlab#1{#1}\fi

\bibitem[{{Aizawa} {et~al.}(2017){Aizawa}, {Uehara}, {Masuda}, {Kawahara}, \&
  {Suto}}]{2017AJ....153..193A}
{Aizawa}, M., {Uehara}, S., {Masuda}, K., {Kawahara}, H., \& {Suto}, Y. 2017,
  \aj, 153, 193

\bibitem[{{Akinsanmi} {et~al.}(2018){Akinsanmi}, {Oshagh}, {Santos}, \&
  {Barros}}]{2018A&A...609A..21A}
{Akinsanmi}, B., {Oshagh}, M., {Santos}, N.~C., \& {Barros}, S.~C.~C. 2018,
  \aap, 609, A21

\bibitem[{{Anglada} {et~al.}(2017){Anglada}, {Amado}, {Ortiz}, {G{\'o}mez},
  {Mac{\'{\i}}as}, {Alberdi}, {Osorio}, {G{\'o}mez}, {de Gregorio-Monsalvo},
  {P{\'e}rez-Torres}, {Anglada-Escud{\'e}}, {Berdi{\~n}as}, {Jenkins},
  {Jimenez-Serra}, {Lara}, {L{\'o}pez-Gonz{\'a}lez}, {L{\'o}pez-Puertas},
  {Morales}, {Ribas}, {Richards}, {Rodr{\'{\i}}guez-L{\'o}pez}, \&
  {Rodriguez}}]{2017ApJ...850L...6A}
{Anglada}, G., {Amado}, P.~J., {Ortiz}, J.~L., {et~al.} 2017, \apjl, 850, L6

\bibitem[{{Arnold} \& {Schneider}(2004)}]{2004A&A...420.1153A}
{Arnold}, L., \& {Schneider}, J. 2004, \aap, 420, 1153

\bibitem[{{Barnes} \& {Fortney}(2004)}]{2004ApJ...616.1193B}
{Barnes}, J.~W., \& {Fortney}, J.~J. 2004, \apj, 616, 1193

\bibitem[{{Barnes} {et~al.}(2011){Barnes}, {Linscott}, \&
  {Shporer}}]{2011ApJS..197...10B}
{Barnes}, J.~W., {Linscott}, E., \& {Shporer}, A. 2011, \apjs, 197, 10

\bibitem[{{Braga-Ribas} {et~al.}(2014){Braga-Ribas}, {Sicardy}, {Ortiz},
  {Snodgrass}, {Roques}, {Vieira-Martins}, {Camargo}, {Assafin}, {Duffard},
  {Jehin}, {Pollock}, {Leiva}, {Emilio}, {Machado}, {Colazo}, {Lellouch},
  {Skottfelt}, {Gillon}, {Ligier}, {Maquet}, {Benedetti-Rossi}, {Gomes},
  {Kervella}, {Monteiro}, {Sfair}, {El Moutamid}, {Tancredi}, {Spagnotto},
  {Maury}, {Morales}, {Gil-Hutton}, {Roland}, {Ceretta}, {Gu}, {Wang},
  {Harps{\o}e}, {Rabus}, {Manfroid}, {Opitom}, {Vanzi}, {Mehret}, {Lorenzini},
  {Schneiter}, {Melia}, {Lecacheux}, {Colas}, {Vachier}, {Widemann},
  {Almenares}, {Sandness}, {Char}, {Perez}, {Lemos}, {Martinez},
  {J{\o}rgensen}, {Dominik}, {Roig}, {Reichart}, {Lacluyze}, {Haislip},
  {Ivarsen}, {Moore}, {Frank}, \& {Lambas}}]{2014Natur.508...72B}
{Braga-Ribas}, F., {Sicardy}, B., {Ortiz}, J.~L., {et~al.} 2014, \nat, 508, 72

\bibitem[{{Brown} {et~al.}(2001){Brown}, {Charbonneau}, {Gilliland}, {Noyes},
  \& {Burrows}}]{2001ApJ...552..699B}
{Brown}, T.~M., {Charbonneau}, D., {Gilliland}, R.~L., {Noyes}, R.~W., \&
  {Burrows}, A. 2001, \apj, 552, 699

\bibitem[{{Carter} \& {Winn}(2010)}]{2010ApJ...709.1219C}
{Carter}, J.~A., \& {Winn}, J.~N. 2010, \apj, 709, 1219

\bibitem[{{Dyudina} {et~al.}(2005){Dyudina}, {Sackett}, {Bayliss}, {Seager},
  {Porco}, {Throop}, \& {Dones}}]{2005ApJ...618..973D}
{Dyudina}, U.~A., {Sackett}, P.~D., {Bayliss}, D.~D.~R., {et~al.} 2005, \apj,
  618, 973

\bibitem[{{Hatchett} {et~al.}(2018){Hatchett}, {Barnes}, {Ahlers}, {MacKenzie},
  \& {Hedman}}]{2018NewA...60...88H}
{Hatchett}, W.~T., {Barnes}, J.~W., {Ahlers}, J.~P., {MacKenzie}, S.~M., \&
  {Hedman}, M.~M. 2018, \na, 60, 88

\bibitem[{{Hedman}(2015)}]{2015ApJ...801L..33H}
{Hedman}, M.~M. 2015, \apjl, 801, L33

\bibitem[{{Heising} {et~al.}(2015){Heising}, {Marcy}, \&
  {Schlichting}}]{2015ApJ...814...81H}
{Heising}, M.~Z., {Marcy}, G.~W., \& {Schlichting}, H.~E. 2015, \apj, 814, 81

\bibitem[{{Kalas} {et~al.}(2008){Kalas}, {Graham}, {Chiang}, {Fitzgerald},
  {Clampin}, {Kite}, {Stapelfeldt}, {Marois}, \& {Krist}}]{2008Sci...322.1345K}
{Kalas}, P., {Graham}, J.~R., {Chiang}, E., {et~al.} 2008, Science, 322, 1345

\bibitem[{{Kipping}(2013)}]{2013MNRAS.435.2152K}
{Kipping}, D.~M. 2013, \mnras, 435, 2152

\bibitem[{{Lecavelier des Etangs} {et~al.}(2017){Lecavelier des Etangs},
  {H{\'e}brard}, {Blandin}, {Cassier}, {Deeg}, {Bonomo}, {Bouchy},
  {D{\'e}sert}, {Ehrenreich}, {Deleuil}, {D{\'{\i}}az}, {Moutou}, \&
  {Vidal-Madjar}}]{2017A&A...603A.115L}
{Lecavelier des Etangs}, A., {H{\'e}brard}, G., {Blandin}, S., {et~al.} 2017,
  \aap, 603, A115

\bibitem[{{Lissauer} {et~al.}(2011){Lissauer}, {Fabrycky}, {Ford}, {Borucki},
  {Fressin}, {Marcy}, {Orosz}, {Rowe}, {Torres}, {Welsh}, {Batalha}, {Bryson},
  {Buchhave}, {Caldwell}, {Carter}, {Charbonneau}, {Christiansen}, {Cochran},
  {Desert}, {Dunham}, {Fanelli}, {Fortney}, {Gautier}, {Geary}, {Gilliland},
  {Haas}, {Hall}, {Holman}, {Koch}, {Latham}, {Lopez}, {McCauliff}, {Miller},
  {Morehead}, {Quintana}, {Ragozzine}, {Sasselov}, {Short}, \&
  {Steffen}}]{2011Natur.470...53L}
{Lissauer}, J.~J., {Fabrycky}, D.~C., {Ford}, E.~B., {et~al.} 2011, \nat, 470,
  53

\bibitem[{{Mamajek} {et~al.}(2012){Mamajek}, {Quillen}, {Pecaut}, {Moolekamp},
  {Scott}, {Kenworthy}, {Collier Cameron}, \& {Parley}}]{2012AJ....143...72M}
{Mamajek}, E.~E., {Quillen}, A.~C., {Pecaut}, M.~J., {et~al.} 2012, \aj, 143,
  72

\bibitem[{{Mandel} \& {Agol}(2002)}]{2002ApJ...580L.171M}
{Mandel}, K., \& {Agol}, E. 2002, \apjl, 580, L171

\bibitem[{{Markwardt}(2009)}]{2009ASPC..411..251M}
{Markwardt}, C.~B. 2009, in Astronomical Society of the Pacific Conference
  Series, Vol. 411, Astronomical Data Analysis Software and Systems XVIII, ed.
  D.~A. {Bohlender}, D.~{Durand}, \& P.~{Dowler}, 251

\bibitem[{{Masuda}(2015)}]{2015ApJ...805...28M}
{Masuda}, K. 2015, \apj, 805, 28

\bibitem[{{Niemczura} {et~al.}(2015){Niemczura}, {Murphy}, {Smalley},
  {Uytterhoeven}, {Pigulski}, {Lehmann}, {Bowman}, {Catanzaro}, {van Aarle},
  {Bloemen}, {Briquet}, {De Cat}, {Drobek}, {Eyer}, {Gameiro}, {Gorlova},
  {Kami{\'n}ski}, {Lampens}, {Marcos-Arenal}, {P{\'a}pics}, {Vandenbussche},
  {Van Winckel}, {St{\c e}{\'s}licki}, \& {Fagas}}]{2015MNRAS.450.2764N}
{Niemczura}, E., {Murphy}, S.~J., {Smalley}, B., {et~al.} 2015, \mnras, 450,
  2764

\bibitem[{{Ohta} {et~al.}(2009){Ohta}, {Taruya}, \&
  {Suto}}]{2009ApJ...690....1O}
{Ohta}, Y., {Taruya}, A., \& {Suto}, Y. 2009, \apj, 690, 1

\bibitem[{{Ortiz} {et~al.}(2015){Ortiz}, {Duffard}, {Pinilla-Alonso},
  {Alvarez-Candal}, {Santos-Sanz}, {Morales}, {Fern{\'a}ndez-Valenzuela},
  {Licandro}, {Campo Bagatin}, \& {Thirouin}}]{2015A&A...576A..18O}
{Ortiz}, J.~L., {Duffard}, R., {Pinilla-Alonso}, N., {et~al.} 2015, \aap, 576,
  A18

\bibitem[{{Ortiz} {et~al.}(2017){Ortiz}, {Santos-Sanz}, {Sicardy},
  {Benedetti-Rossi}, {B{\'e}rard}, {Morales}, {Duffard}, {Braga-Ribas}, {Hopp},
  {Ries}, {Nascimbeni}, {Marzari}, {Granata}, {P{\'a}l}, {Kiss}, {Pribulla},
  {Kom{\v z}{\'{\i}}k}, {Hornoch}, {Pravec}, {Bacci}, {Maestripieri}, {Nerli},
  {Mazzei}, {Bachini}, {Martinelli}, {Succi}, {Ciabattari}, {Mikuz},
  {Carbognani}, {Gaehrken}, {Mottola}, {Hellmich}, {Rommel},
  {Fern{\'a}ndez-Valenzuela}, {Campo Bagatin}, {Cikota}, {Cikota}, {Lecacheux},
  {Vieira-Martins}, {Camargo}, {Assafin}, {Colas}, {Behrend}, {Desmars},
  {Meza}, {Alvarez-Candal}, {Beisker}, {Gomes-Junior}, {Morgado}, {Roques},
  {Vachier}, {Berthier}, {Mueller}, {Madiedo}, {Unsalan}, {Sonbas}, {Karaman},
  {Erece}, {Koseoglu}, {Ozisik}, {Kalkan}, {Guney}, {Niaei}, {Satir},
  {Yesilyaprak}, {Puskullu}, {Kabas}, {Demircan}, {Alikakos}, {Charmandaris},
  {Leto}, {Ohlert}, {Christille}, {Szak{\'a}ts}, {Tak{\'a}csn{\'e} Farkas},
  {Varga-Vereb{\'e}lyi}, {Marton}, {Marciniak}, {Bartczak}, {Santana-Ros},
  {Butkiewicz-B{\c a}k}, {Dudzi{\'n}ski}, {Al{\'{\i}}-Lagoa}, {Gazeas},
  {Tzouganatos}, {Paschalis}, {Tsamis}, {S{\'a}nchez-Lavega},
  {P{\'e}rez-Hoyos}, {Hueso}, {Guirado}, {Peris}, \&
  {Iglesias-Marzoa}}]{2017Natur.550..219O}
{Ortiz}, J.~L., {Santos-Sanz}, P., {Sicardy}, B., {et~al.} 2017, \nat, 550, 219

\bibitem[{{Osborn} {et~al.}(2017){Osborn}, {Rodriguez}, {Kenworthy}, {Kennedy},
  {Mamajek}, {Robinson}, {Espaillat}, {Armstrong}, {Shappee}, {Bieryla},
  {Latham}, {Anderson}, {Beatty}, {Berlind}, {Calkins}, {Esquerdo}, {Gaudi},
  {Hellier}, {Holoien}, {James}, {Kochanek}, {Kuhn}, {Lund}, {Pepper},
  {Pollacco}, {Prieto}, {Siverd}, {Stassun}, {Stevens}, {Stanek}, \&
  {West}}]{2017MNRAS.471..740O}
{Osborn}, H.~P., {Rodriguez}, J.~E., {Kenworthy}, M.~A., {et~al.} 2017, \mnras,
  471, 740

\bibitem[{{Parviainen}(2015)}]{2015MNRAS.450.3233P}
{Parviainen}, H. 2015, \mnras, 450, 3233

\bibitem[{{Protassov} {et~al.}(2002){Protassov}, {van Dyk}, {Connors},
  {Kashyap}, \& {Siemiginowska}}]{2002ApJ...571..545P}
{Protassov}, R., {van Dyk}, D.~A., {Connors}, A., {Kashyap}, V.~L., \&
  {Siemiginowska}, A. 2002, \apj, 571, 545

\bibitem[{{Rappaport} {et~al.}(2012){Rappaport}, {Levine}, {Chiang}, {El
  Mellah}, {Jenkins}, {Kalomeni}, {Kite}, {Kotson}, {Nelson},
  {Rousseau-Nepton}, \& {Tran}}]{2012ApJ...752....1R}
{Rappaport}, S., {Levine}, A., {Chiang}, E., {et~al.} 2012, \apj, 752, 1

\bibitem[{{Sanchis-Ojeda} \& {Winn}(2011)}]{2011ApJ...743...61S}
{Sanchis-Ojeda}, R., \& {Winn}, J.~N. 2011, \apj, 743, 61

\bibitem[{{Sanchis-Ojeda} {et~al.}(2013){Sanchis-Ojeda}, {Winn}, {Marcy},
  {Howard}, {Isaacson}, {Johnson}, {Torres}, {Albrecht}, {Campante}, {Chaplin},
  {Davies}, {Lund}, {Carter}, {Dawson}, {Buchhave}, {Everett}, {Fischer},
  {Geary}, {Gilliland}, {Horch}, {Howell}, \& {Latham}}]{2013ApJ...775...54S}
{Sanchis-Ojeda}, R., {Winn}, J.~N., {Marcy}, G.~W., {et~al.} 2013, \apj, 775,
  54

\bibitem[{{Santos} {et~al.}(2015){Santos}, {Martins}, {Bou{\'e}}, {Correia},
  {Oshagh}, {Figueira}, {Santerne}, {Sousa}, {Melo}, {Montalto}, {Boisse},
  {Ehrenreich}, {Lovis}, {Pepe}, {Udry}, \& {Garcia
  Munoz}}]{2015A&A...583A..50S}
{Santos}, N.~C., {Martins}, J.~H.~C., {Bou{\'e}}, G., {et~al.} 2015, \aap, 583,
  A50

\bibitem[{{Schlichting} \& {Chang}(2011)}]{2011ApJ...734..117S}
{Schlichting}, H.~E., \& {Chang}, P. 2011, \apj, 734, 117

\bibitem[{{Schmitt} {et~al.}(2014){Schmitt}, {Agol}, {Deck}, {Rogers}, {Gazak},
  {Fischer}, {Wang}, {Holman}, {Jek}, {Margossian}, {Omohundro}, {Winarski},
  {Brewer}, {Giguere}, {Lintott}, {Lynn}, {Parrish}, {Schawinski}, {Schwamb},
  {Simpson}, \& {Smith}}]{2014ApJ...795..167S}
{Schmitt}, J.~R., {Agol}, E., {Deck}, K.~M., {et~al.} 2014, \apj, 795, 167

\bibitem[{{Schneider}(1999)}]{1999CRASB.327..621S}
{Schneider}, J. 1999, Academie des Sciences Paris Comptes Rendus Serie B
  Sciences Physiques, 327, 621

\bibitem[{{Thompson} {et~al.}(2017){Thompson}, {Coughlin}, {Hoffman},
  {Mullally}, {Christiansen}, {Burke}, {Bryson}, {Batalha}, {Haas},
  {Catanzarite}, {Rowe}, {Barentsen}, {Caldwell}, {Clarke}, {Jenkins}, {Li},
  {Latham}, {Lissauer}, {Mathur}, {Morris}, {Seader}, {Smith}, {Klaus},
  {Twicken}, {Wohler}, {Akeson}, {Ciardi}, {Cochran}, {Barclay}, {Campbell},
  {Chaplin}, {Charbonneau}, {Henze}, {Howell}, {Huber}, {Prsa}, {Ramirez},
  {Morton}, {Christensen-Dalsgaard}, {Dotson}, {Doyle}, {Dunham}, {Dupree},
  {Ford}, {Geary}, {Girouard}, {Isaacson}, {Kjeldsen}, {Steffen}, {Quintana},
  {Ragozzine}, {Shporer}, {Silva Aguirre}, {Still}, {Tenenbaum}, {Welsh},
  {Wolfgang}, {Zamudio}, {Koch}, \& {Borucki}}]{2017arXiv171006758T}
{Thompson}, S.~E., {Coughlin}, J.~L., {Hoffman}, K., {et~al.} 2017, ArXiv
  e-prints, arXiv:1710.06758

\bibitem[{{Weiss} {et~al.}(2013){Weiss}, {Marcy}, {Rowe}, {Howard}, {Isaacson},
  {Fortney}, {Miller}, {Demory}, {Fischer}, {Adams}, {Dupree}, {Howell},
  {Kolbl}, {Johnson}, {Horch}, {Everett}, {Fabrycky}, \&
  {Seager}}]{2013ApJ...768...14W}
{Weiss}, L.~M., {Marcy}, G.~W., {Rowe}, J.~F., {et~al.} 2013, \apj, 768, 14

\bibitem[{{Zuluaga} {et~al.}(2015){Zuluaga}, {Kipping}, {Sucerquia}, \&
  {Alvarado}}]{2015ApJ...803L..14Z}
{Zuluaga}, J.~I., {Kipping}, D.~M., {Sucerquia}, M., \& {Alvarado}, J.~A. 2015,
  \apjl, 803, L14

\end{thebibliography}
\newpage \renewcommand{\arraystretch}{1.12}
\begin{landscape}\thispagestyle{empty}
{\setlength{\tabcolsep}{2pt}\footnotesize\setlength{\LTleft}{1cm plus -1fill}
\footnotetext[1]{Values from {\it Kepler} Object of Interest (KOI) Catalog Q1-Q17 DR 25(https://exoplanetarchive.ipac.caltech.edu/)}
\footnotetext[2]{\small FP=Possible False Positive (https://exofop.ipac.caltech.edu/cfop.php); GD = Gravity Darkening (B.1); Evap = Evaporating planet (B.2); Spot=Spot Crossing (B.3); Bad Fold = incorrect data folding (B.4); Small = non-significant signal (B.5); Others = B.6; }
\begin{longtable}{c c c c c c c c c c}\caption{Parameters and statistics of 13 systems with $p<0.001$}\label{final_table1} \\ \hline
 KOI& {\it Kepler}&$P^{1}_{\rm orb}$ (day)&$t_{\rm damp}$ (Gyr)& $(R_{\rm out}/R_{\rm p} )_{\rm upp, \;Aligned}$ & $(R_{\rm p}/R_{\star})_{\rm ringless}$& $p$ & ($\chi^{2}_{\rm ringless, min}$, $\chi^{2}_{\rm ring, min}$, $N_{\rm bin}$) & $(S/N)$& Comment$^{2}$
\endfirsthead
KOI& {\it Kepler}&$P_{\rm orb}$ (day)&$t_{\rm damp}$ (Gyr)& $(R_{\rm out}/R_{\rm p} )_{\rm upp, \;Aligned}\;$ & $(R_{\rm p}/R_{\star})_{\rm ringless}$& $p$ & ($\chi^{2}_{\rm ringless, min}$, $\chi^{2}_{\rm ring, min}$, $N_{\rm bin}$) & $(S/N)$& Comment$^{2}$ \endhead \hline
2.01 & 2 b & 2.20 & 2.60e-05 & 1.11 & 0.07755 $\pm$ 2e-05 & 5.02e-07 & (534.22, 494.94, 500) & 4357.03 & GD\\ 
3.01 & 3 b & 4.89 & 8.76e-04 & 2.75 & 0.05886 $\pm$ 3e-05 & 2.60e-09 & (2011.23, 1820.95, 500) & 2403.33 & Spot\\ 
13.01 & 13 b & 1.76 & 7.92e-06 & 1.63 & 0.064683 $\pm$ 4e-06 & $<$1e-10 & (8830.73, 5904.78, 500) & 6359.84 & GD\\ 
63.01 & 63 b & 9.43 & 5.18e-03 & 2.44 & 0.06481 $\pm$ 4e-05 & $<$1e-10 & (1222.04, 1078.89, 500) & 732.29 & Spot\\ 
102.01 & \nodata & 1.74 & 7.01e-05 & 5.51 & 0.02810 $\pm$ 6e-05 & 9.78e-05 & (554.11, 525.66, 500) & 432.74 & Bad Fold\\ 
676.01 & 210 c & 7.97 & 5.35e-03 & 7.20 & 0.0520 $\pm$ 5e-04 & 2.00e-10 & (693.78, 620.69, 500) & 302.61 & Spot\\ 
868.01 & \nodata & 236.00 & 2.95e+01 & 6.65 & 0.144 $\pm$ 1e-03 & 7.88e-06 & (202.04, 171.63, 203) & 182.23 & Others\\ 
971.01 & \nodata & 0.53 & 9.00e-10 & 8.13 & 0.1 $\pm$ 1e+00 & 3.54e-04 & (319.16, 304.58, 500) & 257.01 & FP \& Others\\ 
1416.01 & 840 b & 2.50 & 3.55e-05 & 1.74 & 0.1459 $\pm$ 2e-04 & 5.14e-04 & (579.31, 553.80, 500) & 919.15 & FP \& Spot\\ 
1539.01 & \nodata & 2.82 & 1.20e-05 & 1.10 & 0.2568 $\pm$ 2e-04 & $<$1e-10 & (982.31, 791.58, 500) & 1364.17 & FP \& Spot\\ 
1714.01 & \nodata & 2.74 & 5.29e-06 & 1.10 & 0.17618 $\pm$ 2e-05 & $<$1e-10 & (6978.85, 2755.22, 500) & 688.33 & FP \& Spot\\ 
1729.01 & \nodata & 5.20 & 2.88e-04 & 1.80 & 0.1764 $\pm$ 3e-04 & $<$1e-10 & (688.30, 610.78, 500) & 816.17 & FP \& Spot\\ 
3794.01 & 1520 b & 0.65 & 3.27e-07 & \nodata & 0.101 $\pm$ 3e-03 & 2.37e-06 & (678.22, 632.70, 500) & 265.86 & Evap\\ 
\hline
\end{longtable}}
\end{landscape}
\begin{landscape}\thispagestyle{empty}
{\setlength{\tabcolsep}{2pt}\footnotesize\setlength{\LTleft}{-1.5cm plus -1fill}
\footnotetext[1]{Values from {\it Kepler} Object of Interest (KOI) Catalog Q1-Q17 DR 25(https://exoplanetarchive.ipac.caltech.edu/)}
\footnotetext[2]{\small FP=Possible False Positive (https://exofop.ipac.caltech.edu/cfop.php); GD = Gravity Darkening (B.1); Evap = Evaporating planet (B.2); Spot=Spot Crossing (B.3); Bad Fold = incorrect data folding (B.4); Small = non-significant signal (B.5); Others = B.6; }
\begin{longtable}{c c c c c c c c c c c c}\caption{Parameters and statistics of 16 systems with $0.001<p<0.05$}\label{final_table2} \\ \hline
 KOI& {\it Kepler}&$P^{1}_{\rm orb}$ (day)&$t_{\rm damp}$ (Gyr)& $(R_{\rm out}/R_{\rm p} )_{\rm upp, \;Aligned}$ & $(R_{\rm out}/R_{\rm p} )_{\rm upp, \;Saturn}$& $(R_{\rm out}/R_{\star})_{\rm upp}$& $(R_{\rm p}/R_{\star})_{\rm ringless}$& $p$ & ($\chi^{2}_{\rm ringless, min}$, $\chi^{2}_{\rm ring, min}$, $N_{\rm bin}$) & $(S/N)$& Comment$^{2}$ 
\endfirsthead
KOI& {\it Kepler} &$P_{\rm orb}$ (day)&$t_{\rm damp}$ (Gyr)& $(R_{\rm out}/R_{\rm p} )_{\rm upp, \;Aligned}$ & $(R_{
m out}/R_{
m p} )_{
m upp, \;Saturn}$& $(R_{\rm out}/R_{\star})_{\rm upp}$& $(R_{\rm p}/R_{\star})_{\rm ringless}$ &$p$ & ($\chi^{2}_{\rm ringless, min}$, $\chi^{2}_{\rm ring, min}$, $N_{\rm bin}$) & $(S/N)$& Comment \endhead \hline
4.01 & \nodata & 3.85 & 1.76e-04 & \nodata & \nodata & \nodata & 0.0394 $\pm$ 3e-04 & 3.44e-02 & (538.39, 525.32, 500) & 148.29 & FP \& Small\\ 
5.01 & \nodata & 4.78 & 6.05e-04 & \nodata & \nodata & \nodata & 0.04 $\pm$ 1e-02 & 3.15e-02 & (481.35, 469.45, 500) & 455.35 & FP \& Small\\ 
12.01 & 448 b & 17.86 & 1.44e-02 & 1.86 & 1.23 & 0.107 & 0.09018 $\pm$ 5e-05 & 1.23e-02 & (548.57, 532.47, 500) & 792.22 & Others\\ 
70.02 & 20 b & 3.70 & 1.23e-03 & 2.47 & 1.27 & 0.022 & 0.01799 $\pm$ 9e-05 & 1.34e-03 & (671.05, 644.41, 500) & 128.93 & Bad Fold\\ 
148.01 & 48 b & 4.78 & 2.46e-03 & 9.51 & 1.44 & 0.026 & 0.0196 $\pm$ 1e-04 & 1.13e-02 & (663.02, 643.29, 500) & 102.75 & Bad Fold\\ 
212.01 & \nodata & 5.70 & 1.04e-03 & \nodata & 1.72 & 0.093 & 0.0649 $\pm$ 3e-04 & 2.14e-02 & (525.10, 511.11, 500) & 159.08 & Small\\ 
214.01 & 424 b & 3.31 & 1.46e-04 & 2.99 & \nodata & \nodata & 0.104 $\pm$ 3e-03 & 2.37e-02 & (446.95, 435.27, 500) & 448.51 & Small\\ 
257.01 & 506 b & 6.88 & 5.14e-03 & 9.25 & 2.22 & 0.033 & 0.0224 $\pm$ 2e-04 & 3.84e-02 & (550.10, 537.05, 500) & 161.18 & Small\\ 
423.01 & 39 b & 21.09 & 2.12e-02 & 3.89 & 1.85 & 0.129 & 0.0890 $\pm$ 6e-04 & 1.66e-02 & (500.13, 486.18, 500) & 225.79 & Small\\ 
433.02 & 553 c & 328.24 & 7.94e+01 & 6.10 & \nodata & \nodata & 0.120 $\pm$ 7e-03 & 2.07e-02 & (250.09, 238.97, 303) & 112.04 & Small\\ 
531.01 & \nodata & 3.69 & 2.97e-04 & \nodata & \nodata & \nodata & 0.096 $\pm$ 4e-03 & 2.44e-02 & (469.53, 457.33, 500) & 183.64 & Small\\ 
686.01 & \nodata & 52.51 & 3.36e-01 & 4.68 & \nodata & \nodata & 0.118 $\pm$ 4e-03 & 4.38e-02 & (46.09, 31.21, 40) & 153.64 & FP \& Small\\ 
872.01 & 46 b & 33.60 & 1.36e-01 & \nodata & \nodata & \nodata & 0.084 $\pm$ 2e-03 & 4.57e-02 & (486.30, 475.20, 500) & 173.00 & Small\\ 
1131.01 & \nodata & 0.70 & 2.08e-07 & 1.02 & 1.02 & 0.211 & 0.2 $\pm$ 7e-02 & 3.84e-02 & (549.24, 536.21, 500) & 727.17 & FP \& Small\\ 
1353.01 & 289 c & 125.87 & 4.46e+00 & 4.23 & 2.00 & 0.153 & 0.1048 $\pm$ 6e-04 & 9.04e-03 & (544.28, 525.28, 442) & 198.14 & Spot \\ 
6016.01 & \nodata & 4.55 & 7.76e-05 & 1.32 & 1.65 & 0.463 & 0.23 $\pm$ 2e-02 & 9.91e-03 & (503.24, 487.04, 472) & 1719.88 & FP \& Spot\\ 
\hline
\end{longtable}}
\end{landscape}
\begin{landscape}\thispagestyle{empty}
{\setlength{\tabcolsep}{2pt}\footnotesize\setlength{\LTleft}{-1.5cm plus -1fill}
\footnotetext[1]{Values from {\it Kepler} Object of Interest (KOI) Catalog Q1-Q17 DR 25(https://exoplanetarchive.ipac.caltech.edu/)}
\footnotetext[2]{\small FP=Possible False Positive (https://exofop.ipac.caltech.edu/cfop.php); GD = Gravity Darkening (B.1); Evap = Evaporating planet (B.2); Spot=Spot Crossing (B.3); Bad Fold = incorrect data folding (B.4); Small = non-significant signal (B.5); Others = B.6; }
\begin{longtable}{c c c c c c c c c c c c}\caption{Parameters and statistics of 139 systems with $p>0.05$}\label{final_table3} \\ \hline
 KOI& {\it Kepler}&$P^{1}_{\rm orb}$ (day)&$t_{\rm damp}$ (Gyr)& $(R_{\rm out}/R_{\rm p} )_{\rm upp, \;Aligned}$ & $(R_{\rm out}/R_{\rm p} )_{\rm upp, \;Saturn}$& $(R_{\rm out}/R_{\star})_{\rm upp}$& $(R_{\rm p}/R_{\star})_{\rm ringless}$& $p$ & ($\chi^{2}_{\rm ringless, min}$, $\chi^{2}_{\rm ring, min}$, $N_{\rm bin}$) & $(S/N)$& Comment$^{2}$ \\\hline
\endfirsthead
\hline
KOI& {\it Kepler} &$P_{\rm orb}$ (day)&$t_{\rm damp}$ (Gyr)& $(R_{\rm out}/R_{\rm p} )_{\rm upp, \;Aligned}$ & $(R_{\rm out}/R_{\rm p} )_{\rm upp, \;Saturn}$& $(R_{\rm out}/R_{\star})_{\rm upp}$& $(R_{\rm p}/R_{\star})_{\rm ringless}$ &$p$ & ($\chi^{2}_{\rm ringless, min}$, $\chi^{2}_{\rm ring, min}$, $N_{\rm bin}$) & $(S/N)$& Comment \\\hline \endhead 
\hline \endfoot \endlastfoot
\thispagestyle{empty} 1.01 & 1 b & 2.47 & 4.20e-05 & 1.11 & 1.01 & 0.127 & 0.1259 $\pm$ 7e-04 & 1.00e+00 & (511.76, 511.76, 500) & 9353.67 & \nodata\\ 
\thispagestyle{empty} 7.01 & 4 b & 3.21 & 3.42e-04 & 5.99 & 1.36 & 0.031 & 0.02445 $\pm$ 7e-05 & 3.41e-01 & (572.68, 566.10, 500) & 284.12 & \nodata\\ 
\thispagestyle{empty} 10.01 & 8 b & 3.52 & 1.11e-04 & 1.78 & 1.20 & 0.109 & 0.0940 $\pm$ 4e-04 & 7.51e-01 & (491.06, 488.39, 500) & 1565.89 & \nodata\\ 
\thispagestyle{empty} 17.01 & 6 b & 3.23 & 9.43e-05 & 3.71 & 1.15 & 0.105 & 0.0932 $\pm$ 1e-04 & 3.60e-01 & (456.00, 450.92, 500) & 2671.28 & \nodata\\ 
\thispagestyle{empty} 18.01 & 5 b & 3.55 & 1.12e-04 & 2.14 & 1.14 & 0.088 & 0.0790 $\pm$ 1e-04 & 1.36e-01 & (482.37, 474.17, 500) & 2054.24 & \nodata\\ 
\thispagestyle{empty} 20.01 & 12 b & 4.44 & 1.69e-04 & 1.81 & 1.20 & 0.138 & 0.1179 $\pm$ 1e-04 & 4.81e-01 & (583.61, 578.28, 500) & 3127.91 & \nodata\\ 
\thispagestyle{empty} 22.01 & 422 b & 7.89 & 1.34e-03 & 1.95 & 1.82 & 0.135 & 0.0956 $\pm$ 2e-04 & 1.33e-01 & (473.51, 465.41, 500) & 1886.93 & \nodata\\ 
\thispagestyle{empty} 42.01 & 410 A b & 17.83 & 9.29e-02 & 8.28 & \nodata & \nodata & 0.0169 $\pm$ 2e-04 & 1.00e+00 & (484.28, 484.20, 500) & 144.80 & \nodata\\ 
\thispagestyle{empty} 46.01 & 101 b & 3.49 & 3.06e-04 & 9.11 & 2.09 & 0.048 & 0.0320 $\pm$ 2e-04 & 2.20e-01 & (451.10, 444.68, 500) & 152.32 & \nodata\\ 
\thispagestyle{empty} 64.01 & \nodata & 1.95 & 2.93e-05 & 2.81 & \nodata & \nodata & 0.04 $\pm$ 1e-02 & 1.45e-01 & (532.93, 524.06, 500) & 286.21 & \nodata\\ 
\thispagestyle{empty} 69.01 & 93 b & 4.73 & 2.94e-03 & 2.78 & \nodata & \nodata & 0.0157 $\pm$ 2e-04 & 1.00e+00 & (525.95, 526.55, 500) & 265.72 & \nodata\\ 
\thispagestyle{empty} 70.01 & 20 c & 10.85 & 1.62e-02 & 9.25 & 1.65 & 0.040 & 0.0289 $\pm$ 1e-04 & 6.40e-01 & (482.40, 479.07, 500) & 225.66 & \nodata\\ 
\thispagestyle{empty} 75.01 & \nodata & 105.88 & 3.63e+00 & 3.20 & \nodata & \nodata & 0.0378 $\pm$ 2e-04 & 7.06e-01 & (575.70, 572.23, 500) & 208.26 & \nodata\\ 
\thispagestyle{empty} 82.01 & 102 e & 16.15 & 6.07e-02 & 5.24 & \nodata & \nodata & 0.0289 $\pm$ 5e-04 & 9.92e-01 & (516.92, 516.39, 500) & 248.53 & \nodata\\ 
\thispagestyle{empty} 84.01 & 19 b & 9.29 & 1.50e-02 & 9.23 & 1.63 & 0.033 & 0.02376 $\pm$ 9e-05 & 8.74e-01 & (490.32, 488.50, 500) & 220.22 & \nodata\\ 
\thispagestyle{empty} 85.01 & 65 c & 5.86 & 3.38e-03 & 6.53 & 1.38 & 0.021 & 0.01652 $\pm$ 5e-05 & 4.95e-01 & (553.43, 548.49, 500) & 191.04 & \nodata\\ 
\thispagestyle{empty} 94.01 & 89 d & 22.34 & 3.70e-02 & 3.71 & 1.40 & 0.089 & 0.0695 $\pm$ 2e-04 & 3.25e-01 & (492.67, 486.86, 500) & 858.57 & \nodata\\ 
\thispagestyle{empty} 94.02 & 89 c & 10.42 & 1.20e-02 & 8.84 & \nodata & \nodata & 0.0255 $\pm$ 1e-04 & 7.58e-01 & (548.33, 545.41, 500) & 165.65 & \nodata\\ 
\thispagestyle{empty} 94.03 & 89 e & 54.32 & 8.66e-01 & \nodata & \nodata & \nodata & 0.0409 $\pm$ 5e-04 & 4.59e-01 & (480.16, 475.61, 500) & 201.68 & \nodata\\ 
\thispagestyle{empty} 97.01 & 7 b & 4.89 & 2.53e-04 & 1.55 & \nodata & \nodata & 0.0823 $\pm$ 1e-04 & 3.32e-01 & (518.47, 512.42, 500) & 1580.40 & \nodata\\ 
\thispagestyle{empty} 98.01 & 14 b & 6.79 & 1.21e-03 & 2.18 & \nodata & \nodata & 0.0455 $\pm$ 1e-04 & 6.91e-01 & (518.91, 515.68, 500) & 588.69 & \nodata\\ 
\thispagestyle{empty} 100.01 & \nodata & 9.97 & 2.14e-03 & \nodata & \nodata & \nodata & 0.055 $\pm$ 3e-03 & 7.82e-01 & (541.86, 539.14, 500) & 188.32 & \nodata\\ 
\thispagestyle{empty} 103.01 & \nodata & 14.91 & 4.75e-02 & \nodata & 1.88 & 0.040 & 0.0271 $\pm$ 2e-04 & 5.61e-01 & (458.09, 454.44, 500) & 114.38 & \nodata\\ 
\thispagestyle{empty} 104.01 & 94 b & 2.51 & 1.93e-04 & 3.03 & \nodata & \nodata & 0.0390 $\pm$ 7e-04 & 8.64e-01 & (520.19, 518.18, 500) & 276.45 & \nodata\\ 
\thispagestyle{empty} 105.01 & 463 b & 8.98 & 8.95e-03 & \nodata & 1.58 & 0.041 & 0.0300 $\pm$ 2e-04 & 3.57e-01 & (535.74, 529.75, 500) & 175.89 & \nodata\\ 
\thispagestyle{empty} 108.01 & 103 b & 15.97 & 4.70e-02 & \nodata & \nodata & \nodata & 0.0212 $\pm$ 5e-04 & 7.25e-02 & (507.09, 496.74, 500) & 109.94 & \nodata\\ 
\thispagestyle{empty} 108.02 & 103 c & 179.61 & 3.21e+01 & \nodata & \nodata & \nodata & 0.0335 $\pm$ 7e-04 & 9.73e-01 & (538.74, 537.79, 500) & 118.02 & \nodata\\ 
\thispagestyle{empty} 111.01 & 104 b & 11.43 & 2.53e-02 & 9.43 & \nodata & \nodata & 0.0208 $\pm$ 6e-04 & 9.95e-01 & (482.47, 482.07, 500) & 149.05 & \nodata\\ 
\thispagestyle{empty} 111.02 & 104 c & 23.67 & 2.16e-01 & \nodata & \nodata & \nodata & 0.0205 $\pm$ 7e-04 & 1.00e+00 & (507.95, 508.06, 500) & 102.89 & \nodata\\ 
\thispagestyle{empty} 111.03 & 104 d & 51.76 & 1.86e+00 & \nodata & \nodata & \nodata & 0.0230 $\pm$ 2e-04 & 6.83e-01 & (506.56, 503.35, 500) & 104.53 & \nodata\\ 
\thispagestyle{empty} 115.01 & 105 b & 5.41 & 2.21e-03 & 4.81 & \nodata & \nodata & 0.0243 $\pm$ 5e-04 & 1.00e+00 & (533.29, 534.26, 500) & 164.82 & \nodata\\ 
\thispagestyle{empty} 119.01 & 108 b & 49.18 & 4.74e-01 & 4.79 & \nodata & \nodata & 0.0403 $\pm$ 5e-04 & 7.28e-01 & (443.98, 441.43, 500) & 146.21 & \nodata\\ 
\thispagestyle{empty} 122.01 & 95 b & 11.52 & 1.92e-02 & \nodata & 1.63 & 0.028 & 0.0203 $\pm$ 1e-04 & 3.48e-01 & (612.05, 605.09, 500) & 124.25 & \nodata\\ 
\thispagestyle{empty} 123.01 & 109 b & 6.48 & 4.52e-03 & 4.25 & \nodata & \nodata & 0.0179 $\pm$ 4e-04 & 9.59e-01 & (567.41, 566.21, 500) & 104.11 & \nodata\\ 
\thispagestyle{empty} 125.01 & 468 b & 38.48 & 1.24e-01 & 6.66 & 1.21 & 0.165 & 0.1396 $\pm$ 5e-04 & 1.71e-01 & (539.81, 531.35, 500) & 586.79 & FP\\ 
\thispagestyle{empty} 127.01 & 77 b & 3.58 & 1.58e-04 & 1.81 & 1.31 & 0.120 & 0.0981 $\pm$ 3e-04 & 5.86e-01 & (497.21, 493.42, 500) & 970.07 & \nodata\\ 
\thispagestyle{empty} 128.01 & 15 b & 4.94 & 3.96e-04 & 1.79 & \nodata & \nodata & 0.1026 $\pm$ 5e-04 & 2.50e-01 & (474.56, 468.17, 500) & 1113.76 & \nodata\\ 
\thispagestyle{empty} 129.01 & 470 b & 24.67 & 4.53e-04 & 2.78 & 1.76 & 0.114 & 0.0805 $\pm$ 4e-04 & 4.45e-01 & (485.79, 481.08, 500) & 211.74 & \nodata\\ 
\thispagestyle{empty} 130.01 & \nodata & 34.19 & 5.36e-02 & 2.95 & \nodata & \nodata & 0.1142 $\pm$ 7e-04 & 6.75e-02 & (457.67, 448.15, 500) & 785.48 & FP\\ 
\thispagestyle{empty} 131.01 & 471 b & 5.01 & 4.38e-04 & 2.65 & 1.61 & 0.105 & 0.0765 $\pm$ 4e-04 & 2.37e-01 & (453.50, 447.26, 500) & 333.34 & FP\\ 
\thispagestyle{empty} 135.01 & 43 b & 3.02 & 8.54e-05 & 1.88 & \nodata & \nodata & 0.0855 $\pm$ 2e-04 & 3.59e-01 & (442.14, 437.21, 500) & 1246.59 & \nodata\\ 
\thispagestyle{empty} 137.01 & 18 c & 7.64 & 3.37e-03 & 6.62 & 1.55 & 0.058 & 0.0426 $\pm$ 1e-04 & 9.21e-01 & (485.86, 484.43, 500) & 425.15 & \nodata\\ 
\thispagestyle{empty} 137.02 & 18 d & 14.86 & 1.90e-02 & 3.51 & \nodata & \nodata & 0.0541 $\pm$ 5e-04 & 4.42e-01 & (460.59, 456.11, 500) & 444.75 & \nodata\\ 
\thispagestyle{empty} 139.01 & 111 c & 224.78 & 4.17e+01 & \nodata & \nodata & \nodata & 0.053 $\pm$ 1e-03 & 2.45e-01 & (471.58, 465.18, 500) & 127.91 & \nodata\\ 
\thispagestyle{empty} 141.01 & \nodata & 2.62 & 1.38e-04 & 6.64 & \nodata & \nodata & 0.055 $\pm$ 2e-03 & 1.83e-01 & (570.48, 561.75, 500) & 325.66 & \nodata\\ 
\thispagestyle{empty} 143.01 & \nodata & 22.65 & 5.32e-02 & 7.41 & \nodata & \nodata & 0.06 $\pm$ 2e-02 & 9.99e-01 & (479.24, 479.01, 500) & 115.81 & FP\\ 
\thispagestyle{empty} 144.01 & 472 b & 4.18 & 9.22e-04 & 2.86 & \nodata & \nodata & 0.0357 $\pm$ 8e-04 & 9.96e-01 & (512.22, 511.84, 500) & 206.94 & \nodata\\ 
\thispagestyle{empty} 148.02 & 48 c & 9.67 & 1.31e-02 & \nodata & 1.57 & 0.038 & 0.0280 $\pm$ 2e-04 & 2.60e-01 & (509.73, 503.01, 500) & 153.76 & \nodata\\ 
\thispagestyle{empty} 149.01 & 473 b & 14.56 & 2.67e-02 & \nodata & \nodata & \nodata & 0.0286 $\pm$ 2e-04 & 9.36e-01 & (530.61, 529.21, 500) & 105.77 & \nodata\\ 
\thispagestyle{empty} 150.01 & 112 b & 8.41 & 1.03e-02 & \nodata & 1.52 & 0.035 & 0.0263 $\pm$ 2e-04 & 6.78e-02 & (512.70, 502.05, 500) & 108.37 & \nodata\\ 
\thispagestyle{empty} 152.01 & 79 d & 52.09 & 6.40e-01 & \nodata & \nodata & \nodata & 0.0506 $\pm$ 6e-04 & 2.86e-01 & (462.65, 456.81, 500) & 166.31 & \nodata\\ 
\thispagestyle{empty} 153.01 & 113 c & 8.93 & 1.12e-02 & \nodata & \nodata & \nodata & 0.031 $\pm$ 1e-03 & 9.01e-01 & (494.06, 492.45, 500) & 132.84 & \nodata\\ 
\thispagestyle{empty} 153.02 & 113 b & 4.75 & 2.13e-03 & \nodata & \nodata & \nodata & 0.0254 $\pm$ 7e-04 & 9.98e-01 & (595.02, 594.71, 500) & 136.46 & \nodata\\ 
\thispagestyle{empty} 156.03 & 114 d & 11.78 & 2.74e-02 & \nodata & \nodata & \nodata & 0.035 $\pm$ 1e-03 & 4.26e-01 & (504.95, 499.90, 500) & 156.89 & \nodata\\ 
\thispagestyle{empty} 157.01 & 11 c & 13.02 & 2.93e-02 & \nodata & \nodata & \nodata & 0.0253 $\pm$ 3e-04 & 6.32e-01 & (495.87, 492.40, 500) & 120.27 & \nodata\\ 
\thispagestyle{empty} 157.02 & 11 d & 22.69 & 1.32e-01 & \nodata & \nodata & \nodata & 0.0275 $\pm$ 3e-04 & 9.20e-01 & (485.51, 484.09, 500) & 121.48 & \nodata\\ 
\thispagestyle{empty} 157.03 & 11 e & 32.00 & 2.72e-01 & \nodata & \nodata & \nodata & 0.0381 $\pm$ 7e-04 & 5.94e-01 & (535.66, 531.63, 500) & 131.42 & \nodata\\ 
\thispagestyle{empty} 161.01 & 475 b & 3.11 & 4.52e-04 & 2.78 & \nodata & \nodata & 0.0309 $\pm$ 7e-04 & 9.80e-01 & (525.60, 524.80, 500) & 173.70 & \nodata\\ 
\thispagestyle{empty} 182.01 & \nodata & 3.48 & 1.15e-04 & 1.61 & 1.17 & 0.154 & 0.1359 $\pm$ 3e-04 & 9.56e-01 & (532.88, 531.71, 500) & 907.22 & FP\\ 
\thispagestyle{empty} 183.01 & 423 b & 2.68 & 5.46e-05 & 1.79 & 1.20 & 0.144 & 0.1240 $\pm$ 2e-04 & 7.58e-01 & (531.38, 528.53, 500) & 1927.67 & \nodata\\ 
\thispagestyle{empty} 186.01 & 485 b & 3.24 & 9.44e-05 & 5.40 & 1.29 & 0.145 & 0.1177 $\pm$ 5e-04 & 7.09e-02 & (463.20, 453.68, 500) & 907.61 & \nodata\\ 
\thispagestyle{empty} 188.01 & 425 b & 3.80 & 1.89e-04 & 2.07 & \nodata & \nodata & 0.1137 $\pm$ 8e-04 & 6.82e-01 & (505.55, 502.34, 500) & 763.67 & \nodata\\ 
\thispagestyle{empty} 189.01 & 486 b & 30.36 & 7.40e-02 & 3.96 & 1.64 & 0.182 & 0.132 $\pm$ 1e-03 & 8.43e-01 & (491.38, 489.33, 500) & 495.16 & FP\\ 
\thispagestyle{empty} 191.01 & 487 b & 15.36 & 1.05e-02 & 2.85 & \nodata & \nodata & 0.1130 $\pm$ 6e-04 & 1.93e-01 & (498.81, 491.33, 500) & 722.16 & \nodata\\ 
\thispagestyle{empty} 192.01 & 427 b & 10.29 & 2.75e-03 & 5.64 & 1.29 & 0.110 & 0.0892 $\pm$ 2e-04 & 6.18e-01 & (575.01, 570.87, 500) & 577.13 & \nodata\\ 
\thispagestyle{empty} 194.01 & 488 b & 3.12 & 7.67e-05 & 1.36 & 1.14 & 0.145 & 0.1346 $\pm$ 4e-04 & 4.36e-01 & (557.03, 551.56, 500) & 894.33 & \nodata\\ 
\thispagestyle{empty} 195.01 & 426 b & 3.22 & 1.05e-04 & 2.31 & \nodata & \nodata & 0.117 $\pm$ 1e-03 & 7.15e-01 & (510.76, 507.73, 500) & 806.33 & \nodata\\ 
\thispagestyle{empty} 196.01 & 41 b & 1.86 & 2.41e-05 & 1.75 & \nodata & \nodata & 0.1001 $\pm$ 5e-04 & 8.59e-02 & (567.27, 556.20, 500) & 932.49 & \nodata\\ 
\thispagestyle{empty} 197.01 & 489 b & 17.28 & 2.22e-02 & 9.19 & 1.52 & 0.124 & 0.0916 $\pm$ 8e-04 & 9.92e-02 & (495.08, 485.81, 500) & 370.23 & \nodata\\ 
\thispagestyle{empty} 199.01 & 490 b & 3.27 & 7.58e-05 & 4.05 & 1.41 & 0.120 & 0.0923 $\pm$ 4e-04 & 7.36e-01 & (506.74, 503.88, 500) & 652.81 & \nodata\\ 
\thispagestyle{empty} 200.01 & 74 b & 7.34 & 9.60e-04 & 2.80 & \nodata & \nodata & 0.0911 $\pm$ 7e-04 & 8.52e-01 & (514.93, 512.85, 500) & 430.98 & \nodata\\ 
\thispagestyle{empty} 201.01 & 491 b & 4.23 & 3.04e-04 & 2.23 & 1.97 & 0.116 & 0.0806 $\pm$ 5e-04 & 8.58e-01 & (441.44, 439.70, 500) & 657.38 & \nodata\\ 
\thispagestyle{empty} 202.01 & 412 b & 1.72 & 1.46e-05 & 1.34 & 1.18 & 0.117 & 0.103 $\pm$ 2e-03 & 7.27e-01 & (448.69, 446.11, 500) & 895.88 & \nodata\\ 
\thispagestyle{empty} 203.01 & 17 b & 1.49 & 8.41e-06 & 1.55 & 1.12 & 0.146 & 0.1323 $\pm$ 1e-04 & 4.32e-01 & (619.80, 613.66, 500) & 3014.03 & \nodata\\ 
\thispagestyle{empty} 204.01 & 44 b & 3.25 & 1.03e-04 & 2.79 & \nodata & \nodata & 0.0802 $\pm$ 8e-04 & 9.14e-02 & (527.39, 517.27, 500) & 348.47 & \nodata\\ 
\thispagestyle{empty} 205.01 & 492 b & 11.72 & 6.63e-03 & 3.53 & \nodata & \nodata & 0.097 $\pm$ 1e-03 & 2.29e-01 & (522.98, 515.67, 500) & 369.77 & \nodata\\ 
\thispagestyle{empty} 206.01 & 433 b & 5.33 & 4.32e-04 & 6.35 & 1.42 & 0.082 & 0.0633 $\pm$ 4e-04 & 6.08e-01 & (533.62, 529.71, 500) & 257.22 & \nodata\\ 
\thispagestyle{empty} 208.01 & 493 b & 3.00 & 9.31e-05 & 2.82 & 1.83 & 0.125 & 0.0865 $\pm$ 5e-04 & 7.29e-01 & (568.67, 565.42, 500) & 202.89 & \nodata\\ 
\thispagestyle{empty} 209.01 & 117 c & 50.79 & 3.85e-01 & 5.41 & 1.55 & 0.094 & 0.0698 $\pm$ 4e-04 & 7.41e-01 & (510.66, 507.82, 500) & 358.63 & \nodata\\ 
\thispagestyle{empty} 209.02 & 117 b & 18.80 & 3.14e-02 & 3.70 & \nodata & \nodata & 0.0466 $\pm$ 5e-04 & 2.40e-01 & (471.61, 465.15, 500) & 237.63 & \nodata\\ 
\thispagestyle{empty} 217.01 & 71 b & 3.91 & 1.60e-04 & 5.25 & 1.32 & 0.166 & 0.1334 $\pm$ 5e-04 & 2.44e-01 & (564.36, 556.70, 500) & 1059.21 & \nodata\\ 
\thispagestyle{empty} 229.01 & 497 b & 3.57 & 3.05e-04 & \nodata & \nodata & \nodata & 0.0505 $\pm$ 5e-04 & 2.48e-01 & (486.02, 479.46, 500) & 117.38 & \nodata\\ 
\thispagestyle{empty} 232.01 & 122 c & 12.47 & 1.25e-02 & \nodata & 1.49 & 0.059 & 0.0438 $\pm$ 2e-04 & 5.17e-01 & (469.98, 465.94, 500) & 269.24 & \nodata\\ 
\thispagestyle{empty} 244.01 & 25 c & 12.72 & 1.63e-02 & 4.23 & \nodata & \nodata & 0.03561 $\pm$ 9e-05 & 2.16e-01 & (522.14, 514.67, 500) & 429.53 & \nodata\\ 
\thispagestyle{empty} 244.02 & 25 b & 6.24 & 4.08e-03 & 8.56 & 1.33 & 0.023 & 0.01875 $\pm$ 6e-05 & 5.76e-01 & (459.81, 456.24, 500) & 223.07 & \nodata\\ 
\thispagestyle{empty} 245.01 & 37 d & 39.79 & 1.17e+00 & \nodata & \nodata & \nodata & 0.0227 $\pm$ 3e-04 & 9.98e-01 & (504.20, 503.92, 500) & 154.71 & \nodata\\ 
\thispagestyle{empty} 246.01 & 68 A b & 5.40 & 3.05e-03 & 7.72 & 1.53 & 0.023 & 0.01688 $\pm$ 4e-05 & 1.00e+00 & (490.53, 498.47, 500) & 249.46 & \nodata\\ 
\thispagestyle{empty} 250.01 & 26 b & 12.28 & 2.19e-02 & \nodata & \nodata & \nodata & 0.0480 $\pm$ 5e-04 & 1.18e-01 & (468.86, 460.52, 500) & 108.72 & \nodata\\ 
\thispagestyle{empty} 251.01 & 125 b & 4.16 & 1.03e-03 & \nodata & \nodata & \nodata & 0.0450 $\pm$ 9e-04 & 9.99e-01 & (522.41, 522.17, 500) & 136.30 & \nodata\\ 
\thispagestyle{empty} 254.01 & 45 b & 2.46 & 4.52e-05 & 1.90 & \nodata & \nodata & 0.1821 $\pm$ 9e-04 & 8.35e-01 & (481.66, 479.60, 500) & 1514.84 & \nodata\\ 
\thispagestyle{empty} 261.01 & 96 b & 16.24 & 5.82e-02 & 6.12 & \nodata & \nodata & 0.0261 $\pm$ 4e-04 & 3.23e-01 & (532.12, 525.82, 500) & 153.96 & \nodata\\ 
\thispagestyle{empty} 277.01 & 36 c & 16.23 & 4.68e-02 & \nodata & \nodata & \nodata & 0.0207 $\pm$ 1e-04 & 9.86e-01 & (527.45, 526.75, 500) & 139.68 & \nodata\\ 
\thispagestyle{empty} 279.01 & 450 b & 28.45 & 1.32e-01 & 6.63 & 1.85 & 0.050 & 0.0349 $\pm$ 2e-04 & 6.91e-01 & (531.31, 528.00, 500) & 206.16 & \nodata\\ 
\thispagestyle{empty} 280.01 & \nodata & 11.87 & 2.98e-02 & 9.31 & \nodata & \nodata & 0.0194 $\pm$ 2e-04 & 8.24e-01 & (527.11, 524.77, 500) & 126.35 & \nodata\\ 
\thispagestyle{empty} 282.01 & 130 c & 27.51 & 2.56e-01 & \nodata & \nodata & \nodata & 0.0236 $\pm$ 7e-04 & 3.94e-01 & (475.67, 470.66, 500) & 111.58 & \nodata\\ 
\thispagestyle{empty} 304.01 & 518 b & 8.51 & 1.15e-02 & \nodata & \nodata & \nodata & 0.0228 $\pm$ 2e-04 & 1.00e+00 & (489.96, 490.58, 500) & 108.97 & \nodata\\ 
\thispagestyle{empty} 314.01 & 138 c & 13.78 & 6.55e-02 & \nodata & \nodata & \nodata & 0.0249 $\pm$ 8e-04 & 1.00e+00 & (448.61, 448.73, 500) & 113.40 & \nodata\\ 
\thispagestyle{empty} 319.01 & \nodata & 46.15 & 3.17e-01 & 9.03 & \nodata & \nodata & 0.051 $\pm$ 9e-03 & 2.25e-01 & (523.15, 515.79, 500) & 137.89 & \nodata\\ 
\thispagestyle{empty} 351.02 & 90 g & 210.60 & 3.43e+01 & \nodata & 1.79 & 0.087 & 0.0597 $\pm$ 4e-04 & 8.91e-02 & (383.31, 374.62, 427) & 111.28 & \nodata\\ 
\thispagestyle{empty} 366.01 & \nodata & 75.11 & 4.79e-01 & 4.14 & \nodata & \nodata & 0.064 $\pm$ 2e-03 & 5.33e-01 & (375.88, 371.57, 367) & 172.90 & \nodata\\ 
\thispagestyle{empty} 367.01 & \nodata & 31.58 & 2.35e-01 & \nodata & 2.01 & 0.062 & 0.0422 $\pm$ 6e-04 & 6.63e-01 & (284.44, 281.38, 310) & 192.72 & \nodata\\ 
\thispagestyle{empty} 398.01 & 148 d & 51.85 & 4.65e-01 & 5.14 & \nodata & \nodata & 0.100 $\pm$ 3e-03 & 3.18e-01 & (493.58, 487.69, 500) & 184.33 & \nodata\\ 
\thispagestyle{empty} 433.01 & 553 b & 4.03 & 5.57e-04 & \nodata & 2.11 & 0.071 & 0.0478 $\pm$ 4e-04 & 9.36e-01 & (494.54, 493.24, 500) & 129.83 & \nodata\\ 
\thispagestyle{empty} 464.01 & 561 b & 58.36 & 8.60e-01 & \nodata & \nodata & \nodata & 0.068 $\pm$ 1e-03 & 4.90e-01 & (461.90, 457.74, 500) & 174.35 & \nodata\\ 
\thispagestyle{empty} 611.01 & \nodata & 3.25 & 1.32e-04 & \nodata & \nodata & \nodata & 0.11 $\pm$ 4e-02 & 3.84e-01 & (413.93, 409.50, 500) & 405.50 & \nodata\\ 
\thispagestyle{empty} 620.01 & 51 b & 45.16 & 3.75e-01 & \nodata & 1.73 & 0.104 & 0.0725 $\pm$ 5e-04 & 8.62e-01 & (525.96, 523.91, 500) & 120.00 & \nodata\\ 
\thispagestyle{empty} 620.02 & 51 d & 130.18 & 5.87e+00 & \nodata & 1.45 & 0.131 & 0.0985 $\pm$ 7e-04 & 5.02e-01 & (288.97, 284.74, 305) & 130.48 & \nodata\\ 
\thispagestyle{empty} 631.01 & 628 b & 15.46 & 1.39e-02 & 8.02 & \nodata & \nodata & 0.0617 $\pm$ 8e-04 & 9.62e-01 & (123.10, 121.88, 111) & 120.26 & \nodata\\ 
\thispagestyle{empty} 674.01 & 643 b & 16.34 & 1.44e-02 & \nodata & \nodata & \nodata & 0.0369 $\pm$ 3e-04 & 9.16e-01 & (454.41, 453.04, 500) & 137.26 & \nodata\\ 
\thispagestyle{empty} 676.02 & 210 b & 2.45 & 2.53e-04 & 7.88 & \nodata & \nodata & 0.0381 $\pm$ 6e-04 & 5.62e-01 & (483.27, 479.43, 500) & 265.21 & \nodata\\ 
\thispagestyle{empty} 680.01 & 435 b & 8.60 & 4.84e-04 & 2.31 & 1.89 & 0.090 & 0.0630 $\pm$ 3e-04 & 9.78e-01 & (504.98, 504.17, 500) & 468.57 & \nodata\\ 
\thispagestyle{empty} 760.01 & \nodata & 4.96 & 3.50e-04 & 9.70 & \nodata & \nodata & 0.112 $\pm$ 3e-03 & 2.52e-01 & (123.17, 117.00, 139) & 179.87 & \nodata\\ 
\thispagestyle{empty} 767.01 & 670 b & 2.82 & 6.55e-05 & 2.08 & 1.67 & 0.166 & 0.1200 $\pm$ 6e-04 & 5.49e-01 & (451.94, 448.26, 500) & 922.58 & \nodata\\ 
\thispagestyle{empty} 802.01 & \nodata & 19.62 & 1.70e-02 & 4.11 & \nodata & \nodata & 0.144 $\pm$ 1e-03 & 9.55e-01 & (407.32, 406.22, 414) & 239.43 & \nodata\\ 
\thispagestyle{empty} 806.01 & 30 d & 143.21 & 8.97e+00 & \nodata & 1.67 & 0.131 & 0.0922 $\pm$ 8e-04 & 4.28e-01 & (472.50, 467.79, 500) & 108.09 & \nodata\\ 
\thispagestyle{empty} 806.02 & 30 c & 60.32 & 5.04e-01 & 6.00 & 1.71 & 0.188 & 0.132 $\pm$ 2e-03 & 6.64e-01 & (484.77, 481.58, 500) & 305.49 & \nodata\\ 
\thispagestyle{empty} 824.01 & 693 b & 15.38 & 1.02e-02 & 5.75 & \nodata & \nodata & 0.121 $\pm$ 2e-03 & 5.43e-01 & (506.00, 501.84, 500) & 122.79 & \nodata\\ 
\thispagestyle{empty} 834.01 & 238 e & 23.65 & 5.62e-02 & 7.56 & \nodata & \nodata & 0.057 $\pm$ 1e-03 & 3.56e-01 & (502.90, 497.27, 500) & 162.09 & \nodata\\ 
\thispagestyle{empty} 841.02 & 27 c & 31.33 & 1.47e-01 & \nodata & \nodata & \nodata & 0.066 $\pm$ 2e-03 & 7.20e-01 & (502.77, 499.83, 500) & 107.50 & \nodata\\ 
\thispagestyle{empty} 880.02 & 82 c & 51.54 & 7.77e-01 & \nodata & \nodata & \nodata & 0.056 $\pm$ 2e-03 & 1.00e+00 & (488.24, 488.50, 500) & 118.13 & \nodata\\ 
\thispagestyle{empty} 883.01 & \nodata & 2.69 & 5.44e-05 & 1.64 & 1.27 & 0.217 & 0.1800 $\pm$ 8e-04 & 7.14e-02 & (493.31, 483.20, 500) & 1462.56 & \nodata\\ 
\thispagestyle{empty} 884.01 & 247 c & 9.44 & 7.18e-03 & \nodata & \nodata & \nodata & 0.0492 $\pm$ 4e-04 & 4.71e-01 & (473.11, 468.71, 500) & 159.90 & \nodata\\ 
\thispagestyle{empty} 889.01 & 75 b & 8.88 & 2.19e-03 & 7.55 & 1.59 & 0.160 & 0.114 $\pm$ 2e-03 & 4.02e-01 & (517.30, 511.93, 500) & 329.94 & \nodata\\ 
\thispagestyle{empty} 918.01 & 725 b & 39.64 & 1.66e-01 & 6.56 & 1.46 & 0.151 & 0.1143 $\pm$ 8e-04 & 5.62e-01 & (544.34, 540.01, 500) & 373.56 & \nodata\\ 
\thispagestyle{empty} 959.01 & \nodata & 12.71 & 2.50e-02 & 2.12 & 1.96 & 0.260 & 0.179 $\pm$ 1e-03 & 8.41e-01 & (92.72, 89.28, 65) & 1216.12 & FP\\ 
\thispagestyle{empty} 961.01 & 42 b & 1.21 & 9.26e-05 & 3.27 & \nodata & \nodata & 0.0446 $\pm$ 3e-04 & 9.61e-01 & (743.83, 742.28, 500) & 107.21 & \nodata\\ 
\thispagestyle{empty} 984.01 & \nodata & 4.29 & 1.41e-03 & \nodata & \nodata & \nodata & 0.031 $\pm$ 3e-03 & 2.26e-01 & (995.18, 981.19, 500) & 180.30 & \nodata\\ 
\thispagestyle{empty} 1074.01 & 762 b & 3.77 & 1.62e-04 & 4.20 & 1.44 & 0.137 & 0.1043 $\pm$ 4e-04 & 8.08e-02 & (568.08, 556.81, 500) & 439.51 & \nodata\\ 
\thispagestyle{empty} 1089.01 & 418 b & 86.68 & 1.92e+00 & \nodata & \nodata & \nodata & 0.083 $\pm$ 2e-03 & 7.55e-01 & (475.60, 473.04, 500) & 145.85 & \nodata\\ 
\thispagestyle{empty} 1426.02 & 297 c & 74.93 & 1.98e+00 & \nodata & \nodata & \nodata & 0.0632 $\pm$ 9e-04 & 6.60e-01 & (392.61, 389.30, 395) & 119.29 & \nodata\\ 
\thispagestyle{empty} 1448.01 & \nodata & 2.49 & 1.50e-05 & 1.68 & 1.27 & 0.230 & 0.1894 $\pm$ 4e-04 & 6.96e-01 & (657.90, 653.84, 500) & 1137.21 & FP\\ 
\thispagestyle{empty} 1456.01 & 855 b & 7.89 & 1.93e-03 & 3.23 & \nodata & \nodata & 0.0754 $\pm$ 6e-04 & 5.40e-01 & (504.84, 500.66, 500) & 226.00 & \nodata\\ 
\thispagestyle{empty} 1474.01 & 419 b & 69.73 & 8.96e-01 & \nodata & \nodata & \nodata & 0.0633 $\pm$ 7e-04 & 3.72e-01 & (458.00, 452.46, 453) & 185.06 & \nodata\\ 
\thispagestyle{empty} 1478.01 & 858 b & 76.14 & 2.86e+00 & \nodata & 1.78 & 0.070 & 0.0489 $\pm$ 2e-04 & 2.39e-01 & (400.00, 392.79, 382) & 122.21 & \nodata\\ 
\thispagestyle{empty} 1545.01 & \nodata & 5.91 & 5.65e-04 & 4.81 & 1.50 & 0.162 & 0.1212 $\pm$ 9e-04 & 5.52e-01 & (506.69, 502.59, 500) & 318.66 & \nodata\\ 
\thispagestyle{empty} 1547.01 & \nodata & 30.69 & 6.38e-02 & \nodata & \nodata & \nodata & 0.126 $\pm$ 2e-03 & 2.09e-01 & (166.69, 157.65, 139) & 110.46 & \nodata\\ 
\thispagestyle{empty} 1781.01 & 411 c & 7.83 & 5.03e-03 & 5.57 & \nodata & \nodata & 0.0420 $\pm$ 6e-04 & 7.47e-01 & (515.99, 513.16, 500) & 173.86 & \nodata\\ 
\thispagestyle{empty} 1784.01 & \nodata & 5.01 & 5.19e-04 & 8.43 & \nodata & \nodata & 0.3 $\pm$ 4e+01 & 3.61e-01 & (558.13, 551.92, 500) & 210.08 & FP\\ 
\thispagestyle{empty} 6969.01 & \nodata & 1.79 & 2.44e-06 & 1.13 & 1.02 & 0.242 & 0.2368 $\pm$ 3e-04 & 3.13e-01 & (527.82, 521.46, 500) & 2074.42 & FP\\ 
\hline \thispagestyle{empty}
\end{longtable}}
\end{landscape}

\end{document}